\documentclass[a4paper,10pt,twocolumn,twoside]{article}

\usepackage{graphicx,natbib,aas_macros}
\usepackage{fancyhdr}

\tt

\title{Novae II. Model, multi-band outburst, bipolar ejecta, accretion disk, relativistic electrons ....}

\author{Nimisha G. Kantharia \\
\normalsize National Centre for Radio Astrophysics, \\ 
\normalsize Tata Institute of Fundamental
Research, \\ \normalsize Post Bag 3, Ganeshkhind, Pune-411007, India   \\
\normalsize \it Email: nkprasadnetra@gmail.com \\
\normalsize \it URL: https://sites.google.com/view/ngkresearch/home}

\date{September 2017}

\begin{document}
\maketitle

\tableofcontents

\vspace{3cm}

\thispagestyle{empty}

\begin{abstract}
The study of novae is continued and a self-consistent updated physical model for classical/recurrent novae 
derived from multi-wavelength observations is presented.
In particular, observations of novae support the origin of the optical continuous emission
in the outburst ejecta, mass-based segregation and clump formation in the ejecta,
origin of the Orion, diffuse enhanced lines and 
dust in the clumps, prompt Fe II line formation in swept-up material, energising of electrons
to relativistic velocities by the explosion and the
existence of a large cool envelope around the accreting white dwarf in quiescence. 
The rapid transfer of thermonuclear energy should be adiabatic which energises and ejects all the particles
in the overlying layers.  Our study results in the following conclusions which are
relevant for novae and other astrophysical systems:  \\
(1) Electrons are instantaneously energised in the explosion alongside the heavier atoms and ions. 
In sufficiently energetic explosions, 
electrons should acquire highly relativistic velocities due to their small mass and emit synchrotron
emission in a magnetic field.  No post-ejection shock acceleration needs to be invoked. \\
(2) Rotation of an incompressible spherical accreting object modifies the effective potential felt by 
infalling particles. 
The combined effect of gravitational and centrifugal forces will lead to a latitude-dependent
potential such that the accretion rate will be maximum at the poles and minimum at the
equator.  The mass accumulation at the poles will exceed that at the equator leading to
the formation of a prolate-shaped or bipolar envelope around the object. 
Energetic expulsion of this envelope will result in a bipolar/prolate shaped ejecta/outflows.
Such outflows cannot be ejected from non-rotating spherical objects.  \\ 
(3) The latitude-dependent accretion rates in a rotating accreting object will also lead to accumulation
of the infalling matter outside the object in the non-polar regions thus forming an accretion disk.   
The angular momentum of the incoming matter plays no role in the formation of an accretion disk.  
Accretion disks cannot form around a non-rotating object. \\

\end{abstract}

\section{Introduction}
\label{intro}
Novae are highly energetic explosions on the accreting white dwarf in a
semi-detached binary system hosting a gaseous companion star.   Such an explosion in a classical
nova ejects matter and brightens the binary by 8-20 magnitudes within a day or so 
whereas it leads to brightening upto 6 magnitudes in a dwarf nova.  While the return to
brightness of the pre-outburst phase takes much longer, the explosion, 
energising and ejection of matter in classical novae are rapidly accomplished supporting
a dominantly adiabatic transfer of energy. 

Most nova outbursts follow the maximum magnitude relation with decline time
(MMRD) which is an inverse relation between the peak V band luminosity of a nova and its decline time
with the latter following an inverse relation with the ejecta velocity derived from the spectral lines.  
This relation should have
dispelled all doubts of a separate origin of the optical spectrum and continuum emission.  However,
literature continues to attribute the origin of the optical continuum to the pseudo-photosphere of 
the white dwarf inflated by the explosion and the spectrum to the ejecta.  
We also recall that the MMRD derived using simple physical arguments by
\citet{2017arXiv170304087K} was able to improve the MMRD fit to existing data on novae,
thus giving strong support to its validity. 
The observed relation between the ejecta velocity $v_{ej}$ and time taken by the light curve
to decline by two magnitudes $t_2$, which is used to calibrate the MMRD
is straightforward to understand if their origin is in 
the ejecta - a rapidly expanding ejecta (large $v_{ej}$) will quickly become optically thin
and find its emission measure dropping rapidly,
so that its light output will also decline faster (smaller $t_2$).  However this
relation is difficult to understand if the continuum emission arises in the pseudo-photosphere
as explained here.  A rapidly expanding ejecta surmised from the spectral lines would also 
indicate a fast expanding pseudo-photosphere and the optical continuum will quickly brighten.
However the decline in the light curve will now be attributed to the contraction of the
photosphere after it has expanded to some maximum size.  
In this case, it is difficult to understand why a fast ejecta should 
lead to a faster contraction of the photosphere and hence a faster drop in the optical continuum
i.e. shorter $t_2$.
One can also resort to a temperature change in the photosphere but again the problem is
why the temperature of the photosphere should change rapidly in a nova in which the ejecta is
rapidly expanding.  To explain the optical continuum from the photosphere requires contrived
explanations.  
The most obvious solution is that both the optical continuum emission and spectrum arise in the ejecta. 

We point to the empirical result that while classical and recurrent novae are found to
obey the MMRD, data on dwarf novae suggests a weak opposite 
correlation between peak luminosity and decline time, if any \citep[Figure 3a in][]{2017arXiv170304087K}.
This supports a different origin location of the optical continuum 
in dwarf novae and a strong contender is the expanded photosphere (outer surface of the
accreted envelope) due to energy injection.  
If the optical continuum was due to the isothermal expansion of a photosphere
then a brighter dwarf nova outburst would indicate a larger photosphere whereas a
faint dwarf nova outburst would indicate a smaller photosphere.  For example, consider a dwarf nova
outburst results in an envelope which is ten times the radius of the white dwarf core ($R_{WD}$)
as compared to another dwarf nova which is inflated to five times $R_{WD}$.
The black body luminosities of the two novae assuming both photospheres are at the same
temperature will differ by a factor of four.  Even if both the photospheres are contracting
at the same rate of say $0.1 R_{WD}$ per day then after 10 days, the
photospheres would have contracted by a $R_{WD}$.  This means that the radius of the envelope which
was $5 R_{WD}$ would be $4 R_{WD}$ i.e. 80\% of the maximum radius whereas for
the photosphere of $10 R_{WD}$, the radius will be $9 R_{WD}$ ie 90\% of the maximum
radius and the luminosities would have decreased to 64\% and 81\% of the peak respectively.  
The contraction of the larger photosphere (i.e. brighter dwarf nova outburst) 
takes longer so that $t_2$ is longer compared to the fainter dwarf nova with the smaller 
photosphere.  This, then, would explain why dwarf novae do not
follow MMRD but show a weak opposite trend.  Moreover the photospheric origin
also explains the observed flat maximum in dwarf novae - the photosphere having expanded
to a maximum size, will stop expanding but remain at that radial extent for some time
before it starts shrinking and the light curve declines. 
Thus, the behaviour of the MMRD in classical and dwarf novae encompass
the physical processes in these systems. 

The existing multi-wavelength observational results offer us an opportunity to coherently 
understand the evolution of a nova outburst and we attempt the same in this paper.    
There exist a range of ill-understood phenomena such as ejecta morphology 
so that some are spherical and others are aspherical, increasing blue-shifted velocity
displacement of absorption lines when the light curve is in the early decline phase,  
only some novae emit radio synchrotron, origin of energetic $\gamma-$rays, 
sites of dust formation, an old nova being brighter than its components 
etc. which we discuss in this paper and then update the existing model used to explain novae.  

Since a nova outburst is representative of an explosive phenomenon in which a huge quantity
of energy is pumped into a system in a short time, we extend our results to other similar
transient explosive systems such as supernovae and jets in active nuclei and find 
parallels so that several phenomena happening there can be explained by extending our nova model.  
All of this is also used to further our
understanding of basic physical processes such as accretion rates and formation of accretion disks, bipolar
outflows; electron acceleration and distribution of electron energies.  The results, some 
supporting existing astrophysical theories and some leading to a paradigm shift in existing
astrophysical theories within the framework of known physics are discussed in the paper. 
Throughout, the aim has been to adhere to the simplest picture that obeys the laws of physics,
explains observational results, is logically consistent and requires the minimum of assumptions 
and personal biases to be introduced. 

We begin by giving a background on novae and the well-defined behaviour of their light curves and spectra.
This is followed by a summary of the existing model used to explain novae and introduction to
white dwarfs.  Then we describe the updated model which is used to understand the 
observations of a few novae and end with conclusions.  

\section{Background on novae}
Soon after novae were identified, there existed suspicions that the star
on which the energetic event occurred and which was a blue star \citep{1938ApJ....88..228H,
1964ApJ...139..457K} was a white dwarf \citep[e.g.][]{1938ApJ....88..228H,1941PA.....49..292M,
1957ApJ...126...23G, 1959ApJ...130..110K, 1962ApJ...135..408K,1965AcA....15..197P} 
but the combination of brighter absolute magnitudes, distinct colours, 
larger estimated masses and lower densities of the blue star \citep{1938ApJ....88..228H, 1941PA.....49..292M} 
compared to isolated white dwarfs seems to have caused the blue star to be misidentified as
a hot sub-dwarf.  However as the number of novae in which a white dwarf could be unequivocally 
identified increased, it became clear that the blue star was a white dwarf.  
From the early studies it was clear that the explosion was very energetic which
led to upto million times increase in the luminosity of the star and several
different origins for the energy were explored ranging from collisions
between two stars to an origin in a thermonuclear explosion 
\citep[e.g.][]{1952MNRAS.112..598M, 1940ApJ....92..321M, 1964ASPL....9..137K,1965AcA....15..197P} 
which is now known to be the cause.  There existed doubts regarding the explosion 
being within the degenerate core of a white dwarf or on in the surface layer.
The energy output if the
explosion occurred within the degenerate core was estimated to resemble the energy release
in a supernova explosion \citep{1952MNRAS.112..598M} and was ruled out and the surface
explosion was favoured \citep[e.g.][]{1965AcA....15..197P,1978ARA&A..16..171G}.  Detection of expanding shells 
around old novae which show enhanced CNO abundance with respect to hydrogen as compared to
other nebulae \citep{1978ApJ...224..171W} have since supported the thermonuclear 
origin of the energy pulse in classical nova outbursts.  It was also found that several
of these explosions were in binary systems \citep{1947PASP...59...87S, 1954AJ.....59..326J,
1954PASP...66..230W,1962ApJ...135..408K}
and this seems to have been established as a basic property of novae when 
Kraft found that seven out of the ten old novae he had examined were unambiguously  
binary systems \citep{1964ApJ...139..457K}.  
It was also suggested that hydrogen fuel was being transferred from the red
companion to the blue star which was a semi-degenerate object on its way to becoming a
white dwarf and this was inducing superficial
outbursts on the blue star \citep{1964ASPL....9..137K} and the emission lines from a nova were
arising in the disk of material formed around the blue star \citep[e.g.][]{1959ApJ...130..110K}.
It is also worth noting that from early observations of novae it was obvious to astronomers that classical,
recurrent and dwarf novae consisted of the same basic components. 
It was suggested that the nature of the observed spectra of nova outbursts can be explained if they arose 
in an expanding sphere of hot gases that cooled as it expanded so that the spectral lines appeared in
absorption or emission and were broadened \citep{1901ApJ....13R.277P}.
From the observed character of the 
light curve, it was also suggested that the nova outburst whatever its cause led to a major ejection of
matter which was followed by continuous ejection in the form of winds \citep[e.g.][]{1943POMic...8..149M}. 
It is instructive to appreciate how the binary nature and its 
member stars, the origin of the outburst energy,  important physical processes etc have gradually
been unravelled from interpretation of observations constrained by physics so that our knowledge
on novae is now extensive.   

Classical novae are identified when in outburst.  The progenitor system can often be identified
in sky images taken before the outburst once its position is known.  The system observed outside
the outburst region is in its minimum state and observations of this state have been useful in
understanding several features of the binary.   We discuss this next. 

\subsection{Novae at minimum}
From a study of 16 old novae, \citet{1938ApJ....88..228H}
established that these were blue stars since all the objects showed
a strong continuum spectrum which extended upto the violet and half of them also showed
emission lines of hydrogen, carbon and neutral or ionized helium. 
No absorption lines were detected.  The temperatures for these blue stars  were estimated to be
between 20000 K and 50000 K similar to O or early B type main sequence stars \citep{1938ApJ....88..228H}.  
\citet{1941PA.....49..292M} followed up this work on novae in
quiescence and derived radii of $\rm \sim 0.1-0.2~R_\odot$ and densities of $\sim 10^3 \rho_\odot$ 
for the blue star using the Stephen-Boltzmann law. 
For comparison, the nominal radii of white dwarfs are $\rm \sim 0.01 R_\odot$ and densities are
$\rm \ge 10^4 \rho_\odot$.  Due to these different physical properties, 
both these authors concluded that the old novae could not be true white dwarfs but were sub-dwarf stars.
Since recurrent novae present quiescent periods between outbursts which represent both
the post-nova and pre-nova phases, these have been studied to understand the blue star.
The spectrum of the recurrent nova RS Ophiuchi from 1937 (after its outburst in 1933) was similar
to its spectrum from 1923 with both showing bright hydrogen and Fe II lines on
a continuous spectrum indicating both were the minimum spectrum - the only difference
seemed to be in the presence of the He II 4686A line
in 1937 which was absent in 1923.  However
the spectrum taken in 1936 was quite different \citep{1941PA.....49..292M} indicating that
the nova was not yet back to its inter-outburst state. 
A spectrum of T Pyxidis taken 14 years after its outburst in 1920 is similar to other
old novae showing a continuous spectrum with a bright He II 4686A line and faint hydrogen lines 
\citep{1941PA.....49..292M}.  Such studies that established the similarity between the 
spectra of recurrent novae in the inter-outburst period and classical novae at minimum
support similar physical processes in classical and recurrent nova outbursts. 
\citet{1941PA.....49..292M} argues that although there are differences
in the spectra of classical novae and U Geminorum stars (dwarf novae); there also are strong resemblances 
which argue for the underlying processes being similar in nature.  These studies helped
infer that pre-nova and post-nova stars have similar physical properties and
ruled out any model which involved destroying the star.  It argued for a superficial explosion on 
a star \citep{1943POMic...8..149M}. 

\subsection{Novae in outburst}

There is a sudden brightening of $8-20$ magnitudes in the optical bands when a nova goes
in outburst making several of them visible to the naked eye before their brightness starts
to decline.  While the rise to maximum is extremely rapid in most novae, they show a 
range in the decline rates using which they are classified into speed classes i.e.
slow and fast novae.  The light curves and spectra of novae evolve after the outburst
and exhibit similar pattern of evolution. 
We describe the observational details on the nova light curves and spectra.

\subsubsection{Optical light curve}

\begin{figure}
\centering
\includegraphics[width=9cm]{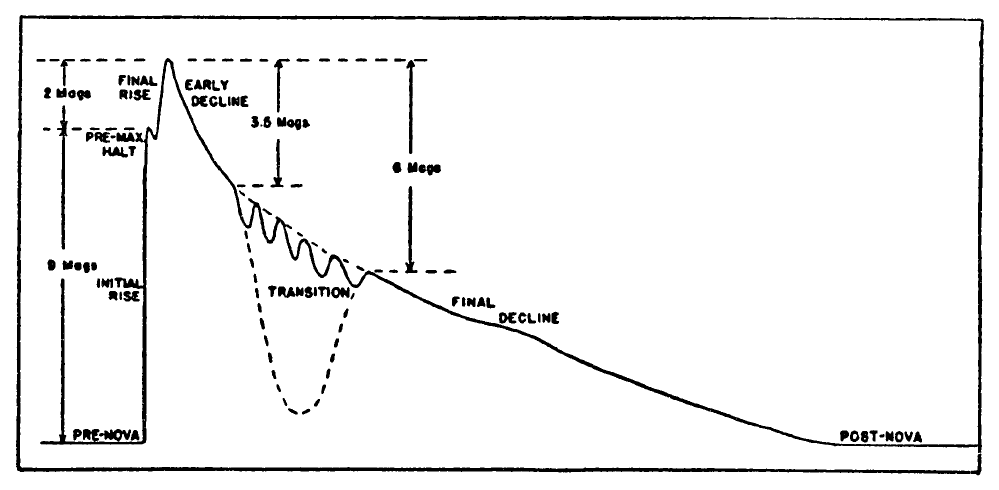}
\caption{A schematic showing the typical optical light curve of a nova reproduced 
from \citet{1943POMic...8..149M}.  The main phases in the evolution of a typical light curve
are labelled in the figure. }
\label{lightcurve}
\end{figure}

A typical optical light curve showing the evolution of light in a
nova outburst is schematically shown in Figure \ref{lightcurve} \citep{1943POMic...8..149M}. 
Light curves of novae were discussed in detail by \citet{1939PA.....47..410M,1939PA.....47..538M}.
Based on his own and other scientists' observational inferences, 
McLaughlin divided the light curve into different evolutionary phases between the pre-nova and
post-nova stages and identified them as the initial rise, pre-maximum halt, final rise, early decline, 
transition and final decline as labelled in Figure \ref{lightcurve}.  
All novae are observed to follow this kind of progression 
with the differences being limited to the presence of the pre-maximum halt
or the transition phase in the form of oscillations or a deep minimum. 
This, then gives a template to understand the underlying physical processes in the nova responsible
for the light evolution.
Most novae show similar spectral patterns in a particular phase of light curve evolution
and these are summarised in Table \ref{tab1} which is copied from \citet{1943POMic...8..149M}.
That most novae adhere to such a well-defined evolution indicates that it 
has an astrophysical origin and 
cannot be attributed to a random, chancy effect.

\begin{table*}
\caption{Table showing the correlated spectral and light curve phases.  Copied from
\citet{1943POMic...8..149M}. }
\begin{tabular}{l|l|c|l}
\hline
{\bf Absorption system} & {\bf Emission system} & {\bf Duration} & {\bf Part of} \\
 &                      & {\bf mag from max} & {\bf light curve}\\
\hline
1.Pre-maximum & Pre-maximum &  -  & Rise(decline) \\
2.Principal  & Principal    &  0.6 to 4.1 & Early decline \\
3.Diffuse enhanced & Diffuse enhanced & 1.2 to 3.0 & Early decline \\
4.Orion      & Orion(hazy bands)  & 2.1 to 3.3   & Early decline\\
5. Nitrogen(Orion) & `4640' (Orion) & 3 to 4.5 & Transition \\
6.  -         & Nebular(principal) & 4 to 11 & Transition-final decline \\
7. -          & Post-nova narrow stellar emissions & 8 to min.  & Final decline and post-nova\\
\hline
\end{tabular}
\label{tab1}
\end{table*}
 
We describe the different stages labelled in Figure \ref{lightcurve}. 
In the {\it initial rise}, the nova rapidly brightens from minimum to about two magnitudes 
below maximum which is a rise of 6-18 magnitudes in a day or so.  Some novae remain
at this stage, which is known as the {\it pre-maximum halt}, for a considerable time 
and this stage is typically observable in slow novae detected before maximum.  The {\it final rise} to
maximum brightness after the pre-maximum halt is slower than the initial rise.  
Most fast novae are detected just before or after the maximum brightness often leading to
uncertainty in the peak magnitude.  Slow novae spend a longer time
near maximum brightness whereas fast novae begin to decline soon after reaching
maximum.   The time a nova takes to decline by two or three magnitudes from peak brightness is known as
the decline times $\rm t_2$ and $\rm t_3$ and are used to identify the speed class of the nova. 
Around 3 to 3.5 magnitudes below maximum, the transition phase sets in when the light curve of some
novae show oscillations in the visible light whereas some slow novae show a steep drop in the
light curve (Figure \ref{lightcurve}).
The transition phase in novae was first identified as the phase in which the spectrum
changes from a stellar (absorption lines) to a nebular (emission lines) type 
\citep{1920MNRAS..80..540S}.
The change from a stellar spectrum to a nebular spectrum was found to be correlated
to the phase of the light curve which showed oscillations or a steep fall and hence referred to 
by the same name i.e. transition phase.  When the nova light stops oscillating or
when it recovers from the steep drop, it is generally $\ge 6$ magnitudes below maximum and into the
nebular phase characterised by emission lines and absence of absorption lines.  
The nova begins its final decline which is very slow in most novae and can take
upto a decade or longer.  
At the end of the final decline, the nova reverts to its pre-nova brightness. 
While all novae pass through the same light curve phases,
the time taken by different novae to evolve through the various phases of the light curve
differ.  Light curves of classical nova outbursts take several years to complete
their decline whereas recurrent novae outbursts get over in a year or so. 
The spectral changes (see Table \ref{tab1}) are generally observed to occur
close to the labelled light curve changes with respect to the maximum.
Such well-determined evolution of novae light curves and spectra indicate well-defined
processes at work.

\subsubsection{Optical spectra}
The evolution of the optical spectra of most novae in outburst is closely related to the light
curve evolution as quantified in Table \ref{tab1} and demonstrated by the efficacy of the MMRD.
The spectrum recorded soon after the outburst consists of absorption dips located on
the blue side of the wide emission bands which are seen to extend both bluewards and redwards of 
the rest frequency i.e.  P Cygni profiles although sometimes only the absorption component is detected.   
The excitation level, profile shape and 
velocity displacement of the absorption lines change as the nova evolves.  The changes were
identified and classified into a few well-defined stages
and the spectral evolutionary sequence of novae was presented  
\citep[e.g.][]{1920MNRAS..80..540S,1937ApJ....85..362M,1942ApJ....95..428M} 
as summarised in Table \ref{tab1}. 
Subsequent observations have given fresh inputs to this basic picture of spectral evolution of novae 
in addition to validating the basic model put forward to explain the nova spectra. 
\begin{figure}
\centering
\includegraphics[width=7cm]{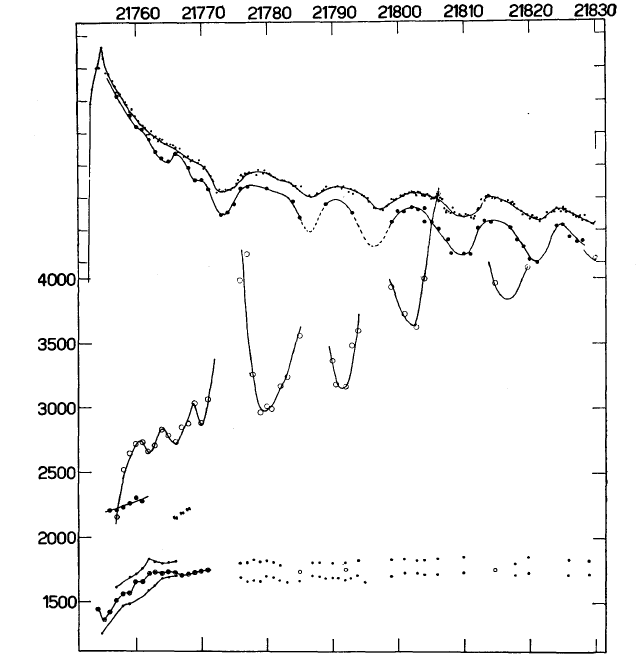}
\caption{Figure showing light curve of nova V603 Aquilae (upper curves) and the change in the velocity of 
absorption features (lower curves).   The velocity curve right below the light curve 
shows the variation in the displacement of the Orion line feature.  Note its anti-correlation
with the light changes.  Figure has been reproduced from \citet{1957gano.book.....G}.}
\label{spectra}
\end{figure}

\citet{1943POMic...8..149M} has described the joint evolution of the optical light curve and line spectrum 
in novae (Table \ref{tab1}).  The nova exhibits characterstics
typical of a supergiant of a spectral type which varies with its light curve evolution.  
In the pre-maximum phase, the nova spectrum resembles an early spectral type like B or A which evolves
to a relatively later type during and after maximum like spectral type F.  So, for example, GK Persei 
exhibited excitation typical of a B9 star in the pre-maximum phase, an A0 type during maximum
and A2 after maximum \citep[e.g.][]{1960stat.conf..585M} whereas DQ Herculis changed from B to A to F0 
\citep{1954ApJ...119..124M}.   An important difference between the nova spectrum
and that of a supergiant is the presence of absorption lines of oxygen and carbon in the nova spectrum
which is not typical of a supergiant star \citep{1943POMic...8..149M}.

McLaughlin divided the optical spectral lines observed from novae into four main systems based 
on the velocity displacement of
the absorption component and excitation of the spectrum namely the pre-maximum, principal, diffuse 
enhanced and Orion spectral systems.  The onset and disappearance of these systems were  
correlated with well-defined phases of light curve evolution as listed in Table \ref{tab1}. 
The following discussion is mainly based on \citet{1943POMic...8..149M} 
and \citet{1960stat.conf..585M}.  
The pre-maximum spectrum is observed before the light curve rises to maximum brightness and the lines
of hydrogen, Ca II, Na I, O I, C I are detected.  The absorption lines are blue-shifted and   
show the lowest velocity displacement compared to the lines which are detected later. 
Just after the maximum in the light curve, the principal system of spectral lines
appear and the absorption component is displaced to a larger blue-shifted velocity than the pre-maximum 
lines.  The pre-maximum system either disappears as soon as the principal system
shows up or at times is detected alongside the principal spectrum for a short while. 
The principal spectrum resembles that a star of type A or F.
The principal spectrum shows a composition similar to the pre-maximum spectrum with a few additional
lines of Mg II, Ti II, Fe II, He I, Si II.  The principal absorptions disappear about 4 magnitudes
below maximum.   Strong absorption lines fade faster so that lines of Mg II, 
O I, C I and Si II are the first to disappear whereas the hydrogen lines are the last to fade.  
Some of the principal absorptions are found to split into double or triple components.  
The principal emission bands are long-lived and are even detected from the 
nebular shells observed to expand around some old novae many years after the outburst.
After the light curve has declined by about a magnitude from maximum, 
the diffuse enhanced system of lines consisting
of hydrogen and metallic lines such as Ca II, Mg II, O I, Na I, Fe II appear.  These  absorption
features show a blue shift which is larger than the principal system by a factor of $1.5-2$. These
lines which are wide and hazy are often seen to split into multiple sharp components.
Around 2 magnitudes below optical maximum, the Orion system of lines generally appear consisting 
of higher excitation lines like O II, N II, He I, N III at similar or larger velocity displacement than
the diffuse enhanced lines.  The presence of higher excitation lines as observed in OB associations
like in the Orion star forming region prompted McLaughlin to name this system as the Orion system. 
An important peculiarity of the Orion system is that hydrogen is often missing in this system of lines unlike
the spectral systems that preceded it.  The velocity displacement of the Orion absorption lines
with respect to principal lines is 1.6 to 3.3 times and the emission components are wide and diffuse.  
These lines are not observed to split into several sharp components
although at times multiple wide components are identified.  The Orion
lines are more evident in slow novae than in fast novae.   The Orion lines are most sensitive
to light curve oscillations.  
The simultaneous existence of the principal, diffuse enhanced and Orion systems in the nova spectrum 
when the light curve is about 1-4 magnitudes below maximum, 
indicates the range of excitation and physical conditions that exist in a nova.
While the principal system of emission lines continue to be present, 
the diffuse enhanced and Orion system of lines disappear in the transition phase which
sets in when the nova has faded by about 3 magnitudes.   The principal absorption lines also disappear
in the transition phase.  The composition and excitation of the emission bands of the 
principal spectrum changes in the transition phase but the velocity displacement remains
the same.  The transition phase defines the changeover from a stellar to a nebular
spectrum i.e. all absorption features disappear leaving behind wide emission bands which also
engulf the absorption velocities. 
In the transition phase, the light curve of several novae show oscillations and alongwith the
light changes, the Orion system lines show correlated changes in excitation levels and 
velocity displacement as shown in Figure \ref{spectra}.  The excitation level and the
blue-shifted velocity displacement of the absorption lines increase when the light
curve undergoes a minimum in the oscillation and vice versa (Figure \ref{spectra}).
The velocity displacement of the principal system of lines are not observed to  
change with the light oscillations.  However in a few novae, the oscillating light is
seen to result in the oscillating detection of high excitation principal lines like
He II and N III such that lines appear at the secondary light minimum and fade at the
secondary maximum.  After the transition phase, the spectrum of the nova evolves to a
nebular nature.  When the nova is
on the final decline, the spectrum continues to be nebular with only emission lines typically
of [O III], N II, He II, [Ne III] and C II being detected with velocity displacements
and widths typically of the principal 
spectral system.   The spectrum also frequently shows the presence of high excitation coronal,
forbidden and Neon lines.  The nebular spectrum, which is fully developed around 7 magnitudes below
optical maximum, closely resembles that of a planetary nebula except for the larger emission linewidths
and some differences in the composition.
The nebular stage lines are detected as long as the expanding ejecta around the nova retains
its identity and is detectable.   When it has been possible to separate the emission from the
central star and the nebula in the later stages of evolution,
it is found that the emission bands arise in the expanding shell
whereas the continuum is predominantly from the central star and as the nebula
fades, only the spectrum of the star is detected \citep{1960stat.conf..585M}.

\begin{figure}
\centering
\includegraphics[width=9cm]{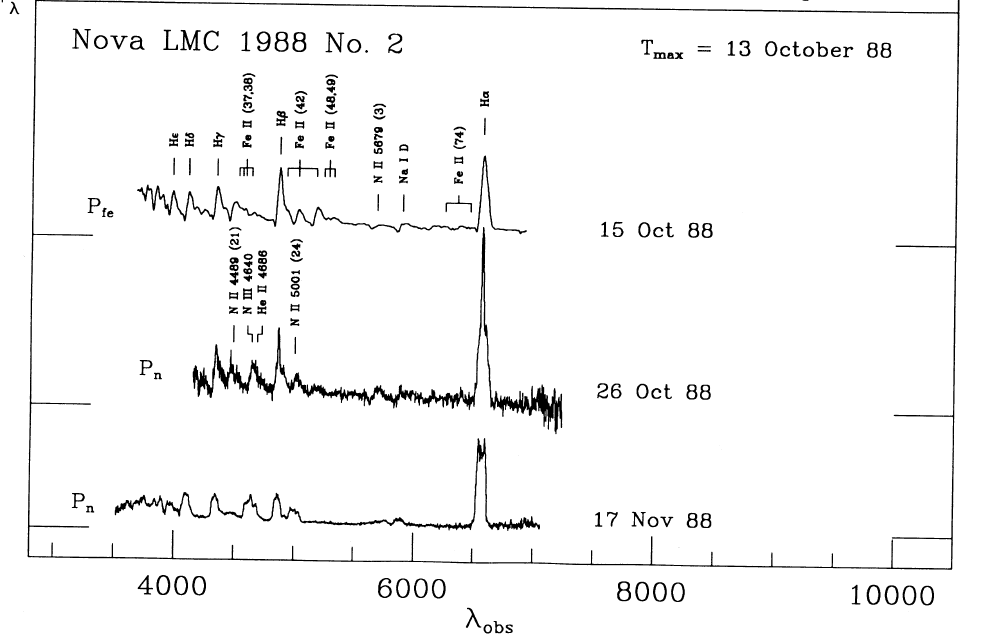}
\caption{ Typical spectra of a hybrid (Fe II and He/N) nova.  Figure reproduced from \citet{1991ApJ...376..721W}.}
\label{novaspectrum}
\end{figure}

\subsubsection{Fe~II and He/N novae}
In addition to the classification of spectral stages and light curve phases in the evolution
of a nova outburst described in the previous section, it was found that
novae could be classified into two types depending on the atomic species other 
than hydrogen which showed the
strongest lines in the optical band soon after maximum \citep{1991ApJ...376..721W, 1992AJ....104..725W}.  
Study of the post-maximum spectra of several novae revealed that the strongest spectral line (other than the
Balmer lines) in the nova spectra were either of Fe II or of helium and nitrogen which led to
the classification of novae into Fe~II and He/N classes \citep{1992AJ....104..725W}.  
Spectra of a nova in the Large Magellanic Cloud (LMC) taken at a few epochs 
following optical  maximum are shown in 
Figure \ref{novaspectrum}.  Several P Cygni profiles are detected in the spectrum taken 
two days after optical maximum which appears to be very different from the spectrum
taken 13 days after maximum indicating changes in the composition of the spectrum, line 
strengths and widths.  The non-Balmer dominant lines are of iron in the spectrum taken
two days after maximum which are replaced by lines of helium and nitrogen in the later spectra. 
The H$\alpha$ profile is seen to change from a gaussian shape to a double component line 
with a narrow feature perched on a broad feature to a double-peaked profile.  
While P~Cygni profiles are noticeable
in the spectrum in the top panel, only emission lines are detected in the spectrum in the lowest panel.   
In Nova LMC 1998 No 2 (see Figure \ref{novaspectrum}), features due to Fe~II and He/N were 
simultaneously detected and subsequently the Fe~II features were replaced by He/N features.
Such novae are termed as hybrid novae \citep{1992AJ....104..725W}.  The vice versa i.e. early
post-maximum detection of He/N and subsequent replacement by Fe~II is generally not observed. 
At this point, it is interesting to note that spectral observations of the slow recurrent nova T~Pyx 
in its pre-maximum phase revealed the presence of high ionization emission lines of 
helium, nitrogen and other metals which were soon replaced by P~Cygni lines of Fe~II near 
maximum \citep{2014AJ....147..107S}.  Since few novae are studied in the pre-maximum phase due to
its rapid rise, this result gives us a rare glimpse into the pre-maximum evolution of the nova
spectrum.  However it does not change the classification into Fe~II and He/N types which is based on
the post-maximum spectrum. 

The Fe~II class of novae display narrower lines 
(HWZI $< 2500$ kms$^{-1}$), P~Cygni profiles, lines of lower excitation and 
the prompt post-maximum light curve decline is generally slower ($t_2$ is long). The He/N class of novae 
are characterised by broad lines (HWZI $>2500$ kms$^{-1}$; often broader then 5000 kms$^{-1}$),
jagged flat-topped emission lines often without P~Cygni absorption, higher excitation and 
the post-maximum light curve decline is faster ($t_2$ small) \citep{1992AJ....104..725W}.
The above classification has also been extended into the near-infrared bands where the main
difference between the two types is found to lie in the presence of neutral carbon lines in
Fe~II type novae and absence of the same in He/N novae which is inferred to indicate the
different excitation conditions in the two types of novae \citep{2012BASI...40..243B}. 
Dust shows a tendency to form more often in Fe~II novae \citep[e.g.][]{2012BASI...40..243B}.

\subsection{Existing astrophysical model for novae}
Observations of novae in outburst have motivated a model 
which is summarised here.  Most of the following is based on \citet{1943POMic...8..149M}. 
\begin{figure}[t]
\centering
\includegraphics[width=8cm]{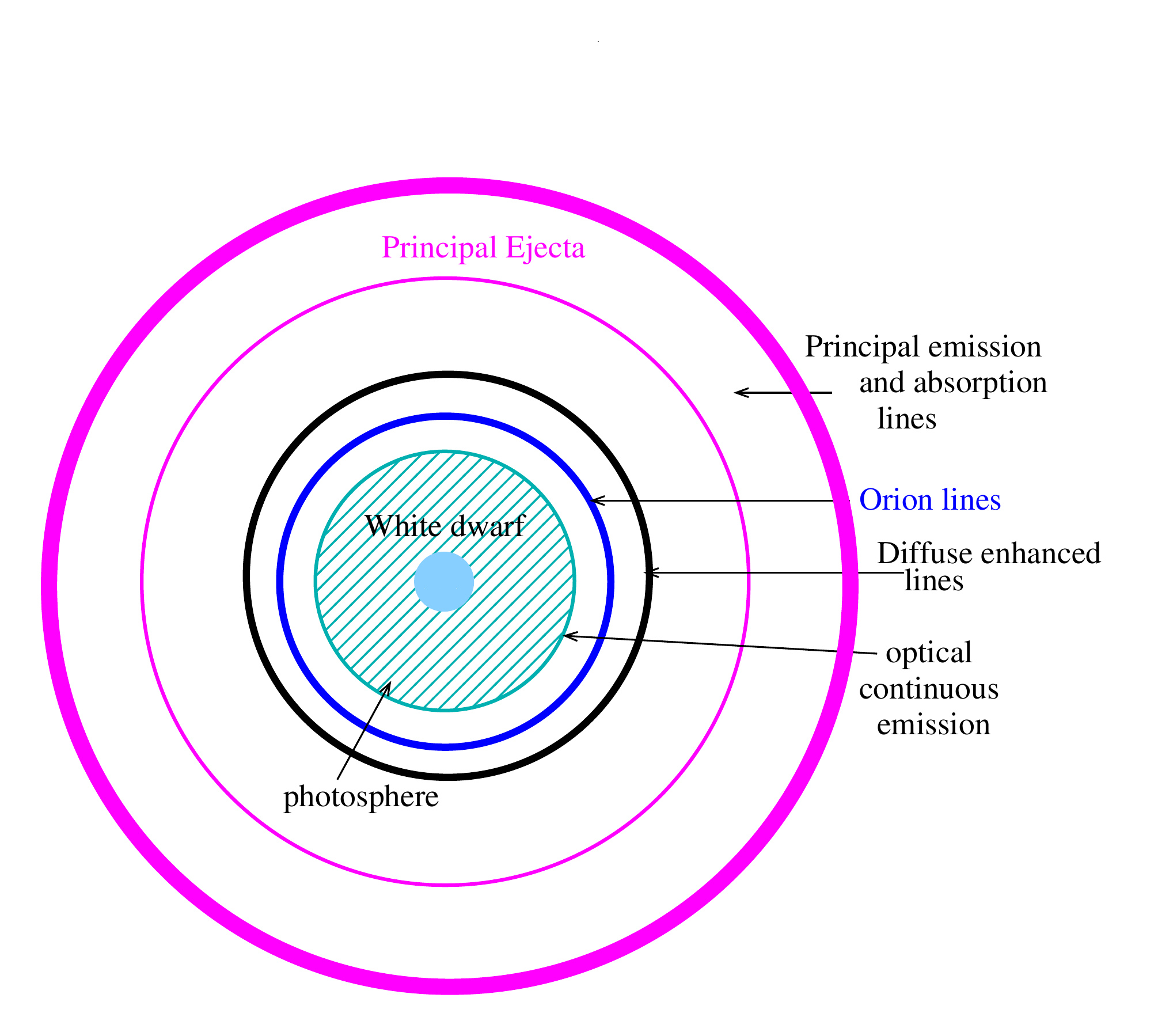}
\caption{Model summarised by \cite{1943POMic...8..149M} to explain observations of novae soon after
outburst upto about 4 magnitudes below optical maximum.  The principal ejecta is enclosed by
the magenta lines.  The diffuse enhanced (black) and Orion (blue) spectra arise in later
faster ejection from the central star.  Optical continuum comes from the expanded
photosphere of the central star. }
\label{novaold}
\end{figure}
It is interesting to note that the paper makes it a point to mention that the velocity displacements 
of the absorption lines and widths of the emission lines are due to the Doppler effect in an expanding
shell of gas as was suggested by \citet{1901ApJ....13R.277P}.  Although this is something we
take for granted today,  it was not obvious then and shows how our knowledge has gradually
built up over time from the sincere efforts of many scientists.
The model describes the outburst by a `main burst' in which most of the mass and energy
are expelled and which is followed by continuous expulsion 
of matter at a lower rate for a long time i.e. a wind.  Since there is no change in
the star, the model supports release of energy below the
superficial layer which then expands outward.  The inner regions of the expanding cloud are
dense and radiate as an effective photosphere while the outer layers become progressively tenuous.
Since the photosphere is an optical depth extent, it lags behind as the ejecta expands.
The observed optical continuous emission in this model arises from 
the photosphere of the white dwarf.   The changing spectral class in this model is explained by
the expanding photosphere which cools and changes the energy distribution in the 
continuum and hence the spectral class. 
At optical maximum, the photosphere and surrounding atmosphere will resemble a supergiant star with a
radius of $\rm \sim 150 R_\odot$.  A rapid drop in the rate
of ejection is inferred just before light maximum and around maximum, the ejecta is taken to detach 
from the photosphere.  The entire ejecta becomes optically thin after the optical maximum.
The continuous spectrum starts to decline after the maximum.  
The post-maximum model consists of a detached 
expanding optically thin ejecta and an attached cloud of a relatively smaller radial
extent around the star which is expanding at a rate which is 
lower than the rate at which the photospheric surface is moving inwards.
This inward motion of the photospheric surface is due to the expansion of the inner shell 
which leads to a decrease in the radial extent of the $\tau=1$ surface i.e. a contracting
photosphere due to decreasing densities.
The photosphere keeps shrinking within the smaller expanding cloud around the central star.
As the photosphere shrinks, the $\tau=1$ surface gets closer to the star and hence hotter. 
The model uses this argument to explain the observed increase in the colour temperature 
as the light declines.  For example, Nova Herculis 1934 and Nova Lacertae 1936,
had colour temperatures of about 6000 K near optical maximum which increased to
$\sim 30000$ K when the light dropped by four magnitudes after maximum \citep{1941PA.....49..292M}. 
McLaughlin suggests that the emergence of the principal spectrum with a larger blue shift 
and disappearance of the pre-maximum spectrum are due to the radiation pressure exerted by
the stellar radiation field on the inner dense layers which are then pushed forward at a 
higher velocity and leads to the entire outer shell being ejected and going optically thin. 
He explains the bright wide bands in the principal shell as being emission from the 
entire shell and the absorption lines being formed in the outer layers of the shell which absorb the light
of gas immediately behind them i.e. the regions of emission and absorption merge 
into one another in the principal ejecta shell.  However an alternate interpretation in literature is
that the absorption lines are formed due to absorption of the continuous light of the star. 
Irrespective of that, as the shell expands, the number of atoms which can absorb
keeps decreasing and most absorption lines weaken and eventually fade.  As densities fall below
the critical densities, forbidden lines appear in the spectra of several novae.  

In this model,  
winds continue to blow from the star well after the main burst which sets forth the ejecta.
The model locates the  diffuse enhanced and Orion spectral lines in these winds. 
These winds are postulated to blow out at even higher velocities than the main ejecta and hence
explain the higher blue shifted velocities of the diffuse enhanced and Orion systems 
(see Figure \ref{novaold}). The appearance of the diffuse enhanced 
P Cygni features happens when sufficient atoms have accumulated in the winds.  
The multiple sharp components which the diffuse enhanced lines often evolve into are explained
as forming in the winds moving outwards in the form of concentric independent shells and the broad 
diffuse absorption is produced
in the outer layers of the envelope.  The formation of Orion lines is explained as being in a later
but faster ejection from the star.  The shrinking photosphere exposes a still hotter surface which is 
inferred as the reason for the higher excitation of the spectral lines of the Orion system. 
It is suggested that the hard radiation field weakens the metallic lines of the diffuse enhanced 
system due to ionization.  The model explains that the multiple lines of the diffuse enhanced 
system  disappear when the shells become thin.
All through the spectral evolution, the photosphere around the white dwarf is contracting and
when the Orion system reaches
its maximum strength, McLaughlin estimates the photosphere to be $\rm \sim 10 R_\odot$.
As the excitation increases, N III absorption and 4640 emissions emerge as part of the Orion system
while the lines He I, O II, N II fade.  Since the diffuse enhanced and Orion systems are produced
in gas located just outside the photosphere in this model,
the larger emission line widths are attributed to enhanced turbulence near the star.
The disappearance of the Orion N III lines is attributed to the increased ionization due to 
the appearance of the hotter inner levels as the photosphere recedes towards the star.
It is suggested that subsequently the radiation field is so hard that the diffuse enhanced and
Orion lines disappear and no further observable lines are produced although
winds continue to be blown by the sub-dwarf star.  At this time,
the principal shell has expanded to several astronomical units and emission lines are 
detected from it as the nebular spectrum.  This model advocates continuous ejection of matter 
over a period of years after the main burst.
McLaughlin says that `The highest velocity of expulsion was reached in the '4640'
stage but the greatest violence (in terms of total momentum of the matter ejected in unit
time) occurred near light maximum in the eruption of the principal `shell' or `main burst.'
In this model, the secondary maxima of light observed after maximum and the oscillations in the transition
phase are seen as secondary outbursts which lead to temporary increase in ejection rates of the wind.
These increase the radius of the effective photosphere with an associated decrease in its
temperature.  The ejected gases also react to the transient change in the temperature of 
the photosphere which explains the observed correlations at the secondary maxima.  
The deep minimum in the transition phase exhibited by some slow novae (see Figure \ref{lightcurve}) 
was suggested to be due to dust formation in the ejecta \citep{1935PA.....43..323M,1937POMic...6..107M}
or presence of molecules \citep{1945MNRAS.105..275S}.  However the dust explanation was favoured since
molecular lines had seldom been detected in the nova spectra. 
The recovery of the light curve was taken to be indicative of dispersion of the dust so that
the obscuring agent had disappeared.  \citet{1943POMic...8..149M} ends his
model exposition by stating `Greater radiation pressure at the higher temperature may be
suggested tentatively as a possible cause of the velocity-magnitude correlation.'  
\citet{1941PA.....49..292M} also inferred that novae were hottest during the final
decline at a few magnitudes above quiescence and that the nebulous shell contributed
to the light curve from about six to nine magnitudes below maximum.  

This model was comprehensive and could explain most of the optical observations. 
A schematic which summarises some points of this model is shown in Figure \ref{novaold}.

\subsection{White dwarf primary}
\label{primary}
The existence of massive objects (white dwarfs) whose gravity
is balanced by the pressure of degenerate electrons was first suggested by \citet{1926MNRAS..87..114F}
and a detailed theoretical treatment which resulted in an upper mass limit for white dwarfs
was derived by \citet{1931ApJ....74...81C}.   A white dwarf consists of two distinct parts -
a degenerate core surrounded by a shallow layer of non-degenerate matter which obeys
the ideal gas laws and has a thickness of $\sim 10\%$ of the core radius \citep{1939isss.book.....C}.
The degenerate electron gas will have a high kinetic temperature ($10^7-10^8$ K) when it is formed.
The high conductivity of this degenerate electron gas ensures that the core is isothermal.  
A white dwarf will radiate, cool and finally settle down as a black 
dwarf at zero temperature unless there is an additional source of energy.  The cooling has no effect
on the degenerate electron pressure which will continue to support the gravity of the black dwarf.
The cooling timescales for isolated white
dwarfs are estimated to be of the order of billions of years due to radiative losses occuring from
the cooler non-degenerate layer.  The white dwarf would have cooled considerably faster if the hot
degenerate core was radiating at 100 million K. 
The non-degenerate layer can contain hydrogen, helium and other metals  leftover from the earlier phase
of evolution of the white dwarf.  A steep radial gradient in the temperature of this shallow layer
has been determined so that the surface temperature is much lower than the core
temperature \citep{1940ApJ....92..321M}.  It was surmised that
the core was bereft of hydrogen and helium and the energy source of white dwarfs was not
thermonuclear in nature otherwise their lifetimes would be much shorter than has been estimated
\citep{1940ApJ....92..321M}.  If hydrogen or helium were present in the core, then the
high temperatures therein ($\ge 10^8$ K) would instantly lead to a thermonuclear explosion 
which in turn would lift the degeneracy \citep{1965stst.conf..297M}.  
\citet{1940ApJ....92..321M} estimated a core temperature of $1.5\times10^7$ K, surface 
temperature of $9\times10^5$ K and a radius of $5.7\times10^8$ cm for the white dwarf Sirius B.  
The radius he estimated was discrepant by a factor
of two with the observational estimate of $13.6\times10^8$ cm by Kuiper. 
It was difficult to understand the discrepancy which was tentatively attributed to a larger 
non-degenerate accreted envelope around Sirius B which is in a binary.  
This issue was resolved when new observations modified the radius of Sirius B to a value
smaller by a factor of two \citep{1971ApJ...169..563G} which matched the theoretical value
estimated by \citet{1940ApJ....92..321M}.  
The surface temperature of Sirius B was measured to be $32000$ K \citep{1971ApJ...169..563G}
which was much lower than the value estimated by \citet{1940ApJ....92..321M}. 
Surface temperatures of white dwarfs in cataclysmic variables have been determined to span a 
wide range - 8500 to 50000 K \citep{1999PASP..111..532S}.

An accreting white dwarf can accumulate a hydrogen-rich surface layer 
in which thermonuclear reactions can ignite if its temperature rises
and which is not improbable since the internal temperature of the white
dwarf is very high \citep{1965stst.conf..297M}.  
If the temperature in the surface layer increased to $2 \times 10^8$ K then the 
hydrogen within would explosively ignite via the Bethe cycle 
especially if trace amounts of carbon, nitrogen and oxygen are present \citep{1939PhRv...55..434B}.
In the CNO (Bethe) cycle,  
the liberated energy is a very sensitive function of temperature since the nitrogen
reaction (capture of protons by $N^{14}$ which is the main process in which energy is
liberated) depends very strongly on temperature ($\propto T^{18}$) \citep{1939PhRv...55..434B}. 
Thus the energy production rises steeply with temperature. 
The thermonuclear reaction when the temperature rises beyond $10^8$ K can be explosive 
and all the hydrogen can be transmuted into helium in matter of seconds 
releasing huge quantities of energy which can eject the overlying layers of matter to infinity
\citep{1965stst.conf..297M}.  We now know that this summarises the nova explosion.
On the other hand, the energy production also drops rapidly as the temperature decreases. 

Novae are semi-detached binaries and by definition the companion star fills its Roche lobe
and hence the size of the companion star gives an idea of the binary separation.
%which determines the binary separation. 
The separation and orbital period are larger when the companion star is
a red giant of radius $\sim $ an astronomical unit as compared to when it is a main sequence star of
radius $\rm \sim R_\odot$. 
For a mass ratio of one i.e. both the white dwarf and the companion
are of the same mass (say $\rm \sim M_\odot$) and for a typical binary separation of 1 $\rm R_\odot$, the
Roche lobe extent will be $\rm 0.207 R_\odot$ and 
the product of the orbital period $P$ and density of the outer parts of the companion star $\rho$ will be
$P\sqrt{\rho} = 0.4168 $ \citep{1983ApJ...268..368E}. For a typical orbital period $P=0.25$ days, 
the density will be $\rho=2.8$ gm cm$^{-3}$.
For a red giant companion with same mass ratio and a binary separation of 100 $\rm R_\odot$,
the Roche lobe extent will be $\rm 20.7 R_\odot$ \citep{1983ApJ...268..368E} and 
for an orbital period of 300 days will result in density $\rho = 1.9\times10^{-6}$ gm cm$^{-3}$.  
The densities in the outer parts of a main sequence
star of a solar mass will be a million times larger than in the outer parts of a red giant star of
the same mass.  The implication for novae would be that the white dwarf is
likely to be immersed in a higher density ambient medium with a main sequence companion as compared to a red
giant companion which could also have some implications on the accretion rates. 

Accretion of matter on the white dwarf will release gravitational energy which can also
contribute to the heating of the accreted hydrogen-rich matter.  The combination of internal
temperature of the white dwarf, gravitational energy and compression of the accreted matter will 
contribute to the heating of the accreted layer on the white dwarf. 

\section{The updated model for novae}

The last section was mainly based on optical observations where 
sufficient data had been collected and analysed to warrant a model. 
We now have at our disposal, data on novae at wavebands ranging from $\gamma-$rays at the high frequency 
end to radio waves at the low frequency end of the electromagnetic spectrum which
arise due to different physical processes in the nova. These can, hence, be used
to modify and update the existing model.
Here we put forward a modified physical model based on the multi-band detections.  
Before we embark on updating the nova model,
we begin with a discussion on accretion of matter by a white dwarf and formation of an accretion disk
alongwith other associated changes in the system expected from basic physical considerations.  This is
included here because this explanation deviates significantly 
from the standard explanation given for the formation of accretion disks in literature.  
Moreover formation of bipolar outflows have continued to intrigue us with our understanding
being limited to some connection with an accretion disk. 
The following discussion is valid for any accreting massive object.  

\subsection{Accretion disk, accreted envelope, bipolar/ellipsoidal ejecta}
\label{accretiondisk}
The existing model attributes the formation of a rotating accretion disk around a massive object to the
angular momentum carried by the infalling matter.  It is believed that the accretion disk
facilitates the loss of angular momentum for the particle prior to being accreted on the white dwarf.
This means that the formation of the accretion disk does not depend on the rotation of
the accretor and will form around both rotating and non-rotating accreting objects. 
In this case, one would expect the rotation axis of the accretion disk to be oriented along the
angular momentum axis of the incoming matter and all the matter should have the same
angular momentum axis to be able to form a massive accretion disk.  This means that 
the accretion disk should have a rotation axis which should bear no relation to 
the rotation axis of the accretor and hence the two axes should be rarely aligned. 
However observations give contrary results that the accretion disk is 
predominantly formed in the equatorial plane of the accretor and shares its rotation axis.  
While one can invent several explanations for this, it should 
be borne in mind that such a strong correlation suggests a significant role of the accretor
in the formation of the accretion disk but which seems to have been largely ignored
in literature.  Additionally it appears contrived to suggest that all the infalling particles
have the same spin axis.  While a fixed angular momentum axis would be expected if a massive object 
like a star remained intact while being accreted, most of the accretion is of particles which are 
likely to have a random spin axis and this is especially true of a white dwarf in a nova. 
These appear to be major irreconciliable problems with the existing theory used to explain the formation 
of the accretion disk.  Unless we can understand it, preferably in the purview of known physics, it
will remain a bottleneck in furthering our understanding of any accretion-related phenomenon.  
It is likely that a physical understanding of this process will also lead to the resolution of
several other perplexing observables. 
We, hence, examine the physics which should dictate the formation of an accretion disk.

\begin{figure}
\centering
\mbox{
\includegraphics[width=3.2cm]{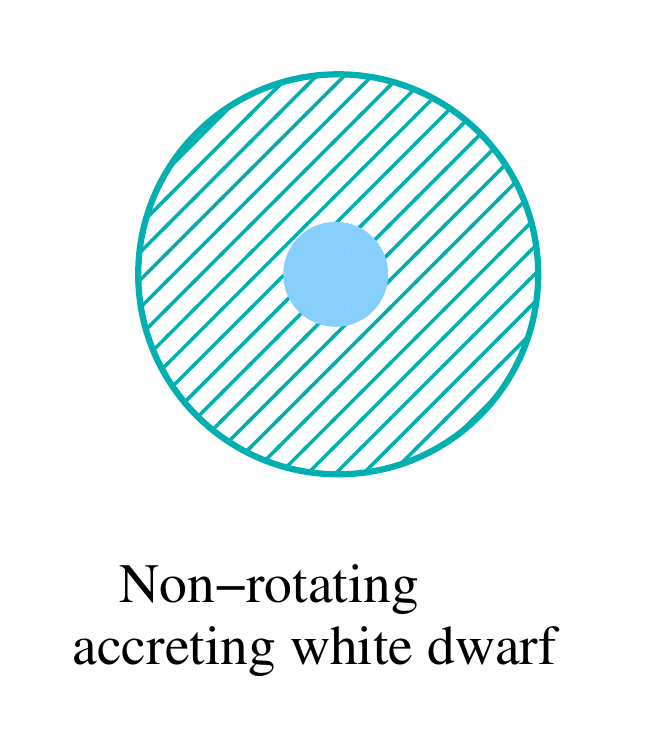}(a)
\includegraphics[width=3cm]{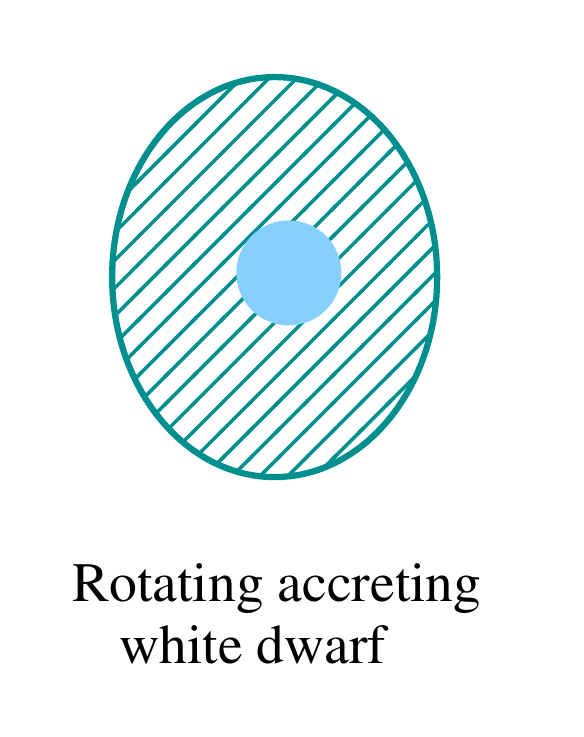}(b)}
\caption{Schematic showing the effect of rotation on the morphology of the accreted envelope. (a)
a non-rotating white dwarf will accrete a spherical envelope around it and a nova outburst will
result in a spherical ejecta. (b) A rotating white dwarf will accrete a prolate-shaped envelope
around it and a nova outburst will result in an elliptical ejecta.}
\label{prolate}
\end{figure}

Rigid rotation of a gravitating spherical object of radius $R$ modifies the 
effective attractive potential felt by the infalling matter.  While at the 
poles the effective potential will be equal to
the gravitational potential, the combined effect of gravitational and centrifugal forces
in the non-polar regions will lead to a latitude-dependent effective potential such that
the attractive potential is lowest at the equator where the effect of the centrifugal
force is maximum.  This indicates that within a given time, more matter should fall in
at the poles than at the equator i.e. a latitude-dependent 
accretion rate should be set up such that it is maximum at the poles and minimum at the equator.  
The extent of the accreted envelope will be proportional to the accreted mass and hence the larger accretion
rate at the poles will lead to the formation of a prolate-shaped envelope (see Figure \ref{prolate}). 
For a uniform rate of infall around the object, the matter in the non-polar
regions will accumulate outside the object due to the lower accretion rates.  This matter should  
form the accretion disk.  An accretion disk, thus formed,
should have the largest radial extent at the equator where the accretion rates are lowest
and it will taper down towards the poles.  
The thickness of the accretion disk should depend on the rotation speed of the object so that it
will be thick for a fast rotating object and thin for a slowly rotating object.  
The matter in the accretion disk will be dragged along by the gravitating object
and will start rotating.  Thus around an accreting rotating body, the formation of an accretion disk
and a prolate-shaped envelope are unavoidable although the extent, thickness of the accretion
disk and the ellipticity of the envelope will depend on the rotation speed of the object.  
Significantly, formation of the accretion disk
has no dependence on the angular momentum of the infalling matter.  In fact considering that
the matter being accreted can have different spin axis, if at all, it does appear far fetched to 
attribute the formation and rotation of the accretion disk to the angular momentum of the infalling matter. 
{\it To summarise, a rigidly rotating accreting body will set up a latitude-dependent accretion 
rate with maximum rates at the poles and minimum rates at the equator.  This will result in
a prolate-shaped envelope which in the extreme case can be bipolar.  The low equatorial accretion
rates lead to matter accumulation in the non-polar regions forming an accretion disk which is
dragged along by the accretor's rotation.  This, then, explains the formation of accretion disks
and prolate-shaped envelopes/outflows around compact massive objects.  Both require that the
accretor is rotating.} 

The outwardly directed centrifugal force acting on a particle of mass $m$ on the surface
of the object at a latitude $i$ will be $ F_{c,i}=mv_i^2/r_i$ and since for a rigidly
rotating body $ v_i=r_i\omega$, it follows that $ F_{c,i} = mr_i\omega^2$.
$\omega$ is a constant and $r_i$ is the radial separation between the rotation
axis and a point on the accreting body at latitude $i$ hence the centrifugal force varies 
as $r_i$.  $r_i=R$ at the equator whereas $r_i=0$ at the poles which quantifies the varying
centrifugal force as a function of latitude and hence the varying effective potential 
that a particle at different latitudes
would experience.  The gravitational force is the same over the entire surface of
radius $R$.  

In a spherically accreting non-rotating white dwarf, matter should be accreted at the same rate over
the entire surface and a spherical envelope around the white dwarf can be expected
(see Figure \ref{prolate}).  No accretion disk will form. 

We summarise the effect of rotation and accretion on compressible and incompressible objects.
If in a compressible object, rotation has led to an originally spherical object
evolving to a prolate-shaped ellipsoid then the accretion rate across the entire object will be
constant.  This will lead to the formation of a prolate-shaped envelope on the object mimicking the
shape of the object, but no accretion disk will form.  
In an incompressible object, rotation cannot modify the shape
of the object and it will remain spherical but the effective potential across the surface will
vary with latitude.  The accretion rates will also be a function
of latitude with highest rates at the poles.  This will lead to the formation of
a prolate-shaped ellipsoidal envelope around the object.  The lower accretion rates
at the equator will lead to accumulation of the infalling matter in an accretion disk. 
Thus in an incompressible object which cannot be distorted, rotation and accretion will lead 
to the formation of a prolate-shaped envelope and an accretion disk. 

An instability, at the base of the accreted envelope on a white dwarf, which injects copious amounts of energy
into the envelope can lead to its ejection.  For a non-rotating white dwarf the spherical
envelope will form a spherical ejecta while for a rotating white dwarf the prolate-shaped envelope will
be observed as a prolate ellipsoidal ejecta.  A bipolar ejecta will be an extreme case of
a prolate-shaped ejecta and would indicate a fast-rotating white dwarf wherein most of the
mass has accumulated at the poles due to extremely low accretion rates at the equator. 
Ellipsoidal ejecta have been observed in several
novae such as DQ Herculis, HR Delphini while there are also novae whose bipolar emission morphologies
in early times evolved to spherical morphology at later times (e.g. RS Ophiuchi).
The presence of an accretion disk could influence the morphology and speed of mass ejection in
the non-polar regions.
An accretion disk might quench the outburst in the equatorial regions either
through its disruption or by absorbing the ejecta energy, in either case, slowing down the ejecta
in the equatorial regions.  Faster polar expansion has been noted in nova ejecta
e.g. HR Delphini.   These are secondary observable effects. 

The main inferences derived in this section are:
(1) Accretion disks are formed around accreting rotating white dwarfs and their
formation is triggered by the latitude-dependent effective potential that the
infalling matter experiences which leads to low equatorial accretion rates.  The
formation of the accretion disk has no dependence on the angular momentum carried by the
accreted material  (2) detection of a bipolar ejecta/jets indicates the presence 
of a prolate envelope which indicates a rotating white dwarf. 

We now discuss novae in quiescence and in outburst where the above discussion will be useful.

\subsection{Novae in quiescence}
\label{quiescent}

We begin by examining properties of novae in quiescence which will be dictated by the
binary components.  These can help us better understand the equilibrium
configurations of the white dwarf and the companion star which would then result in an
initial condition for the outburst.

\begin{figure}[t]
\centering
\includegraphics[width=8cm]{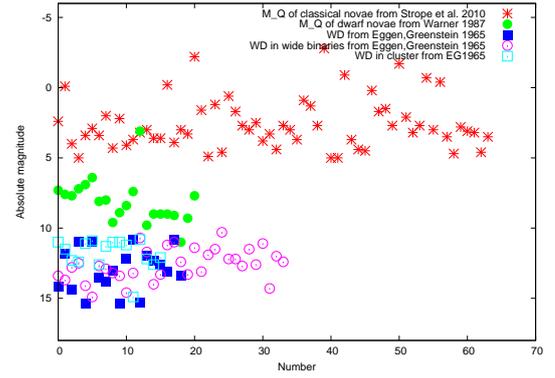}(a)
\includegraphics[width=8cm]{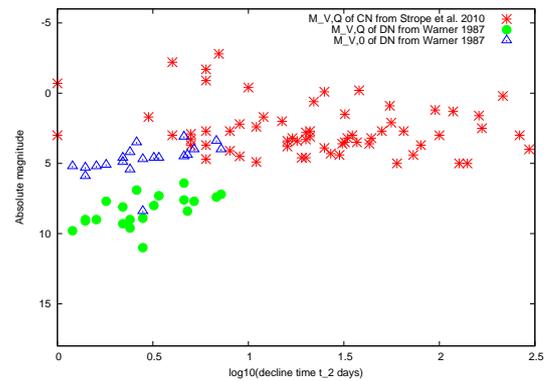}(b)
\caption{(a) Figure shows the absolute magnitudes of classical novae, dwarf novae 
in quiescence and white dwarfs.  
(b)The quiescence magnitudes of classical novae and dwarf novae and outburst maximum of
dwarf novae are plotted as a function of the decline time $t_2$. 
The required data have been taken from  \citet{2010AJ....140...34S}
\citet{1987MNRAS.227...23W}, \citet{1965ApJ...141...83E}, \citet{2017arXiv170304087K}. }
\label{minDN}
\end{figure}

The quiescence absolute magnitude in the V band $\rm M_{V,q}$ of classical, recurrent, dwarf novae and
of isolated white dwarfs are plotted in Figure \ref{minDN}a with data on novae taken from 
\citet{2010AJ....140...34S}, \citet{1987MNRAS.227...23W} and on white dwarfs from
\citet{1965ApJ...141...83E}.
In Figure \ref{minDN}b, the data on novae are plotted as a function of the decline time and
data on the outburst peak $\rm M_{V,0}$ of dwarf novae are also included.
The $\rm M_{V,q}$ of classical novae are estimated from the difference
between the peak absolute magnitude ($\rm M_{V,0}$)
determined using the MMRD calibration in \citet{2017arXiv170304087K} and 
the outburst amplitude (and $\rm t_2$) taken from \citet{2010AJ....140...34S}.
The three different symbols used to denote $M_V$ of white dwarfs in Figure \ref{minDN} indicate 
the different methods used to determine the distance by \citet{1965ApJ...141...83E} 
based on the environment of the white dwarf.  Trigonometric parallaxes for isolated white dwarfs, wide binary and 
Galactic cluster membership is indicated by the different symbols.  A few inferences 
can be drawn from Figure \ref{minDN}: (1) M$_{V}$ of white dwarfs is significantly lower than 
$\rm M_{V,q}$ of novae especially classical and recurrent novae (2) the $\rm M_{V,q}$ of dwarf novae 
is systematically lower than that of classical novae.  The outburst peak luminosity of dwarf novae merges 
with the $\rm M_{V,q}$ of classical novae as seen in Figure \ref{minDN}b and  (3) $\rm M_{V,q}$
of classical novae shows no dependence on the decline time unlike dwarf novae.
 
The mean quiescent absolute magnitude of classical novae is $\rm <M_{V,q}>=2.59\pm0.23$ magnitudes,
for dwarf novae is  $\rm <M_{V,q}>=8.37\pm0.25$ magnitudes whereas the mean outburst peak luminosity of
dwarf novae is $\rm <M_{V,0}>= 4.75\pm0.25$ magnitudes. The absolute magnitudes of
white dwarfs were found to divide into two types with one type following the
relation $M_V = 11.65 + 0.85(U-V)$ magnitudes and another fainter population following a
relation which is steeper \citep{1965ApJ...141...83E}.
The quiescent luminosity of the system is a combination of the
white dwarf and the companion star and hence is expected to be brighter than an isolated white dwarf.  
Since both classical and dwarf novae are
semi-detached binaries in most cases hosting either a late type main sequence star or sub-giant or giant companions,
one would have expected similar distribution of quiescence luminosities.
The observed difference in their quiescence brightness is, hence, perplexing and needs to be understood
since it indicates differences in their equilibrium physical properties. 
We note that a few classical novae like GK Persei 1901 now undergo dwarf nova outbursts. 
The secondary star in GK Persei is a subgiant of type K2IV type \citep{1983MNRAS.205..265S}
and the system has a $\rm M_{V,q}=4$ magnitudes for a distance of 337 pc \citep{1987MNRAS.227...23W}.
A K0 subgiant is expected to have an absolute magnitude of $+3.2$ magnitudes \citep{1973asqu.book.....A}
indicating that in GK Persei, the companion star likely dominates the combined luminosity.
In other cases like V446 Her, the system showed dwarf nova-like eruptions a few years before
the classical nova outburst \citep{1975AJ.....80..515R}.  These results also argue for 
similar binary components in dwarf novae and classical novae and hence
the distinct distribution of their quiescence magnitudes is puzzling. 
We note that literature tends to attribute luminosity changes in novae at times other than outburst to 
the companion star suggesting that the luminosity of the nova in quiescence is dominated
by the companion star.  We investigate this further. 

The companion stars in novae have been identified to be either late main sequence or subgiant 
or giant stars.   Most systems host a faint main sequence companion as inferred from their
short orbital periods of a few hours and the difficulty encountered in detecting their spectroscopic or
photometric presence.  It is easier to detect a subgiant or red giant companion.
The visible band absolute magnitude of main sequence stars of types G0 to M0 
range from $+4.6$ to $+9$ magnitudes \citep{1973asqu.book.....A}.  
A sub-giant of spectral type K0 is expected to have $\rm M_{V}=+3.2$ magnitudes.  A main sequence star
of spectral type F0 to F5 will have $\rm M_{V}$ of $+2.6$ to $+3.4$ magnitudes
\citep{1973asqu.book.....A}.  The $\rm M_{V,q}$ of most classical novae lie between 0 and 5 magnitudes  
(Figure \ref{minDN}).  If the observed $\rm M_{V,q}$ of the novae is dominated by the
companion star then in classical novae, the companion stars have to be 
main sequence or subgiant stars of types A0 to G4 or giants while in dwarf novae the
companions can only be main sequence stars of types K7 and later.  While this is possible,
such a difference seems not to have been noticed and also seems to be difficult to reconcile with the
observed duality of a few novae in showing both dwarf nova and classical nova outbursts.    
We could not find a study devoted to this aspect of novae in literature.  It
would be useful to examine data for any systematic difference between the companion
stars in classical and dwarf novae.  
For the current study, we suggest that the distint $M_{V,q}$ of classical and dwarf novae with statistically
similar types of companion stars argue for differences in the properties of the white dwarf.   

The $\rm M_{V,q}$ of recurrent novae plotted in Figure \ref{minDN} show the largest values - for
example, RS Ophiuchi has $\rm M_{V,q} \sim -2.8$ magnitudes, T CrB has $\rm M_{V,q} \sim -1.7$ magnitudes,
U Sco  has $\rm M_{V,q}\sim -0.7$ magnitudes and T Pyx has $\rm M_{V,q}\sim 0.1$ magnitudes. 
The companion stars in the four cases have been identified to be a red giant, red giant, sub-giant 
and a main sequence star respectively.  So while it appears that the companion star could dominate the
quiescent luminosity in case of RS Oph, T CrB and U Sco it would need to be a hot late 
B type star in case of T Pyx to explain its large minimum luminosity.  Since the companion
star in T Pyx is deduced to be a low mass main sequence star like the sun, it suggests that the
quiescent brightness of the system might be because of a brighter-than-usual white dwarf. 
Thus while the quiescence luminosity, in some cases, is indeed dominated by the companion star, it is 
not universally the case.  This supports our earlier conclusion that the accreting white dwarfs
in novae can have different properties which contribute to the luminous quiescent states. 

\begin{figure}
\centering
\includegraphics[width=9cm]{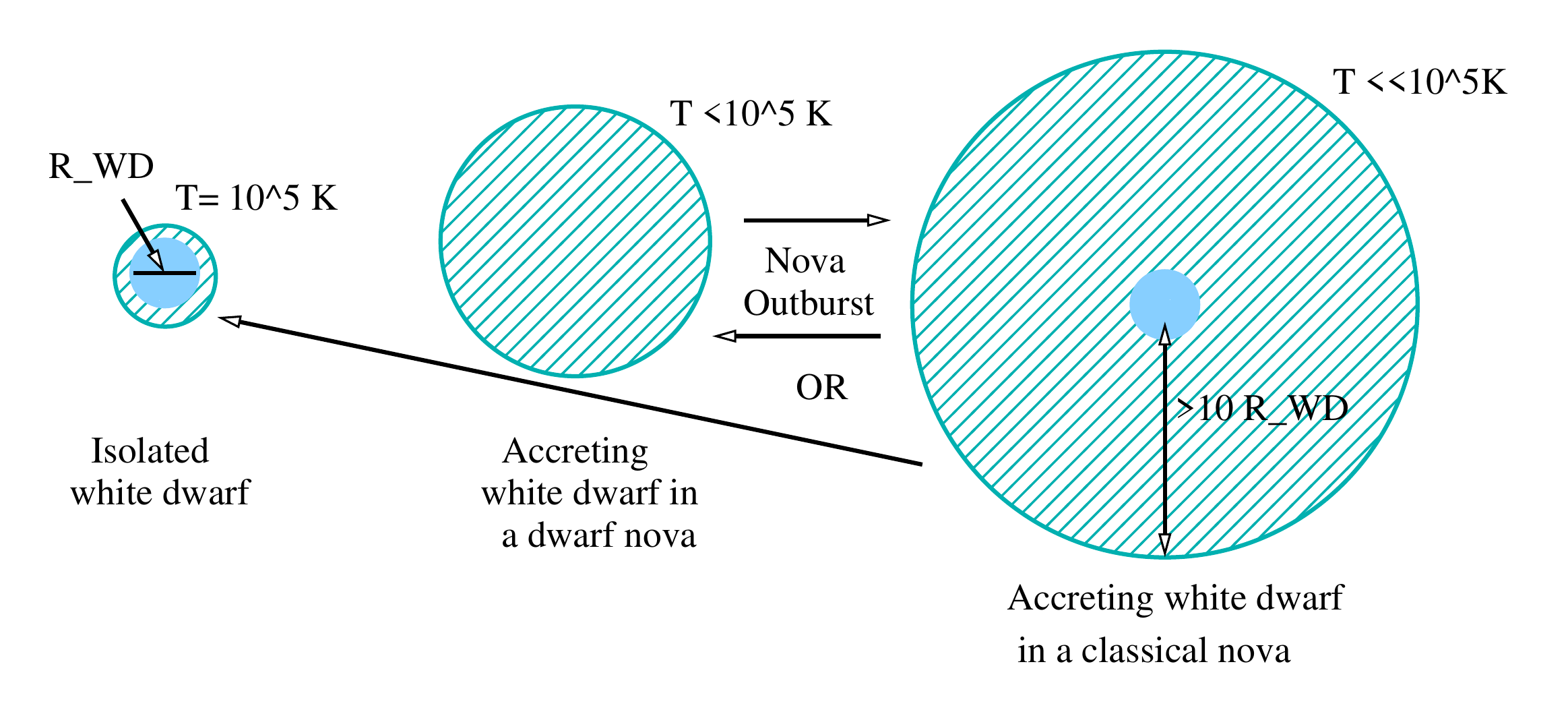}
\caption{Schematic comparing the varying sizes of the accreted envelopes around an
isolated white dwarf (left) and accreting non-rotating white dwarfs in a dwarf nova (middle) and classical nova
(right) in quiescence.   It also 
pictorially demonstrates our suggestion that the white dwarf in dwarf nova (middle) has a smaller envelope
than a classical nova (right).  The envelope size following an
outburst in either might resemble the other since dwarf nova inflates its envelope whereas
a classical nova ejects it.  Alternatively classical nova can eject the entire envelope and resemble
an isolated white dwarf for a short time before accretion restarts. }
\label{V458}
\end{figure}

We now examine the contribution of the white dwarf to the quiescent luminosity of novae.  
Classical novae appear brighter by $\ge 8$ magnitudes
than an isolated white dwarf.  We recall that a white dwarf consists of a degenerate
core surrounded by a thin layer of non-degenerate matter.  While the radius of the degenerate
core will shrink with increasing mass, the non-degenerate layer will expand
with increasing mass.  Thus, an accreting white dwarf
will increase the thickness of the layer of non-degenerate matter (see Figure \ref{V458}) 
as it accretes matter from the companion star.  The increase in the radial extent of the
outer layer should be accompanied by a drop in surface temperature assuming there is no
change in the heating source.  However it should be kept in mind that the gravitational
energy released due to accretion would contribute to heating the non-degenerate layer as will
compression in the lower layers.  Thus, two important
differences in an accreting white dwarf is a larger envelope and a lower surface temperature.
Assuming the white dwarf emits as a black body,
the bolometric luminosity can be estimated as $\rm L_b = 4 \pi R^2 \sigma T^4$ where
T is the surface temperature, R is the radius of the white dwarf and 
$\rm \sigma = 5.67 \times 10^{-8} W m^{-2} K^{-4}$ is the Stefan-Boltzmann constant. 
To convert the luminosity to the magnitude scale we use $\rm L_\odot = 3.826\times 10^{26}$ Watts and
4.75 magnitudes \citep{1973asqu.book.....A} as the bolometric absolute magnitude 
of the sun in the formula $\rm M_b - M_{\odot} = -2.5 log_{10} L_b/L_{\odot}$.  
The bolometric correction (BC) which is a function of the temperature is then used to convert 
this to V band magnitude i.e. $\rm M_V = M_b - BC$.
BC$=-0.08$  for the sun giving $\rm M_V=4.83$ magnitudes \citep{1973asqu.book.....A}.  
For a typical isolated white dwarf of $R=6000$ km and temperature of $10^5$ K, 
$\rm L_b=2.56\times10^{27}$ Watts and $\rm M_b=2.69$ magnitudes. For $10^5$ K, BC$=-7$  
\citep{1973asqu.book.....A} which gives $\rm M_V=9.69$ magnitudes.  A white dwarf
with $\rm R=6000$ km and $T=10^4$ K will have $\rm M_V=13.04$ magnitudes since BC = $-0.36$ 
\citep{1973asqu.book.....A}. 
While the hotter white dwarfs can explain the minimum luminosity of dwarf novae when not dominated
by the companion star, 
%if the companion star does not dominate it,
none of the white dwarfs with surface temperature between $10^4$ to $10^5$ K and radius
of 6000 km will be bright enough to explain the $\rm M_{V,q}$ of classical novae. 
We now include the effect of an inflated envelope due to accretion.  An expanded envelope of surface temperature 
$10^4$ K and $R=10\times6000$ km $ \sim 0.1 R_\odot$ will result in a much brighter $M_V=8.04$ magnitudes
as compared to $\rm M_V=13.04$ magnitudes estimated for a white dwarf of radius 6000 km. 
A larger increase in the size of the accreted envelope would then result in a still
more luminous white dwarf in a classical nova.   This result which follows from incorporating
the expected effect of accretion of matter on the white dwarf prompts us to suggest that the differences in the
$\rm M_{V,q}$ of classical and dwarf novae are due to the differing extents of the
accreted envelopes around the white dwarf.  The accreted envelope is smallest in an isolated
white dwarf, larger in a dwarf nova and largest in a classical nova (see Figure \ref{V458}).  
This could, amongst other factors, reflect the different accretion rates that have 
been surmised for a dwarf nova (lower) and a classical nova (higher).
The argument can be taken further to be suggestive of still larger envelopes in recurrent novae which
have higher accretion rates than most classical novae. 
We recall that to explain the luminosity of old novae, \citet{1941PA.....49..292M} estimated a radius of
$\rm 0.1 R_\odot$ and density of $\rm 1000 \rho_\odot$ for the nova star using similar
arguments.  However this preceded the identification of the binary nature of novae and
the white dwarf as the primary star.  In fact, ironically these physical parameters were the reason that a white
dwarf was ruled out as a nova star. 
We find ourselves arriving at similar dimensions for the white dwarf but with the added knowledge that
these are accreting white dwarfs in close binaries which support the existence of
such inflated lower density envelopes on white dwarfs.

\begin{table}
\centering
\caption{T denotes the surface temperature of the white dwarf, R denotes the radius of the white
dwarf in units of $\rm R_{WD}=6000$ km, $\rm L_b$ denotes the bolometric luminosity in units of
$\rm L_\odot$.  $\rm M_{V,\odot}= 4.83$ magnitudes is used.  Note how the white dwarf brightens
as the accreted envelope increases in size. }
%An optically thick ejecta expanding at 2000 kms$^{-1}$ will traverse a distance of $1.7\times10^{11}$ m (i.e. of
%the order of an AU which is $1.496\times10^{11}$ m) in a day which is the cause of the rapid
%brightening of the system. So for a white dwarf with a radius of 100 $R_{WD}$ which has $M_V=3.03$
%can brighten to $-6.96$ when the envelope is adiabatically energised and 
%isothermally blasted out to $10^4 R_{WD}$ i.e. about an AU.  }
\begin{tabular}{c|c|c|c|c}
\hline
{\bf T} & {\bf R} & {$\rm \bf L_b$} & {$\rm \bf M_b$} &  {$ \rm \bf M_V$} \\
K  & $\rm R_{WD}$ & $\rm L_\odot$ & mag &  mag \\
\hline
$10^4$  & 1  & 0.00067  & 12.68  &  13.04 \\ 
(BC$=-0.36)$    & 10 & 0.067   & 7.68  &  8.04  \\
        & 20 & 0.27  & 6.17  &  6.53  \\
        & 100 & 6.75 & 2.68  &  3.03 \\ 
        & $10^4$ & $6.75\times10^4$ & $-7.32$ & $-6.96$ \\
$10^5$  & 1  & 6.67  &2.69 &  9.69 \\
(BC=7)  & 2  & 26.7 & $1.18$ &  8.18 \\
        & 4  & 106.8  & $-0.32$ &  6.68  \\
        & 10 & 667.5 & $-2.3$ &   4.69 \\
\hline
\end{tabular}
\label{tabWD}
\end{table}

While most novae show similar pre-nova and post-nova luminosities, there do exist
a few cases wherein they differ.  We suggest these can be attributed to changes
in the size or temperature of the accreted envelope. 
For example, V533 Her recorded $\rm m_V=14.2$ magnitudes well before the outburst which brightened to 12 
magnitudes over a 2 year period before it exploded as a classical nova in 1963 \citep{1975AJ.....80..515R}.
A possible explanation, which needs to be verified, is that the brightening in the pre-nova phase 
was caused by an increase in the radius of the envelope attributable to enhanced radiation pressure exerted
by an increase in the temperature at the base of the envelope.
The possibility of an increase in the luminosity of the white dwarf before a classical nova outburst has 
been considered in literature \citep[e.g.][]{1987MNRAS.227...23W}. 
In the cases where the post-nova is brighter than the pre-nova, it could indicate that the
white dwarf in the post-nova has not yet declined to its pre-nova
luminosity.  This could then be attributed to the higher temperature of the post-nova white
dwarf following the outburst.  Thus, we find that changes in the accreted envelope around the white dwarf can 
explain observations. 

To quantify the above, we list the expected V band absolute magnitudes of white dwarfs for 
combinations of its surface temperature $T$ and radius $R$ estimated using the Stefen-Boltzmann law
assuming black body emission from the white dwarf in Table \ref{tabWD}.  
Two surface temperatures of $10^4$ K and $10^5$ K
are used and most white dwarfs in cataclysmic variables have been noted to have temperatures between
8500 K and 50000 K \citep{1999PASP..111..532S} so the two values for which we estimate the
emission from white dwarfs define the lower and upper limits.  We estimate the bolometric luminosity
for white dwarf radius ranging from $\rm R_{WD}$ to $\rm 10^4 R_{WD}$ where $\rm R_{WD} = 6000$ km for $
\rm T=10^4$ K and for radius ranging from $\rm R_{WD}$ to $\rm 10 R_{WD}$ for $T=10^5$ K. 
This is done keeping in mind that the temperature of the white dwarf stripped of its accreted envelope
in an outburst will be higher.  The post-outburst white dwarf will be hotter as also
been observationally found and the remaining envelope will be much smaller since
the quiescent state envelope would have been ejected (see Figure \ref{V458}).  
If the envelope is made very large then the
temperature is likely to reduce or it will become optically thin and not emit as a black body. 
Thus for the higher temperature we only consider radius upto $\rm 10 R_{WD}$.  
Some of the inferences we can draw from the table are: (1) For a white dwarf with radius of $R_{WD}=6000$ km,
$\rm M_V$ will lie between 13.04 to 9.69 magnitudes for surface temperature between $10^4$ K and
$10^5$ K.  We note that this encompasses the range of observed absolute magnitudes of isolated
white dwarfs. (2) The quiescent mean $M_V$ of dwarf novae can be explained by a white dwarf of
$T\sim 10^4$ K with an envelope of radius $< 10 R_{WD}$ or $T \sim 10^5$ K with envelope of
radius $< 2 R_{WD}$.  It appears that the white dwarf in a quiescent dwarf nova
has temperature ranging from about $10^4$ K to $10^5$ K and a radius of the accreted envelope
ranging from a few $R_{WD}$ to $R_{WD}$. (3) Classical novae in quiescence have
an envelope with radius $\ge 10 R_{WD}$.  If the surface temperature is low ($\le 10^4$) then the
envelope can even be $\ge 100 R_{WD} \sim R_\odot$. 
Since the accretion rates in dwarf novae are lower 
($\rm \sim 10^{-10} M_\odot yr^{-1}$) compared to classical novae ($\rm \sim 10^{-8} M_\odot yr^{-1}$),
this might be one of the causes for the different envelope dimensions.  However there could also
be other factors.  If the accreted mass distributed 
in an envelope of radius $100 R_{WD}$ on the white dwarf is $\rm 10^{-5} M_\odot$ then the hydrogen number densities 
will be $\rm \ge 10^{16} cm^{-3}$.  
The discussion here assumes spherical accretion on a white dwarf which leads to the formation of
a spherical/ellipsoidal envelope and ignores the contribution of the accretion disk to
the luminosity of the white dwarf.

Old novae which contain an accreting white dwarf were expected to be strong emitters of soft X-rays.
Observations have shown otherwise so that most old novae rarely emit X-rays 
and if X-rays are occasionally detected they are generally with energies $>5$ keV, 
\citep[e.g.][]{1981ApJ...245..609C,1982ApJ...262L..53F}. 
We find that this result is as expected from the discussion above wherein the cool accreted
envelope will hide the hot X-ray emitting surface of the white dwarf and no soft X-rays should be detectable
from the nova in quiesence.  In an outburst the cool envelope is ejected
exposing a hot surface with a temperature  $\ge 10^5$ K 
which will emit soft X-rays explaining their detection during an outburst. 

There are other observational results which support the establishment and existence of inflated envelopes
around the accreting white dwarf in quiescent novae. 
For example, a decrease in excitation was noted towards the end of the final decline in the light curve 
in a few novae like Nova Aquilae 1918, Nova Cygni 1920 and it was suggested that
the last 1-1.5 magnitudes decline in the light curve is due to the lowering of the temperature
of the superficial layers of the star \citep{1953ApJ...117..279M}.  
In Nova Geminorum 1912, observations indicated a decrease in excitation about 21 years 
after outburst and observations showed that the excitation definitely declined between 1917 to 1933
\citep{1953ApJ...117..279M}.  The luminosity of Nova Geminorum 1912 was found to decrease alongwith
the decreasing excitation so that the apparent magnitude was  
11.3 in 1914, 12 in 1916 and 14.5 magnitudes in 1933 \citep{1953ApJ...117..279M}. 
It is also deduced that the excitation in RS Ophiuchi is lower in the inter-outburst period 
than near the end of the decline from maximum \citep{1953ApJ...117..279M}. 
The $\rm B-V$ colour of DQ Herculis indicated a temperature of $\sim 10^4$ K for
the star \citep{1956ApJ...123...68W}
and a radius of $0.1 R_\odot$ has been estimated \citep{1960stat.conf..585M}. 
All these results support the formation of a lower density cooler envelope of several
times the radius of the core on the white dwarf and changes in its physical properties
especially radius and temperature.

\subsubsection{Eclipsing novae}

\begin{table*}[h]
\centering
\caption{Estimating the absolute magnitude of the white dwarf and companion star using the
eclipse depth $\Delta m$ under the assumption of complete occultation of the white dwarf during
the eclipse.  Bolometric absolute magnitudes required for estimating radius of the white
dwarf $r_{WD}$ are estimated from $\rm M_{V,WD}$ using the bolometric correction at the two temperatures
and it is assumed that the white dwarf emits like a black body.}
\begin{tabular}{l|c|c|c|c|c|c|c|c|c|c|c}
\hline
{\bf Nova} & $\rm \bf m_{V,q}$ & $\rm \bf m_{V,0}$ & $\rm  \bf A_V$ & $\rm \bf \Delta m$ & $\rm \bf t_2$ & $\rm\bf M_{V,0}$ & $\rm \bf M_{V,q}$ & $\rm \bf M_{V,sec}$ & $\rm \bf M_{V,WD}$ & \multicolumn{2}{|c}{$\rm \bf r_{WD}$ ($\rm \bf R_{\odot}$)}   \\
     & mag   & mag    & mag  & mag & days  & mag  & mag & mag & mag & $10^4$ K & $10^5$ K \\
\hline
DQ Herculis 1934 & 15   & 1.4    & 13.6  &  1.3$^1$ &  67 & $-6.9$ & 6.7 & 8.0 & 7.1 & 0.13 &  0.0008\\
U Scorpii$^*$ &  17.6  & 7.5  &  10.1 &  1.3$^2$ &  1  & $-10.8$  & $-0.7$ & 0.6 & $-0.31$ & 4 & 0.74 \\
U Geminorum$^{**}$ & 14.6 & 9.4 &  5.2  &    0.7$^3$ &  -   & 4.3  & 9.6 & 10.3 & 10.4 & 0.03 & 0.000038  \\
V1494 Aql &   17.1  &  4.1 &  13 &  0.35$^4$ & 8 & $-8.9$ & 4.1 & 4.4  &  5.5 & 0.3 & 0.004 \\
BT Monocerotis 1939 &  15.4$^5$ & - & & 2.7$^5$ & - & - & 4.0$^5$ & 6.7 & 4.1  & 0.5 & 0.01 \\
\hline
\end{tabular}

{\small $^1$ From \citet{1956ApJ...123...68W};
$^2$ From \citet{2010ATel.2452....1S};
$^3$ From \citet{1965ApJ...142.1051K};
$^4$ From \citet{2004PASJ...56S.125K};} \\
{\small $^5$ From \citet{1982ApJ...254..646R}; $^*$ Recurrent nova; $^{**}$ Dwarf nova }

\label{tab4}
\end{table*}

We estimate the luminosities of the binary members from the quiescence light curve 
of a white dwarf in a few eclipsing systems.  The main assumptions in doing so are that the primary eclipse is total
and that the white dwarf is eclipsed by the secondary star.
The former assumption will introduce some errors in the luminosities if the eclipse is partial.
%The eclipse is believed to be due to the companion star obscuring the white dwarf in almost edge-on binaries.  
Higher order emission lines of the Balmer series in hydrogen which are detected in the non-eclipsing spectrum of
DQ Herculis disappear during the eclipse \citep{1956ApJ...123...68W}.  Since high excitation lines 
are associated with the white dwarf, such results indicate that the primary eclipse is due to the
occultation of the white dwarf by the companion star.  Moreover the 71 second light modulation observed
in DQ Herculis and deduced to be indicative of the rotation period of the white dwarf vanishes during part 
of the primary eclipse \citep{1961ApJ...134..171W}.  Thus both assumptions are reasonable.
The method we use is as follows.  We estimate the peak absolute magnitude $\rm M_{V,0}$ 
of the classical/recurrent nova outburst
using the MMRD calibration of \citet{2017arXiv170304087K} and then use the amplitude of the outburst $A_n$ 
and estimate the quiescent absolute magnitude of the nova $\rm M_{V,q}$.  
$t_2$,  $A_n$ and the observed eclipse depth $\Delta m$ of the novae are taken from literature.  
The luminosity of the nova at deepest point of the primary eclipse will correspond to the luminosity
of the secondary star.  This is estimated from the difference between $\rm M_{V,q}$ and $\Delta m$ 
(see Table \ref{tab4}).  
Magnitudes are converted to the luminosity scale from
$ M_V - M_{V,\odot} = -2.5 log_{10} L_V / L_{V,\odot}$. The V band magnitude of the sun
$\rm M_{V,\odot}$ is 4.83 magnitudes and $L_V$ is estimated in units of $L_{V,\odot}$.  
The luminosities of the nova and secondary star are determined and their
difference corresponds to the luminosity of the white dwarf which can be converted to the magnitude scale
as above.  The resultant absolute magnitudes of the binary components are listed in Table \ref{tab4}.
The white dwarf and the companion star are of comparable V band luminosities.  
Both the components of the recurrent nova U Scorpii are bright whereas both components of the
dwarf nova U Geminorum are faint.  In most cases, the white dwarf is the brighter member of the
binary.   We convert the $M_V$ to the bolometric magnitude using the bolometric correction and
estimate the radius of the white dwarf assuming it emits as black bodies at temperatures $10^4$ K and $10^5$ K. 
These are listed in the last two columns of Table \ref{tab4} and inform us of the range of radius
of the white dwarf required to explain its luminosity.  It is clear that for temperatures between
these two extremities, most of the white dwarfs in classical/recurrent novae require an inflated 
envelope to explain their brightness and the data on eclipsing novae have given strong support
to the presence of these envelopes.  As a next step, the measured surface temperatures of the white dwarf
in eclipsing novae can be used to get a better estimate of the size of the envelope for particular novae. 
This seems to be one of the few methods, applicable only to eclipsing novae, available to us to 
study the physical properties of the binary components of a nova. 

We move to the next section where we discuss novae in outburst and update the existing model.
To summarize the initial conditions we start with:
(1) The white dwarf in a nova is several times brighter than an isolated white dwarf and the main
reason is attributed to the presence of an inflated accreted envelope.
(2) One important difference between classical and dwarf novae is in the size of the
accreted envelope explaining the difference in their quiescence luminosity.
(3) Rotation of the white dwarf leads to the formation of prolate-shaped ellipsoidal envelope
and equatorial accretion disks, both due to the latitude-dependent accretion rate. 

\subsection{Novae in outburst}

\begin{figure}
\centering
\includegraphics[width=8cm]{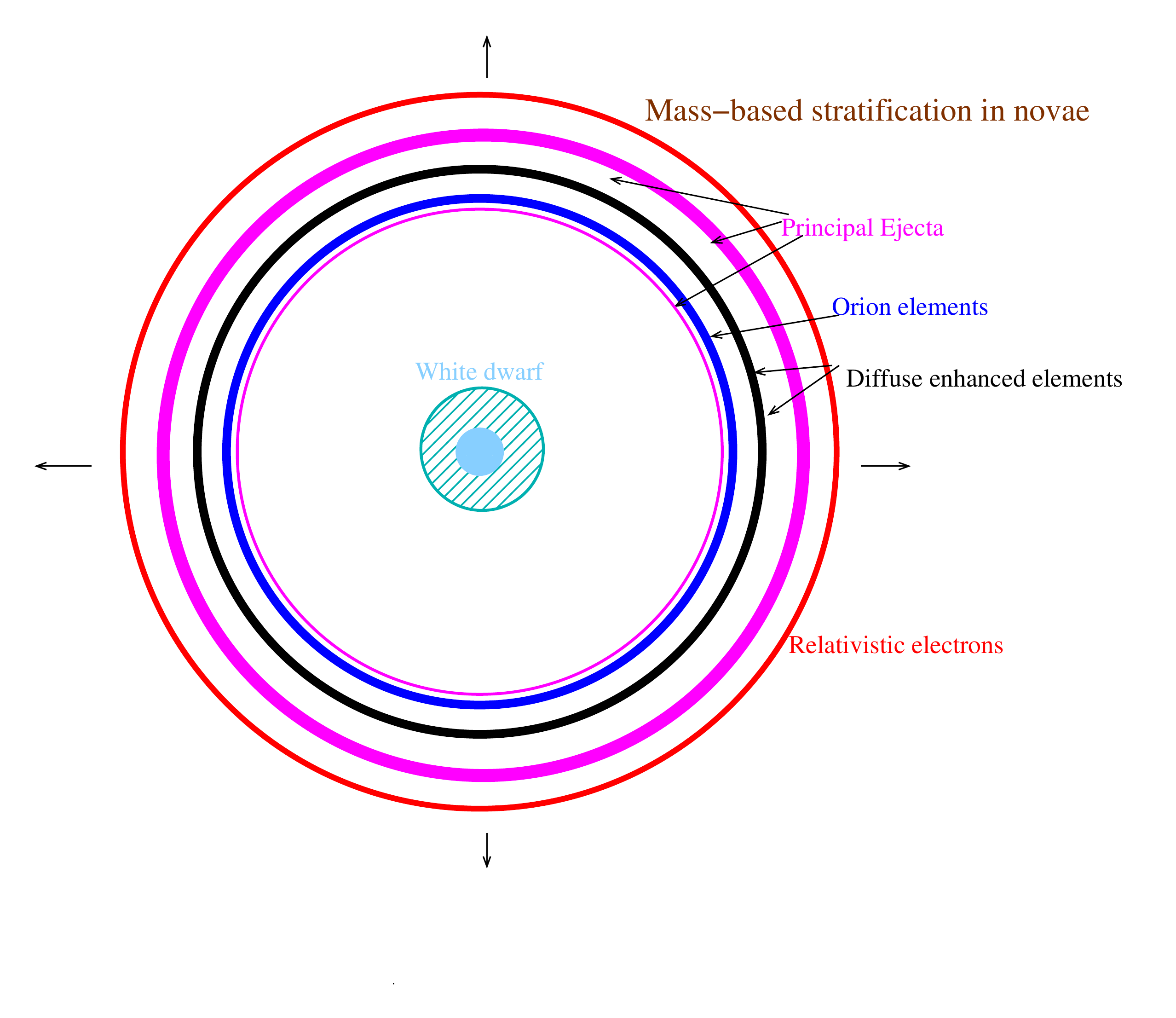}(a)
\includegraphics[width=8.5cm]{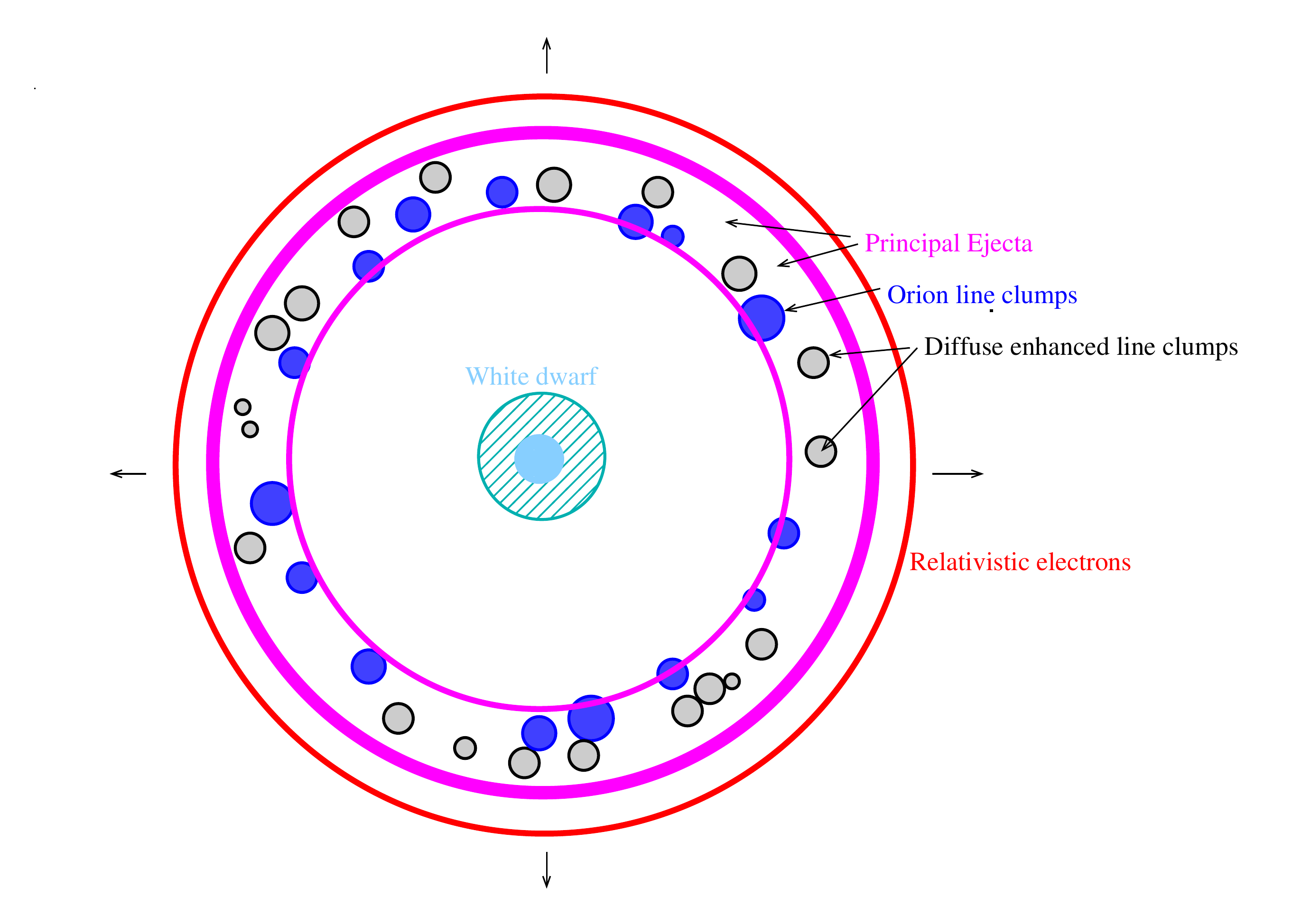}(b)
\includegraphics[width=8cm]{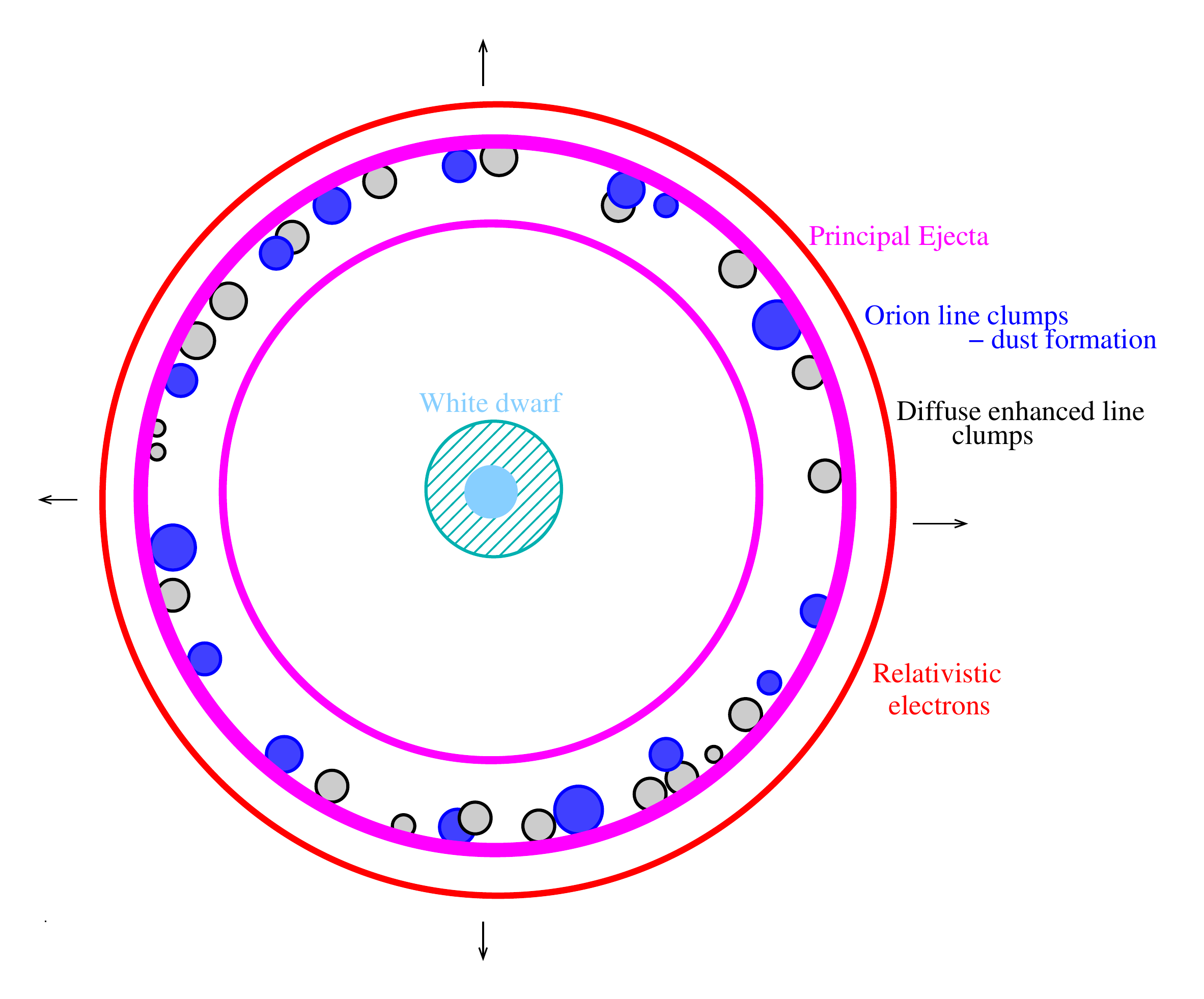}(c)
\caption{Schematics showing some points of the updated model to explain nova observations.  
The magenta lines enclose the main ejecta while the outermost red circle shows the location of
the relativistic electrons. 
(a) Mass-based segregation of elements in the ejecta. 
The heavier atoms/ions accumulate in the inner parts of the ejecta while lighter elements
like hydrogen lead the ejecta.  (b) Clumps form in the inner part of the ejecta 
where metals have accumulated.  Diffuse enhanced and Orion lines form in these clumps.  
(c) The clumps are accelerated by radiation pressure from
the hot white dwarf and move ahead in the principal shell.   Dust forms in
the Orion clumps.  }
\label{shock}
\label{novanew}
\end{figure}

In this section, we update the nova outburst model.  
It includes known and fresh aspects which coherently explain the outburst and the model is
schematically summarised in Figure \ref{shock}.
We refer to Figures \ref{lightcurve} and \ref{spectra} for the observed changes in the light
curve and spectra.  To preserve continuity the known and fresh aspects of the model are not
explicitly distinguished in the explanation but
the expert should face no difficulty in identifying these.  One can also refer to
the details of the existing model in section 2.   
The updated model incorporates the knowns of a nova system namely that it consists of 
a binary with an accreting white dwarf and
gaseous companion and when the physical conditions in the accreted non-degenerate envelope
of matter are appropriate, a thermonuclear runaway ignites and injects copious amounts of energy to the
envelope leading to its expulsion. 
We first summarise some of the common observational features of a nova 
outburst mainly in the optical (already detailed earlier in the paper) 
which any model should self-consistently explain and then we address the different
ill-understood aspects of the nova outburst which helps build the comprehensive model:

\begin{itemize}
\item Sudden optical brightening of a star and evolution of its light with time including
peculiarities such as light oscillations and steep drop in the light curve.
\item Continuously evolving optical spectrum of the star and simultaneous detection of
the  pre-maximum, principal, diffuse enhanced and Orion line systems with different velocity 
displacements and ionization/excitation levels. 
\item Correlated changes between the evolution of the optical light curve and 
spectrum shown by most novae.  
\item The existence of dominant Fe II or He/N spectral lines in the post-maximum spectrum. 
\item Detection of multi-band emission at particular times and durations with respect
to the optical maximum.
\end{itemize}

The model can be summarised as follows:
Accreted matter accumulates in an envelope around the white dwarf.  The base of this envelope
is compressed and heated.  When it heats to $\ge 10^8$ K,  an
explosive thermonuclear ignition of the hydrogen-rich normal matter exhausts the fuel in a short time  
and leads to an energy pulse.  There appears no need to invoke degeneracy in this matter which is not
likely to be present since the accreted matter will lie above the non-degenerate layer of matter
that a white dwarf is born with. 
This energy can lead to the expansion of the same highly pressured layer and it can adiabatically 
transmit the energy to the overlying layers increasing the latter's internal energy
which can be in the form of mechanical energy imparted to each chemical constituent/particle.   
Noting that the layer in which the explosion occured will be
at $10^8$ K ($T_h$) and the overlying layers will be much cooler at $\le 50000$ K ($T_c$), the energy 
transfer efficiency based on the Carnot cycle ($(T_h + T_c)/T_h$) can be 99\%.  Thus most of the 
energy can be adiabatically injected
into the overlying layers as mechanical energy, ejecting them to infinity and also
imparting a random motion to all the particles therein.
Depending on the total energy injection and mass which is ejected, each particle will acquire
some random velocity component and a forward expansion component.   
While the forward motion component will necessarily have to be larger than the escape velocity
of the white dwarf if the ejecta is to leave the system, the random velocity of the particle
will be indicative of the remnant energy.
The ejected matter consisting of electrons, atoms, ions will start expanding with similar velocities.  
Electrons can acquire random velocities which are relativistic in sufficiently energetic explosions.  
Sometimes the relativistic electrons can also precede the main
ejecta.  The nova and expanding ejecta give rise to multi-band emission. 
We explain in detail below.

\paragraph{Optical emission from ejecta:}
There exist several points of evidence which suggest that the sudden optical brightening of the
nova is due to the increasing emission from the ejecta.  In the short duration that the accreted
envelope is getting energised by the burst of thermonuclear energy and before it is ejected,
one can consider the photosphere as
defining the outer boundary of the ejecta.   Once the envelope is ejected, evidence suggests that
the origin of the optical continuous emission like the line spectrum is in the ejecta and not 
in the photosphere of the white dwarf.  The strongest evidence for the common
origin of the optical continuous emission and spectral lines in the ejecta comes from
(1) the inference that most novae follow the
maximum magnitude relation with decline time (MMRD) wherein the maximum visible luminosity of the nova
outburst is strongly correlated with the decline time of light and the ejecta velocity
(also see Section \ref{intro})
and (2) the correlated evolution of the optical light curve and line spectrum in a nova outburst.
The simplest explanation for the observed nature of the light curve is that
as the ejecta expands, the optical
emission reaches a maximum when the optical depth drops to one and then the emission fades as the
ejecta expands further and its emission measure declines.
Slow novae expand at a slower rate than fast novae so the change in its emission measure (or density)
is also gradual and the light curve continues to hover near the maximum for
a longer time.  A drop of two magnitudes in the peak V band luminosity would
require the luminosity to change by a factor of 6.3 which will be due to a drop in
the emission measure of the ejecta.  A slow nova
will take longer to achieve the drop then a fast nova if we assumed that similar masses were ejected.
Thus the observed evolution of the light curve is trivially explained if it arises in
the ejecta.  {\it Thus we conclude that the optical continuous emission from the nova outburst
is dominated by the ejecta near the maximum and weakens as the ejecta expands. }

We now discuss the optical emission from the ejecta and the light curve evolution.
The typical temperatures of the white dwarf in cataclysmic variables has been measured to be
between 8500 K and 50000 K \citep{1999PASP..111..532S}. The explosion will inject energy into
this envelope and eject this hot material.
Electron temperatures around $5000-10000$ K are measured for the ejecta indicating that the
outburst energy is sufficient to retain these temperatures and the ejecta starts to radiate in the optical
and ultraviolet bands through the free-free and free-bound processes.
The optical emission is observed to quickly (within a day or so) strengthen by several
magnitudes before it is halted around two magnitudes below
maximum.  Some novae show a short plateau at this point and we suggest this is because the
optically thick ejecta is shining entirely as a result of the energy input
from the explosion.  Around this time, 
radiation of the hot white dwarf will also start contributing to the physical state of the ejecta
and to the light curve.  The radiation field of the white dwarf can exert radiation pressure on
the optically thick ejecta and once the ejecta is optically thin, the
white dwarf can become optically
visible through most of the ejecta and if it is optically brighter compared to quiescence, then it will
contribute to the nova brightness in the optical band.
We suggest that the white dwarf radiation contributes the final $\le 2$ magnitudes rise of brightness
to optical peak in many novae.  Thus, in some novae it might contribute as much as 2 magnitudes
and in some novae, the white dwarf might
not contribute anything especially if a large fraction of the ejecta continues to obscure the
white dwarf radiation field.  An increase by two magnitudes can be contributed by a post-outburst
white dwarf which emits 6.3 times its quiescence optical luminosity.
The brightness contribution to the light curve by the white dwarf, if any, will remain constant
over a longer duration while the contribution by the ejecta will keep declining as it expands.
This inference is supported by the observation that the colour temperature
estimated from multifrequency optical light curves
increases as the light curve declines ie the light
gets bluer.  If the fractional contribution of the cooler ejecta to the optical emission is
reducing and that of the hot white dwarf is increasing, the colour should get bluer as is observed.
The excitation of the expanding ejecta is constantly changing in the initial phases so that it
resembles a hot B/A type star
before the optical maximum and then evolves to later A/F types at/after maximum.  These
will the combined effect of the explosion energy, expansion of the ejecta and the
white dwarf radiation field.

\paragraph{Generation of relativistic electrons:}
The source of energy in a nova has been shown to be in a thermonuclear runaway reaction. 
When the temperature at the base of the accreted hydrogen/helium envelope 
rises to $\sim 10^8$ K due to compression and accretion heating, then even the presence of 
trace amounts of carbon, 
nitrogen and oxygen can set up an explosive CNO reaction cycle \citep{1939PhRv...55..434B} 
which in a matter of seconds can burn the accreted hydrogen with the net reaction products being:
$\rm 4 H^1 + 2e^- \rightarrow He^4 + 2 \nu_e + 3 \gamma + 26.7$ MeV.  $\nu_e$ are neutrinoes
and $\gamma$ are $\gamma-$ray photons.  The energy liberated
in the CNO cycle is  $\propto T^{18}$  \citep{1939PhRv...55..434B}.
Since the proton-proton chain can ignite at lower temperatures, it could be the process that happens
in some novae, especially dwarf novae where the energy released is lower. 
The energy liberated in a proton-proton chain is $\propto T^4$ i.e. is not so
sensitive to temperature changes unlike the CNO cycle.  This sudden energetic event 
will release copious quantities of neutrinoes, $\gamma-$ray photons and heat energy.
Since the observational properties of the pre-nova
and post-nova are similar, the explosion does not affect the degenerate
core of the white dwarf.  The released particles are energised to velocities
higher than escape velocity and erupt outwards from the white dwarf.  
The escape velocity from a white dwarf of
mass $0.5 M_\odot$ and radius $0.015 R_\odot$ will be 3600 kms$^{-1}$, from a 
mass $1 M_\odot$ and radius $0.01 R_\odot$ will be 6300 kms$^{-1}$ and from a mass
$1.3 M_\odot$ and radius $0.005 R_\odot$ will be 10200 kms$^{-1}$. 
It appears likely that a thermonuclear burst is also the mechanism for a dwarf nova outburst but
which only leads to expansion of the accreted envelope and not ejection.  This origin
can explain the observed light curves of dwarf novae which show a rapid rise, a
flat top and then a gradual decrease - all accomplished within a month or so as being due to the
expansion and subsequent contraction of the envelope.  An outburst amplitude upto 
five magnitudes which is typical of dwarf novae is possible if the 
accreted envelope isothermally expands to $\rm \le 0.1 R_{\odot}$ which as discussed in the previous
section will be similar to the extent of the envelope of the white dwarf in a quiescent classical nova.  

All the matter in the envelope - atoms, ions, electrons is adiabatically energised by the thermonuclear energy
which will result in two velocity components - an outward directed radial velocity and a random 
velocity.  In classical novae, the envelope will detach once it acquires an 
outward velocity which is of the order of the escape velocity.   
If we assume equal energy is imparted to all particles
and that all particles acquire similar expansion velocity then the electrons
can acquire very high random velocities whereas the random velocity component will be smaller
for the relatively heavier ions and atoms.  It is also likely that the electrons can also acquire
higher expansion velocities and we note that imaging observations of radio synchrotron emission
seem to support this hypothesis wherein the radio synchrotron emission arises from a region ahead
of the thermally emitting region which is the main ejecta. 
Since the mass ratio of a proton and an electron is 1846, the velocities can differ
by a factor $\le \sqrt{1846} \sim 42$.  If the bulk ejecta velocity in a classical
nova is $\rm v_{ej} = 3600$ $\rm kms^{-1}$ then each proton will acquire this velocity which 
translates to an energy of $1.0368\times10^{-14}$ Joules per proton.  The electrons will also
be imparted this energy and it can be translated to a velocity using the relativistic formula for energy 
namely $E_e = (\gamma-1) m_e c^2$.  This gives $\gamma = 1.127$.  Since $\gamma=1/\sqrt{1-v_e^2/c^2}$,
this gives $v_e=0.4885c$. 
If $v_{ej}=6300$ kms$^{-1}$ then electrons have energy of $\gamma=1.3877$ and $v_{e} = 0.8167c$ and
if $v_{ej}=15000$ kms$^{-1}$ then electrons can have energies of $\gamma=3.2$ i.e. $v_e=0.9499c$.
These $\gamma$ factors could indicate the mean energy of the electrons or the highest energy and
it is difficult to comment further on this.  We can only say that electrons are ejected in a nova
outburst with relativistic velocities especially for large bulk expansion velocities. 
All nova ejections have been reported to expand with velocities $<15000$ kms$^{-1}$ which could
indicate that the electrons in a nova outburst have energies around $\gamma=3.2$. 
The electrons are relativistic and can emit synchrotron emission in a magnetic field. 
This demonstrates that a population of relativistic electrons should be energised alongwith the 
main ejecta when expansion velocities are $\ge 6500$ kms$^{-1}$ and definitely for ejecta velocities
$>10000$ kms$^{-1}$ and these novae are likely to be synchrotron emitters.
Since it has been noted in some novae that high velocities recorded before the pre-maximum
halt decline rapidly; it should be kept in mind that there might be several novae in which the
initial expansion velocities indicative of the outburst energy are sufficiently high to 
generate a relativistic electron population.
Since few novae are detected in the initial rise phase which is extremely fast, it
remains a difficult task to determine the original expansion velocity otherwise that would be
able to predict the possible detection of radio synchrotron emission from the system. 
This explains the detection of radio synchrotron emission from the fast recurrent novae - 
RS Ophiuchi and V745 Scorpii both of which recorded early ejecta velocities $>10000$ kms$^{-1}$. 
We believe that radio synchrotron emission from nearby fast classical novae 
($\rm v_{ej} \ge 10000$ kms$^{-1}$) should be detectable immediately after the outburst since
%since intense winds are not expected from low mass white dwarfs hosted in classical novae and hence 
the foreground obscuration which delays the detection of radio synchrotron emission especially in
recurrent novae is likely to be low.  We note that since magnetic field is required it can either
be frozen in the ejecta if the relativistic electron population is coexistent with the ejecta
or an ambient medium in which the field is frozen is required for the electrons to radiate. 
In GK Persei 1901, the early ejecta velocities were recorded to 
be only $1000-1500$ kms$^{-1}$ but we detect radio synchrotron emission from its remnant even today
indicating that the outburst was highly energetic and injected a relativistic electron
population into the circumstellar medium.  We do know that GK Persei exploded within 
its planetary nebula and the high ambient densities could have contributed to the rapid deceleration of 
the main ejecta which might have erupted at much higher velocities than recorded.
This is suggested to explain the observed radio synchrotron emission.
The nuclear reaction also emits
$\gamma-$ray photons.  While the photons released before the 
relativistic electrons have been ejected will leave the system, the photons emitted later can
form the seed photon population which can be inverse Compton boosted to energies $\ge 100$ MeV 
which have been detected from several classical novae.  
In a typical nova explosion which generates a relativistic electron population with
$\gamma \sim 3.2$, detection of  
$\gamma-$ray photons of energy $\ge 100$ MeV can be explained as being due to inverse Compton boosting
of seed photons of energy $\sim 7.3$ MeV.  Since the CNO
reaction will emit $\gamma-$ray photons of such energies, this explanation appears to
be a plausible one for the origin of high energy $\gamma-$rays detected soon after the outburst. 
The duration over which these photons are detected then define
the duration of the thermonuclear reactions on the surface of the white dwarf. 
%or the lifetime of the relativistic electron population.  
Once the reaction dies down,  low energy $\gamma-$ray photons are extinguished 
and so are the $\ge 100$ MeV photons.  This is supported by the detection of $\gamma-$rays 
only for a few days near the optical maximum.  
{\it To summarise the discussion till now, we have pointed out that a relativistic electron
population and hence synchrotron radio emission and energetic $\gamma$-rays are all expected from the
explosive nuclear reaction on the surface of the white dwarf and there is no need to invoke
any post-explosion shock acceleration.  }

\paragraph{Mass-based segregation in ejecta:}
We shift our attention to the bulk ejecta and its evolution.  The explosion energises
and ejects the overlying layers and the expansion velocity has to be $\ge$ escape velocity of the 
white dwarf.  If all novae explosions release similar energies then
a more massive envelope should acquire a lower expansion velocity while a lower mass ejection
would be faster.  However a detailed correlation is not expected since the energy released in
the CNO cycle is a very sensitive function of temperature beyond $10^8$ K ($\propto T^{18}$) and
which should lead to a range of values for the energy released in novae explosions. 
As mentioned in the previous item, if equal energy is imparted to each 
particle, it will result in a forward expansion velocity component and a random component.   
For particles of mass $\ge$ mass of proton, the random velocity component appears to be
a small fraction of the expansion velocity. 
Since hydrogen is generally the most abundant element in the ejecta, the average expansion velocity 
should be dictated by the velocity acquired by hydrogen atoms.  
Observations detect different line widths for emission lines of
hydrogen (larger) and iron (smaller) which could indicate slightly different expansion velocities 
and which can cause the heavier atoms to lag behind in the ejecta. 
This, then, can lead to mass-based segregation in the main ejecta so that lower mass elements will 
lead while heavier elements will lag behind in the ejecta (see Figure \ref{shock}).  
For simplicity, we assume that this rearrangement within the ejecta does not
change the thickness of the ejecta which continues to be of
the quiescent envelope size i.e. nominally $\ge 0.1 R_{\odot}$.  
Within the main ejecta, the highest velocities are acquired by hydrogen atoms followed
by helium atoms, carbon atoms, nitrogen atoms, oxygen atoms and so on.  

Segregation, based on mass of the emitting atom, in the ejecta appears to be supported
by the following observational results. 
Distinct line profiles are detected for different elements 
which suggests that various emitting species have distinct spatial distributions 
and/or there exist ionization gradients in the ejecta \citep[e.g.][]{2012ApJ...755...37H}.
Another point of support emerges from the different types of dust
such as amorphous carbon, silicates and hydrocarbons which are detected in the same nova indicating
abundance gradients within the ejecta \citep{2012BASI...40..213E}. 
We discuss implications of such mass-based segregation and how it successfully explains 
several observational results which increases our confidence in such a phenomenon. 

\paragraph{Swept-up matter and Fe II, He/N lines in novae:}
The ejecta is expanding with supersonic velocities and will set up a shock and 
entrain (heat and accelerate) the matter it encounters on its expansion path.  
Thus, the spectral signatures from a nova after outburst can be a combination 
of the accreted envelope which is ejected,  matter native to the white dwarf which is ejected and
ambient matter which the ejecta sweeps up.   While the spectra soon after outburst
should be dominated by the lines from the accreted envelope which is ejected, 
there are likely to be signatures of the other two components.  It might be difficult to differentiate these 
in many cases, but it does appear possible to occasionally separate this from their observational 
signatures.  For example, in DQ Herculis 1934,
when the light curve suffered a steep drop and long after most spectral lines had disappeared, 
emission lines of hydrogen, Fe II and Ca II continued to be detectable.  Since the obscuration had to arise 
either in front of or within the ejecta, this observation    
indicated that these lines were forming in the outermost part of the ejecta at that time 
and which was not yet obscured.  While hydrogen could
be from the ejected material, it seems more likely that the Fe II and
Ca II lines were arising in the swept-up material of the companion star as the ejecta expanded, 
and hence located in the outermost part of the ejecta. 
Mass-based segregation in the ejecta would have led to the ejected iron and calcium lagging to
the rear of the ejecta. 
This suggestion is strengthened when we realise that the atmospheres of late type stars 
(later than F) are characterised by iron and calcium lines \citep[e.g.][]{1973asqu.book.....A}.
Another point of support for some of the detected iron lines forming in the swept up material of the
companion star comes from the observations of symbiotic stars wherein the Fe II lines are found to
show a distinct radial velocity behaviour with orbital phase as compared to other spectral lines.
This observation indicated the origin of the iron lines in the companion star and 
was instrumental in establishing the binary nature of symbiotic stars.  
Unless we encounter serious observational contradictions to the hypothesis of mass-based segregation
within the ejecta, we assume that this is indeed the case and try to 
understand observations in its context.  At no point we resort to contrived explanations to support
our hypothesis since we believe that requiring such an approach means that the hypothesis is incorrect
and non-physical.  To our satisfaction, the hypothesis works within the ambit of known parameters. 

We suggest that the distinction of novae into the Fe II and He/N classes based on the post-maximum
spectrum is indicative of the dominance of lines forming in the swept up material and the ejecta
respectively.  A Fe II type nova shows narrower P Cygni like lines more often than He/N types which show 
broad jagged flat-topped lines.  As discussed in an earlier point, the optical continuum
from a nova is predominantly from the ejecta till about the nebular phase.  The
P Cygni absorption features in Fe II lines support their formation in the outer parts of
the ejecta so that the cool ions absorb the emission lines or the continuum emission arising behind them. 
This, in turn, supports their formation in the swept-up ambient material.
With passage of time, this material entrained by the leading part of the ejecta
(or shock) will also sink to the rear part of the ejecta due to lower expansion velocities acquired
by these heavier ions.
On the other hand, the He/N novae show emission lines with larger widths and seldom
P Cygni absorption so that these must arise in the bulk of the hot ejecta 
with little continuum emission to absorb or cool atoms which can absorb.  
This is also supported by the observation that in hybrid novae the
early detections of Fe II lines are replaced by He/N lines but never vice versa.  Since
swept-up material should be present in all novae, the reasons for lines from it being detectable  
in some novae and not in others should be a function of other properties 
such as the detailed physical conditions in the ejecta, its expansion velocity and the ambient densities. 
We also note that there is a predominance of main sequence
companions in the Fe II type novae and of sub-giants or red giants in the He/N type novae.
For example, V1534 Sco, V1535 Sco and Master OT J010603.18 are classified as He/N type novae and
the companion stars are identified as giant, giant and sub-giant whereas V2949 Oph,
V3661 Oph, TCP J18102829-2729590 and ASASSN-16ma are classified as Fe II novae 
\citep{2017MNRAS.469.4341M} and the companion star is likely to be a main sequence star.  Moreover the
recurrent novae with main sequence companions - T Pyx and IM Nor are Fe II novae 
\citep{2012ApJ...746...61D}.  The companion star will be the main source of the ambient
medium in the nova.  We recall that
the matter densities in the outer parts of the Roche lobe of a main sequence star are almost a million
times higher than that of a red giant (see Section \ref{primary}).  Thus, the
shock/ejecta can sweep up more matter in case of a main sequence companion than a giant star
and can explain the presence of sufficient swept-up matter for the Fe II lines to be detectable.
As the nova ejecta evolves, several of these become hybrid novae with the detection of
strong He/N lines from the ejecta.  
The expanding ejecta soon clears the companion star and hence the swept up matter does not 
increase indefinitely with time.  Moreover, the swept-up matter composed of heavy atoms will sink to 
the rear part of the ejecta and will soon be indistinguishable from the ejected matter.

Before we proceed, we enumerate the main points of the model discussed till now:
\begin{itemize}
\item It is shown that the quiescence luminosity of the nova is high 
due to the presence of a white dwarf with a large accreted envelope ($\ge 0.1 R_{\odot}$ 
for a classical nova).  Eclipsing novae give quantitative support to this hypothesis. 
\item The common origin of the sudden optical brightening of a nova outburst and spectral lines is
in the ejecta which 
explains the efficacy of the MMRD and the correlated changes in the optical light curve and
spectra.  The hotter white dwarf contributes less than two magnitudes to the luminosity after
the pre-maximum halt. 
\item It is suggested that even dwarf nova outbursts are due to a thermonuclear explosion on the
white dwarf.  However the energy released is much lower than in classical novae and only  
leads to an expansion of the envelope in dwarf novae.
\item The explosion sets up a blast wave, energises electrons to relativistic velocities
and ejects the envelope supersonically, all of which propagate outwards.  
\item It is suggested that there is mass-based segregation within the ejecta 
such that the heavier elements having acquired slightly 
lower velocities occupy the rear parts of the ejecta and the lighter elements lead the ejecta. 
\end{itemize}

\paragraph{Spectral systems and clump formation in the ejecta:}
In this point, we examine observational results on the four main spectral systems namely the 
pre-maximum, principal, diffuse enhanced and Orion (see Table \ref{tab1}) which can be
identified in a nova spectrum as the light curve evolves.
In the pre-maximum phase of the light curve, a spectrum consisting of blue-shifted absorption 
lines of low excitation are detected at a certain velocity displacement.  In the new model, the energy 
input to the ejecta till the pre-maximum is entirely due to the explosion energy. 
The velocity of the pre-maximum spectrum, if constant since ejection will
signify the energy output of the explosion if the ejected mass is known.
However some novae like DQ Herculis have shown a rapid decline in this velocity likely due to
dense ambient medium.
At the optical light curve maximum, a new spectral system is detected in which the absorption lines 
appear at a higher blue-shifted velocity compared to the pre-maximum spectrum and show associated 
wide emission bands.   This is referred to as the principal spectrum and which replaces the pre-maximum 
spectrum.  Since the major energy and mass release in the nova outburst occurs in a single episode,
this subsequent increase in the velocity displacement observed in many novae and indicative of an 
increase in the expansion velocity of the ejecta has been intriguing.  We have suggested that
the white dwarf radiation field acts on the ejecta after the pre-maximum halt which is supported
by observations.  The ejecta is dense and optically thick before the optical maximum. 
Thus the radiation field of the white dwarf can exert a radiation pressure on the ejecta and
accelerate it to higher velocities and the absorption lines appear at a higher blue-shifted velocity 
and the emission bands become wider.
After this final push by the white dwarf radiation field, most of the ejecta becomes optically thin 
and no further push to the entire ejecta is possible, which is supported by the observation
that the emission lines of the principal spectrum are never replaced by another line system.  In fact,
the principal spectrum lines and velocities survive through the evolution of the
outburst and are also detected in nova shells observed several years following the outburst. 
We do note that in the earlier models, radiation pressure had been invoked
to explain some observations e.g. the detachment of the envelope at the optical maximum from the 
white dwarf.  There also exist models in which the entire nova outburst is attributed to
the effect of radiation pressure.

About a magnitude below maximum, another independent system of absorption lines at still bluer
velocities and associated wider emission bands appears in the nova spectrum and are 
referred to as the diffuse enhanced lines.
About a couple magnitudes below maximum another independent system of absorption
lines at still bluer velocities and wider emission bands are identifiable in the optical spectra of several
novae.  These lines are of higher excitation than those of the diffuse enhanced system and are
referred to as the Orion system.  In the existing model, these two systems have been attributed 
to faster and newer mass loss from the white dwarf.  In the updated model, we argue that these
lines are also formed in the same ejecta and no fresh ejection of matter from the
white dwarf is warranted.   We refer to mass-based segregation of elements in the ejecta so
that heavier elements lag behind lighter elements as shown in Figure \ref{shock}. 
We suggest that the diffuse enhanced and Orion lines are formed in these regions. 
In Figure \ref{shock},
it is schematically suggested that the elements giving rise to the diffuse enhanced lines accumulate
near the centre of the ejecta and those giving the Orion lines occupy the rear of the ejecta.
The justification for this arrangement is the chronology of their detection, the
differences in the spectral line properties and their simultaneous detection. 
If a large number of heavy atoms accumulate in the inner half of the ejecta then their mutual 
gravity may lead to enhanced attraction, coalescence and formation of clumps 
consisting of mostly non-hydrogen elements.  From this explanation it follows that dense clumps should 
form in the inner parts of the ejecta closest to the white dwarf
as shown in Figure \ref{novanew} and it is suggested that the diffuse enhanced and Orion lines
form in these clumps.  The clumps can be optically thick unlike rest of the ejecta.
The radiation field of the hot white dwarf shining on the clumps can 
trigger either or both of the following: (1) exertion of radiation
pressure can accelerate the clumps outwards in the ejecta and (2) 
increase in the ionization and excitation of the clumps.  
Under the effect of the radiation pressure exerted by the white dwarf, 
the clumps move outwards at a velocity  larger than the average ejecta velocity (i.e.
principal line forming regions) thus explaining
the larger blue shifts observed in the absorption lines of the diffuse enhanced and Orion absorption 
systems.  We note that radiation pressure is a function of the temperature and can be estimated
as $1/3~a T^4$ where $\rm a=7.565\times10^{-16} J m^{-3}K^{-4}$ is the radiation constant.  
A white dwarf of surface temperature $10^5$ K
will exert a radiation pressure $\rm \sim 2.5\times10^4~ J m^{-3}$.
Clumps of different sizes could suffer different acceleration due to the radiation pressure and
if sufficient absorbing and emitting atoms exist, multiple lines at distinct
velocities would form.  If these lines are sufficiently closely spaced, they would appear wide
and would explain the wide diffuse lines that are observed.  The
diffuse enhanced lines are frequently observed to break up into multiple sharp components which
would indicate the decreasing influence of radiation pressure on the disintegrating clumps.
We note that the development of and detection of diffuse enhanced and Orion line systems 
is generally noticeable in slow novae whereas their presence is not always discernible in
fast novae.  Since clump formation in the ejecta involves several steps which require finite time,
%starting with mass-based segregation of elements in the ejecta, 
the entire process of clump formation is likely to be 
more efficient in slow novae which expand slowly and hence higher densities persist longer.  
In fast novae, the fast expansion can rapidly drop the densities and it is likely that there is
not sufficient time for the clumps to form and hence the diffuse enhanced and Orion systems are not
always detected. 
It has been noted that hydrogen is not always detected in the Orion system and which we think lends
strong support to the formation of clumps in the ejecta due to mass-based segregation
at the rear end of the ejecta which can be depleted of hydrogen. 
This would indicate that the Orion clumps are mostly composed of the heaviest elements in
the ejecta whereas the diffuse enhanced clumps contain intermediate mass elements and hydrogen and this
is a justification for their relative location in Figure \ref{novanew}.
From the observed velocity displacements, we can infer that the diffuse enhanced clumps and 
Orion clumps are moving with velocities which are $2-3$ times the average
expansion velocity of the ejecta quantified by the velocity of the principal system of lines. 
The clumps would hence be propelled to  
the leading parts of the ejecta after they are formed (see Figure \ref{novanew}) and before
they are destroyed.  Both these system of lines are short-lived 
and disappear when the light curve has fallen by 4 magnitudes from the maximum and the nova has
entered the transition phase of evolution.  This also gives support to the formation of these
lines in the clumps which evaporate or disintegrate as the ejecta expands.  After the disappearance
of the diffuse enhance and Orion lines and the absorption component of the principal lines, 
absorption lines disappear from the spectrum indicating the lack of absorbing material in the ejecta.  

\paragraph{Transition phase, dust formation and light oscillations:}
Light curves of a few novae have shown a steep fall in their brightness starting
about 3.5 magnitudes below maximum (see Figure \ref{lightcurve}).  This has been
shown to be due to dust formation in the nova ejecta which absorbs and obscures the optical emission
causing the light curve to plunge to a minimum.  However it is not known where and how
the dust forms in the ejecta.   From the above discussion, it appears that the Orion clumps 
provide an ideal site for dust formation.  The clumps contain heavy elements and 
the inner parts of the clumps are shielded from the hard radiation field of
the white dwarf thus facilitating dust formation.  
The Orion clumps move ahead under the influence of the radiation pressure.  If these clumps
have formed large quantities of dust and disintegrate or have a large filling factor 
in the mostly uniform density ejecta, then they
can obscure most of the optical continous emission from the ejecta and the white dwarf,
thus causing the steep fall in the brightness of the nova like DQ Herculis.  The dust eventually
disperses and the continous emission becomes visible again. 
Generally it is noted that the light curve begins its drop when the Orion system of 
lines are detectable and the light curve
revives when the ejecta is in the transition phase or has entered the nebular phase when
the Orion system of lines have disappeared.  This is generally $\ge 6$ magnitudes below maximum.
This signals the end of the transition period in which the stellar type spectra containing absorption lines
evolves into a nebular type spectra containing only emission lines.
Some correlation between detection of Orion system of lines like O II, C II,  N II and 
the deep minimum in the light curve of a nova had also been noted \citep{1945MNRAS.105..275S} which
also lends support to our hypothesis. 

While such a steep decline in brightness is a characteristic of slow novae which supports
obscuration due to the Orion clumps, 
some relatively faster novae show light oscillations of 1-2 magnitudes in the 
transition phase of the light curve (see Figure \ref{lightcurve}).  These light oscillations 
are accompanied by anti-correlated changes in the velocity displacement, excitation
of the Orion lines and colour of the optical light such that a secondary light minimum is
accompanied by a larger velocity displacement,  higher excitation in the Orion lines
and a higher colour temperature {\bf check} and vice versa.  Occasionally a correlated change is observed in
the principal system of lines and the diffuse enhanced lines
but neither are as dramatic as changes in the Orion lines (see Figure \ref{spectra}). 
We suggest that the oscillations are induced by the varying
usage of the white dwarf radiation by the Orion clumps so that the fraction of radiation that
reaches becomes variable.  We recall that in our model 
the hot white dwarf contributes $\le 2$ magnitudes to the optical light curve.  If some of this radiation 
is used up in exerting radiation pressure on the Orion clumps then it
will lead to a minimum in the light curve and a corresponding higher velocity displacement of the Orion
absorption lines.  If this light is used up in exciting an atom in the Orion
clump then the secondary minimum will coincide with a higher excitation Orion line.
One can speculate on why the light varies in an oscillatory fashion but we note that there
exist novae light curves where the recorded light does not oscillate but 
shows randomly spaced light variations which are always noted to be less than 2 magnitudes and
which show correlation with the changes in the Orion lines. 
The most obvious explanation could be that once the changes that the light is capable of effecting on the 
Orion clumps have been done, the light curve recovers and the Orion system reverts to 
its original state in which
case the light can again be absorbed by the Orion clump and the light curve dips.  This
can continue through the transition period for some novae i.e. for the time that
the radiation field can have such an effect on the Orion clumps.  
In some cases, it has also
been noted that a spectral line disappears from the Orion velocity and appears at the principal 
velocity and vice versa.  This is not surprising and merely indicates the changing effect of radiation
pressure on a clump so that its expansion velocity varies. 
About 3 magnitudes below maximum the diffuse enhanced system disappears, around 4 magnitudes below optical
peak, the principal absorption features disappear and by
4.5 magnitudes below optical maximum the Orion system of lines have disappeared. 
After this only emission lines are detected in the nova spectrum and the nova is said to
enter the nebular phase of evolution.   
No spectral lines showing the widths characteristic of the diffuse enhanced or Orion systems
are detected again.  All lines are found to show the widths characteristic of the principal system. 
Thus all the material in the ejecta expands with the velocity of the principal system. 
The optical continuus emission from the nova continues to decline and get bluer as it 
enters the nebular phase which is expected since the contribution of the expanding ejecta
to the continuous emission is decreasing and that of the white dwarf is increasing.

\paragraph{Final decline:}
Most classical novae take several years to revert back to the
brightness and spectrum of the pre-nova stage (e.g. DQ Herculis took 15 years).  A large fraction 
of the time is spent in the final decline phase of the light curve which follows the transition
phase of the light curve.  The luminosity of
the system declines by the last few magnitudes in the final decline to settle down at the pre-nova brightness. 
The brightness of the ejecta will steadily decrease as it expands and the contribution by
the white dwarf will change as accretion restarts and a low temperature envelope forms around it. 
In some novae, the ejecta is observed as a separate expanding shell around the central binary
so that the contributions of the nebula and the white dwarf can be separated.
If the light curve of the central object has not reverted to the pre-nova brightness in this case then it
gives evidence to the contribution by the white dwarf to the light curve.  Similarly the contribution
of the nebula to the light curve can also be estimated at that time.  However such novae are rare since
they have to be very closeby to enable the resolution into two components soon after the outburst.  
Since the white dwarf in the pre-nova phase is characterised by a large accreted envelope 
it appears that the time taken for the cooler envelope to form on the hot white dwarf
will define the time that the nova will take to revert
back to its pre-nova state and this could be several years depending on the accretion
rates.  This time being in decades and not in hundreds of years indicates that
the accretion restarts soon after the outburst, the accretion rates are sufficient to
build up the pre-nova envelope in decades and that further accretion only goes to
increase the density of this envelope with little change in its size. 

Recurrent novae which evolve faster on all counts as compared to classical novae
also revert to the pre-nova optical state much earlier than classical novae. 
For example, RS Oph and T CrB  
reverted back to quiescence brightness in about 95 days after their last outburst and U Sco
in about 65 days after its outburst in 2010. 
Any difference noted between the pre-nova and post-nova brightnesses
can then be traced to remnant changes in temperature or extent of this envelope 
which is disrupted by the outburst.  

\begin{figure}[h]
\centering
\includegraphics[width=6cm]{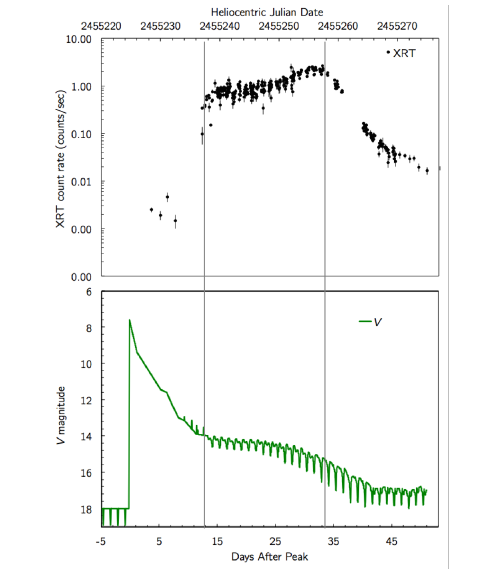}
\caption{Figure showing X-ray and optical light curves of U Sco after its
outburst in 2010 reproduced from \citet{2015ApJ...811...32P}. 
The top curve shows the X-ray (0.3-10 keV) light curve
and bottom one shows the behaviour of the V band light curve.  Note that the eclipses in
the optical light curve revive around day 13 which is when the X-rays are detected.  No radio
detection has been reported for this nova.}
\label{USco}
\end{figure}

\paragraph{Soft X-rays, radio thermal emission and optical: } 
The thermonuclear explosion on the white dwarf in a nova outburst could result in one of the
following: (1) the entire accreted envelope is energised and ejected (2) part of the
envelope is ejected while part of it expands and eventually falls back on the white dwarf (3) the
envelope is not ejected and only expands.  Observations seem to suggest that (1) is typical of classical and
recurrent nova outbursts whereas (3) describes dwarf novae.  However it is possible that either
(1) or (2) would describe classical novae depending on the energy released and the mass of the
accreted envelope which would determine whether all the matter is accelerated beyond the
escape velocity or not.  We note that in the existing model it was believed that the
optical emission arose from the photosphere of the white dwarf i.e. the expanded envelope which
had not been ejected.  Since we show that the optical emission arises in the ejecta, we have
to search for the evidence of an inflated photosphere, if it exists, in other observations. 

If we can assume that the detection of soft X-rays which has to be from matter at
temperatures $\ge 10^5$ K implies that the entire accreted envelope has been
ejected in the explosion exposing the hot surface of the white dwarf then it would lend support
to point (1) made above and hence to the absence of a cooler remnant photosphere from the outburst.  
We recall that the quiescence surface temperature
of the white dwarf has been estimated to be between 8500 and 50000 K
\citep{1999PASP..111..532S} which will not emit soft X-rays and thus the detection of soft X-rays
after the outburst signalling the ejection of the entire envelope is a plausible one.  However we note that 
the non-detection of soft X-rays does not necessarily imply the presence of a remnant photosphere 
(point 3 above) since the reasons could be the optically thick ejecta
absorbing the soft X-rays and/or accretion restarting so that the hot surface of 
the white dwarf is beginning to be obscured.

It is commonly observed that hard X-rays, especially from fast novae are detected
soon after the nova outburst \citep[e.g.][]{2011ApJS..197...31S} while soft X-ray emission is 
detected later e.g. it was detected in U Scorpii and V339 Delphini when the optical light 
curve had dropped by about 6 magnitudes below maximum. 
In Table \ref{tab2}, the onset and end times for the soft X-ray phase taken from 
\citet{2011ApJS..197...31S} are listed in columns 2 and 3 while the decline in the 
V band luminosity of the nova from maximum at 
those times as estimated by eye from the AAVSO light curves are listed in columns 4 and 5.  
The values in the last two columns should 
be considered as rough estimates.  As listed in Table \ref{tab2}, soft X-rays are generally
detected in the transition phase or beginning of the nebular phase i.e. when the peak luminosity
has dropped by $\ge 4.5$ magnitudes.  The similarity in the evolutionary stage of the optical light curve
when soft X-ray is detected in novae gives strong support to its detection being tied up
with the entire ejecta being ionized and becoming transparent to soft X-rays.  It rules out
the detection epoch of soft X-rays corresponding to a change in the white dwarf photosphere i.e.
it contracting sufficiently so that the inner hot portions are detectable which is often cited
as the reason in literature.  
Detection of atomic hydrogen in the ejecta of V339 Delphini till about day 40 post-maximum 
\citep{2014A&A...569A.112S} and the detection of soft X-rays around that time indicates how
the opacity of the ejecta to X-rays can be high and be responsible for the delay in its detection.
This, then, supports the point (1) listed above in which it is suggested that the nova outburst ejects
the entire accreted envelope which leaves behind a hotter white dwarf which should start
emitting soft X-rays as soon as the ejecta is adiabatically energised and detached from the white dwarf.

However this does not rule out a thin layer of material, possibly the matter synthesised in
the thermonuclear outburst, remaining on the 
white dwarf - it only shows no evidence for an inflated photosphere of several solar
radii.  The black body radiation of the hot white dwarf will also contribute to the last $\le 2$ 
magnitudes rise in the optical light curve in some novae.  The turn-off times
show a larger variation with the slow novae detectable in soft X-rays for longer periods till
the nova brightness drops significantly (see Table \ref{tab2}).  
The turn-off of soft X-rays from novae should indicate onset of obscuration of the
hot surface of the white dwarf as the accreted material begins to accumulate on the white dwarf. 
Thus, it appears that slow novae take longer for the accreted matter to envelope the hot
white dwarf which could be indicative of late resumption of accretion or low accretion rates. 
In nova T Pyxidis, soft X-ray emission was detected between 7.5 hours and 12 days after
which it faded \citep{2011ATel.3285....1K}.  A long-lived soft X-ray phase began from
day 117 \citep{2011ATel.3549....1O} when the nova had faded by about 5 magnitudes from the visible peak. 
These observations lend strong support to the soft X-rays from the hot white dwarf being
emitted as soon as the matter is ejected but being detectable by us only when the atomic hydrogen in the
ejecta is ionized in addition to suggesting that matter in the ejecta which if ionized on
ejection, recombines to form atomic hydrogen as it expands. 

Several novae show a slower or no decline in the V band light curve during the soft X-ray phase
(see Figures \ref{USco},\ref{v339del}) strongly implying a connection between the two.   
Radio thermal emission is generally detected at similar epochs as soft X-rays with an onset  which 
precedes detection of soft X-rays by a short interval (see Figure \ref{tpyx}).  
These correlations which are observed in several novae signify a common system of changes in the nova 
which affect the two wavebands.  Such correlations can be instrumental in improving our
understanding of novae.  We note that soft X-rays which arise on the white dwarf can become detectable 
when the ejecta is optically thin to the radiation i.e. fully ionized.  
Radio thermal emission is due to the physical process of free-free emission in the ionized ejecta and
the near simultaneous detection of emission at the two distinct wavebands implies 
that the increasing ionization of the ejecta leads to a rise in the radio thermal emission and also
to the ejecta becoming transparent to the soft X-ray emission.
In many cases, the radio thermal emission shows the frequency-dependent onset due to
optical depth effects.  A temperature and hence ionization gradient in the ejecta such that
the inner parts are at a higher temperature and ionized and the leading parts are cooler and partially ionized, 
will delay the onset of the thermal radio emission due to opacity. 
When the entire ejecta is ionized and isothermal, no further absorption can happen and 
the radio thermal rapidly becomes detectable.  The plateau observed in the
optical light curve during the soft X-ray phase means that an additional source of 
optical emission has been added in the nova system.  This could indicate an increase in the temperature
of the entire ejecta and hence an increase in the free-free and free-bound emission.  One can think of
the increased optical emission being due to the
contribution from the hot white dwarf to the optical bands or contribution from an ionized band around
the companion star ionized by the X-rays from the white dwarf.   However these reasons require the
ejecta to be optically thick to optical wavelengths till the soft X-ray phase. 
The optical emission appears bluer in the soft X-ray phase.
Since we suggest that the hot white dwarf contributes $\le 2$ magnitudes to the optical emission
near the maximum, no further rise due to the white dwarf radiation is expected unless parts of the ejecta
remain optically thick to the white dwarf emission.   We realise that there is some degeneracy in these
scenarios and independent data are required to break it. 

In the 2010 outburst of the eclipsing recurrent nova U Scorpii, soft X-rays were detected between day $\sim 13$
and day 33 during which the optical light curve showed a plateau
and the dips in the optical light curve due to the eclipse of the white dwarf, which had not been
present after the outburst, also resumed around day 13 \citep{2015ApJ...811...32P} (see Figure \ref{USco}). 
The optical light curve had dropped by about 6 magnitudes from the peak. 
Even in its 1999 outburst, U Sco showed a plateau in the optical emission between days 10 and 33
after peak \citep{2010ApJS..187..275S} and it was detected in soft X-rays (0.2-2 keV) 
when it was observed 19-20 days after the optical peak \citep{1999A&A...347L..43K}.  Similar
multi-band behaviour in successive outbursts
implies that similar mass is ejected in all the outbursts so that the ejecta properties are 
similar and the nova evolves on similar timescales.  Thus the ejecta becomes transparent to
soft X-rays at similar epochs in each outburst.   
Soft X-ray emission was detected at similar epochs following the optical peak in the 
1985 and 2006 outbursts of RS Ophiuchi also.  This could have been suggested from the 
very similar optical light curves which are noted for recurrent novae in all outbursts but
it is always better to get observational evidence of the same before any theory is hypothesised.  
It is also worth noting here that differences in the densities of the ambient medium around
RS Ophiuchi was noted from radio synchrotron studies of the two outbursts.  Inspite of this,
similar soft X-ray and optical light curve behaviour indicates that these emissions predominantly
depend on the ejecta properties and not the ambient matter. 
Ultraviolet emission (2600 A) was found to be stronger than the visible band during the 
the soft X-ray phase and returned to normal around day 35 when the soft X-rays were extinguished. 
\citep{2015ApJ...811...32P}. 
\begin{table}
\centering
\caption{X-ray onset/end times from \citet{2011ApJS..197...31S}.  The difference of
the optical emission with respect to the peak on the onset/end days of X-rays are
listed in the last two columns.  These
are visual estimates from the V band light curves downloaded from AAVSO website and hence the values
are only approximate. }
\begin{tabular}{l|c|c|c|c}
\hline
Nova  &   \multicolumn{2}{c|}{\bf X-ray}   &   \multicolumn{2}{c}{\bf Optical}    \\ 
      &    {\bf onset} &  {\bf end}  &         \multicolumn{2}{c}{\bf below max} \\
      &      &    &       onset  & end      \\   
      &      days  &    days   &       mag  & mag      \\   
\hline
KT Eri 2009   &    71 & 280   & $\sim 4.5$  & $\sim 7.5 $ \\
RS Oph 2006    &    35 & 70   & $\sim 4.5$   &  $\sim 5.2 $ \\
U Sco 2011      &   23  & 34  & $\sim 6.5$   &  $\sim 7.5 $ \\
V1494 Aql 1999  &   217 & 515 & $\sim 6.5$   &  $\sim 10 $ \\
V1974 Cyg 1992   &  201 & 561 & $\sim 5.5$   &   $\sim 9 $ \\
V2491 Cyg 2008  &   40  & 44  & $\sim  6$    &  $\sim 6.5 $ \\
V407 Cyg 2010   &   15  & 30  & $\sim  2$    &  $\sim 2.5 $ \\
V4743 Sgr 2002   &  115  & 634 & $\sim 4.5-5$  & $\sim 8.5 $ \\
\hline
\end{tabular}
\label{tab2}
\end{table}

This above behaviour observed in U Sco can be explained as follows.
The optical emission is predominantly from the ejecta which quickly expands beyond the binary extent
and hence no eclipse signatures are observed in the light curve.   However if the 
white dwarf had contributed upto 2 magnitudes to the light curve near maximum then the light curve
should have shown eclipse signatures due to the periodic occultation of this contribution.   
However data show no such signature upto day 13 indicating that there was no contribution of the
white dwarf radiation to the light curve till day 13 or that it was present for a short time
near the maximum and absent since.  
Noting that recurrent novae  often have optically thick winds blowing from the massive
white dwarf due to the high accretion rates, the winds will increases
the ambient densities.  This matter will be swept-up by the ejecta as it expands and will increase its opacity.  
A second reasoning is what is generally favoured in literature that the white dwarf after the outburst is 
surrounded by an inflated photosphere which encloses the companion star and only after it has
deflated to a size smaller than the binary separation that the eclipses can resume and an inner hot
surface which emits soft X-rays is revealed explaining the simultaneous detection of emission at
both wavebands. 
We prefer explaining the plateau and the increasing depth of the eclipses in the optical light 
curve of U Sco due to the increasing contribution of the white dwarf to the light curve as the ejecta gets 
transparent and reveals the binary (see Figure \ref{USco}).  
On a general note then, the plateau or reduced rate of decline in optical emission that often 
accompanies the detection of soft X-rays indicates an enhanced contribution to the light curve
either from the binary or from increased emission measure of the ejecta.
Once the X-ray emission declines, the optical light curve also starts to decline indicating
that the extra source of light is also fading.

The V-band eclipse is about 1.3 magnitudes deep in quiescence (Schaefer 2010).
We use the detailed eclipse data that has been gathered during the 2010 outburst (Schaefer et al. 2011) 
to understand the evolution of this system.
The eclipse was much shallower when it was first detected after day 13 and a secondary eclipse
was also detected which is not 
observed in the V band during quiescence.   Both the primary eclipse 
and secondary eclipse kept increasing in depth during the soft X-ray phase.  
After about day 32 when the soft X-ray phase was ending, the secondary eclipses disappeared
and the primary eclipse appeared to be similar in depth to that in quiescence (Schaefer et al. 2011).    
However peculiar behaviour was noted when the second plateau in the optical
light curve began around day 41 after outburst.  The primary eclipse became shallower by 0.2-0.3
magnitudes and light variations upto 0.6 magnitudes were seen in the non-eclipsing parts of
the orbit.  This behaviour continued till day 67 when it is believed that the nova returned
to quiescence.   The primary eclipse seems to have been deepest between days 32 to 41. 
We note that a scatter of 0.4 to 0.6 magnitudes seems to be present in the non-eclipsing
parts even in the quiescent B and I band light curves and a secondary eclipse of
magnitude 0.3 is detected in the I band (Figure 45,47 in Schaefer 2010).
The outburst behaviour of the eclipsing light curve can be explained in our model as follows.   
We recall that in this model the outburst ejects the entire accreted low temperature envelope
exposing the hot inner surface on the white dwarf and hence the white dwarf should emit
soft X-rays soon after ejection.  However the soft X-rays will only be detectable when
the foreground ejecta is fully ionized and hence transparent to the X-rays.   
These X-rays can heat and ionize the outer atmosphere of the companion star especially the parts
lying close to the orbital plane which could enhance the emission from the companion star. 
The plateau in the light curve indicates a fresh source of optical emission which adds to
the declining ejecta emission.  The likely sources could be the white dwarf or the irradiated
companion star.  That secondary eclipses are seen alongwith the primary eclipse after day 13 in
U Sco indicates that the emission from the companion star has increased by about 0.2
magnitudes in the V band due to irradiation by the X-rays and which is eclipsed by the white dwarf
during the orbital motion.  Such transient emission from the companion star can contribute
to the observed slowing down of the light decay rate.  When the X-rays fade, the companion should
revert to its quiescent state and the secondary eclipse should fade.   That the secondary eclipse is 
not observed after around day 32 supports the hypothesis. 
Coming to the primary eclipse, in the first detection after day 13, the eclipse is shallow
indicating that the contribution of the white dwarf to the light curve is lower than seen in quiescence. 
As time passes, the eclipse keeps getting deeper with it being deepest between days 32-41 after
the end of the supersoft phase indicating that the V band emission from the white dwarf has
returned to its quiescent value which in this case appears to be more than during
the outburst.  Recall that the end of
the soft X-ray phase is hypothesized to be due to accretion restarting and the accreted material 
forming a cooler envelope around the white dwarf which does not emit in soft X-rays.    
Thus, it could be that such an envelope was formed after day 32 - ending the soft X-ray phase and
returning the eclipsed light curve to the quiesent depths.  
However the light curve after day 41 shows several peculiarities 
with light variations upto 0.6 magnitudes in amplitude detected 
throughout the orbital motion and the primary eclipse being shallower
than it was between days 32 and 41.  This was also the start of another plateau in the light curve
indicating another source of light being added to the light curve.  The easiest explanation seems
to be a varying source of light contributing to the light curve and giving rise to the light variations
which continued all the way to quiescence. 
Since in quiescence, U Sco shows variations upto 0.5 magnitudes in the B and I bands
(Schaefer 2010), this behaviour after day 32 could be due to the same reason.

While the soft X-ray phase in most novae is found to be $<3$ years (Schwarz et al. 2011),
there do exist novae which are detectable in soft X-rays for a decade or longer.
One of them is the fast nova V1500 Cyg which recorded an outburst in 1975.   
Its progenitor star had B = 21.5 magnitudes. The nova has taken 30-35 years after outburst
to decrease to 19 magnitudes \citep{2010ApJ...708..381S} which is still brighter than quiescence.  
The soft X-ray phase has also been longer and soft X-rays, albeit faint, are still
detectable from the nova \citep{2010ApJ...708..381S}.  This supports our model
wherein the soft X-rays fade when the cooler accreted envelope forms around the white dwarf and
and which is also responsible for the nova brightness in the optical bands
reverting to the pre-nova values.   The long duration of the soft X-ray phase 
could indicate extremely low accretion rates.  
Another nova with an exceptionally long soft X-ray phase which lasted for more than 15 years was V723 Cas 
(Schwarz et al. 2011).  A decade-long soft X-ray phase was observed in GQ Muscae which recorded an
outburst in January 1983.  Soft X-rays were detected in 
April 1984 (Ogelman et al. 1984) upto 1993 (Shanley et al. 1995).
The nova had $t_3\sim40$ days but the light curve entered a plateau after that 
which lasted for about a year.  The soft X-rays turned on when the light curve was within this plateau 
in 1984.  However this nova was somewhat different from the typical case in that the light curve resumed its
faster decline later in 1984 when soft X-rays were still detectable from the system. 

We summarize the above:
\begin{itemize}
\item The brightening upto the pre-maximum halt and the pre-maximum spectrum are entirely
due to the energising of the ejecta matter by the explosion and contains no energy contribution
from the white dwarf.
\item The final brightening by $\le 2$ magnitudes to the optical peak is probably
a combination of the rapid isothermal expansion of the optically thick ejecta
due to the radiation pressure exerted by the hot white dwarf and
contribution from the hot white dwarf radiation.
The absorption features of the principal spectrum detected at the light curve maximum has a velocity 
displacement which is larger than pre-maximum lines which is attributed to the radiation pressure exerted on the
optically thick ejecta by the white dwarf radiation field.
\item We suggest that mass-based segregation occurs in the ejecta so that the heavier elements
accumulate in the inner parts of the ejecta.  These facilitate the formation of clumps in which
the diffuse enhanced and Orion spectral systems characterised by higher velocity
displacements and excitation are suggested to form.   All higher velocity displacement
features detected post-maximum are attributed to the effect of the radiation pressure on the
optically thick clumps.
\item It is suggested that dust forms within the Orion clumps.
As the Orion clumps move outwards in the ejecta under the influence of radiation pressure,
the dust can obscure the optical continuous emission and the light curve can show a rapid fading like in
DQ Herculis.
\item Some novae show oscillations of amplitude 1-2 magnitudes in the light curve with
correlated changes in the Orion system of lines.  We suggest this is direct evidence
to the contribution of the white dwarf ($\le 2$ magnitudes) to the light curve and effect of the radiation 
field of the white dwarf on the Orion clumps.
\item The final decline in the light curve to pre-nova brightness takes several years
in many classical novae.  We suggest that this timescale is the time needed for the
accreted cooler envelope to form around the white dwarf so that its brightness reverts to pre-nova
state.   
\item Soft X-rays arise on the hot surface of the white dwarf and should be present
soon after the energetic ejection of the envelope.  However detectability depends on the
foreground ejecta becoming transparent which happens when it is fully ionized thus
delaying the onset of soft X-ray phase to the transition phase of the light curve.  The end of the soft X-ray
phase signals the restart of accretion and accumulation of a lower temperature envelope
on the white dwarf.  The soft X-ray phase is generally accompanied by a plateau in the optical
light curve indicating an additional contribution to the light.   

\end{itemize}

This completes the primary discussion on the updated model which explains the evolution of nova outburst. 
In the following, we discuss a few observational results on novae 
in light of the updated model and determine simple
parameters of the nova from the model. This also helps us elucidate and fine-tune our model. 
The list is not exhaustive and more observational results should be examined and it should be possible
to explain those with the model.  This will help better understand the model and also fix any
problems that remain.  As with rest of the paper, no effort is spared in keeping the discussion
bound by physics, observations, common sense and consistency. 

\paragraph{Emission measure of the ejecta:} In the existing model, the optical
continuum is believed to arise in the initially expanding and post-maximum contracting
photosphere of the white dwarf while in our model the main contributor to the optical continuous
emission is the ejecta and the white dwarf contributes
only the last couple magnitudes between the pre-maximum halt and maximum and possibly to 
the plateau in the light curve.  It is also likely that the X-ray irradiated companion star
contributes to the light curve during the plateau phase.  It appears that optical radiation from
central binary in some novae starts contributing to the light curve only at the onset of the 
plateau and till then the ejecta blocks it.  In fact, it appears that the novae
which show the coincidence of a plateau in the optical light curve with the appearance of
soft X-rays should not show a pre-maximum halt before the maximum light.  If it did show
a pre-maximum halt then it would mean that light from the binary did penetrate the ejecta.
An early transparency of the ejecta followed by an opaque phase might also happen in some novae.
Thus, while on a general basis we suggest that the explosive energy is mainly responsible for
the sudden brightening of the nova upto the pre-maximum halt and the last $\le 2$ magnitudes
is due to the white dwarf, we believe that there will be exceptions to this case since a range
of physical conditions exist in the nova shell. 

\begin{figure}
\centering
\includegraphics[width=6cm]{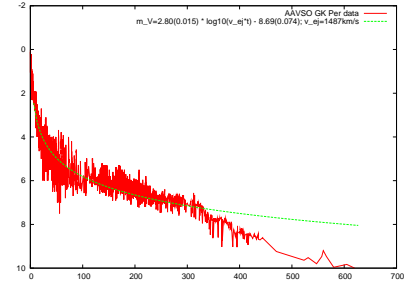}(a)
\includegraphics[width=6cm]{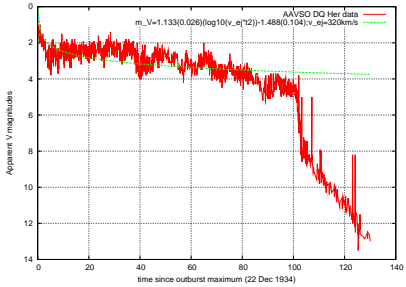}(b)
\includegraphics[width=6cm]{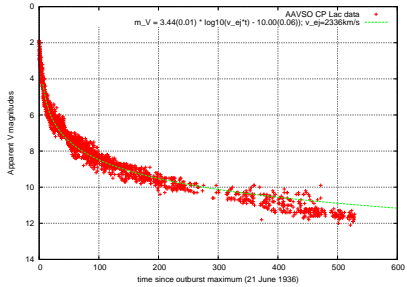}(c)
\caption{The V band light curves of three novae downloaded from the AAVSO website: DQ Her(a),
GK Per(b), CP Lac(c).  The smooth line shows the fit to the light curve estimated from decreasing 
emission measure of the ejected shell.    
The principal absorption line velocities are used as proxy to ejecta velocities
and are taken from \citet{1960ApJ...131..739M,1954ApJ...119..124M}. }
\label{EMnova}
\end{figure}

After the maximum, the light curve begins 
to decline.  In our model, the decrease in the optical flux is due to the decreasing densities and
hence emission measure in the expanding shell.  
For simplicity, we assume that drop in emission measure is dominated by the radial expansion of the 
shell and its thickness is constant.  An increasing thickness will lead to a more rapid decline 
in the emission measure.  We hence 
model the post-maximum light curve with the decreasing emission measure of the expanding ejecta of
constant thickness.  We assume that the density in the expanding ejecta shell falls off as $1/r^n$ i.e.
$n_e \propto 1/r^n$ where $r$ is the radial separation of the shell from the central star.  
In the optically thin phase, the V band emission at any given time $t$ after optical maximum will be
proportional to the emission measure i.e.
$ L_{V}(t) \propto n_e(t)^2 R$ where $R$ is the shell thickness which we have assumed to be constant.
%and assuming that the electronic and ionic densities are the same.   
Since $r= v_{ej}~t$, we have $1/r^n = 1/(v_{ej}~t)^n$.  
Thus, $L_{V}(t) \propto 1/(v_{ej}~t)^{2n}$ and hence $ M_V = -2.5 log_{10} L_{V} + K   =
5n~log_{10}(v_{ej}~t) + K$.  We use $v_{ej}$ in kms$^{-1}$ and $\rm t_2$ in days.  
These function fits to the post-maximum light curve data of three novae
are shown in Figure \ref{EMnova}.  The light curve data have been downloaded from the
AAVSO website. The rate at which the density of the ejecta falls as it expands i.e. $n$ is hence
determined.  We find that $n$ is 0.56, 0.23, 0.69 for GK Persei, DQ Herculis and CP Lac respectively.  
The slowest nova (DQ Herculis) shows the smallest index i.e. the slowest change in density with time,
as expected.  Thus if we assume that all novae start with the same initial electron densities
then at some $t$ after optical peak, the density in a fast ejecta will be lower than in a slow nova and so will
the emission which is another way of saying that $t_2$ for a fast nova is smaller
than for a slow nova when due to the declining emission measure in an expanding ejecta.  
After about 7 magnitudes below maximum in GK Persei, 3 magnitudes below maximum
in DQ Herculis and 9 magnitudes below maximum in CP Lac, the light curve appears to
decline more rapidly than expected from the estimated density index $n$ (see Figure \ref{EMnova}). 
This appears to be the case for the smooth i.e. S type novae shown in Strope et. al. (2010).
In case of DQ Herculis, the decline is due to dust obscuration of the continuum emission.  
This exercise supports the dominant contribution to the 
optical luminosity near maximum being from the ejecta and the decline in the light curve being due 
to the expansion of the ejected shell. 

\begin{figure}[t]
\centering
\includegraphics[width=7cm]{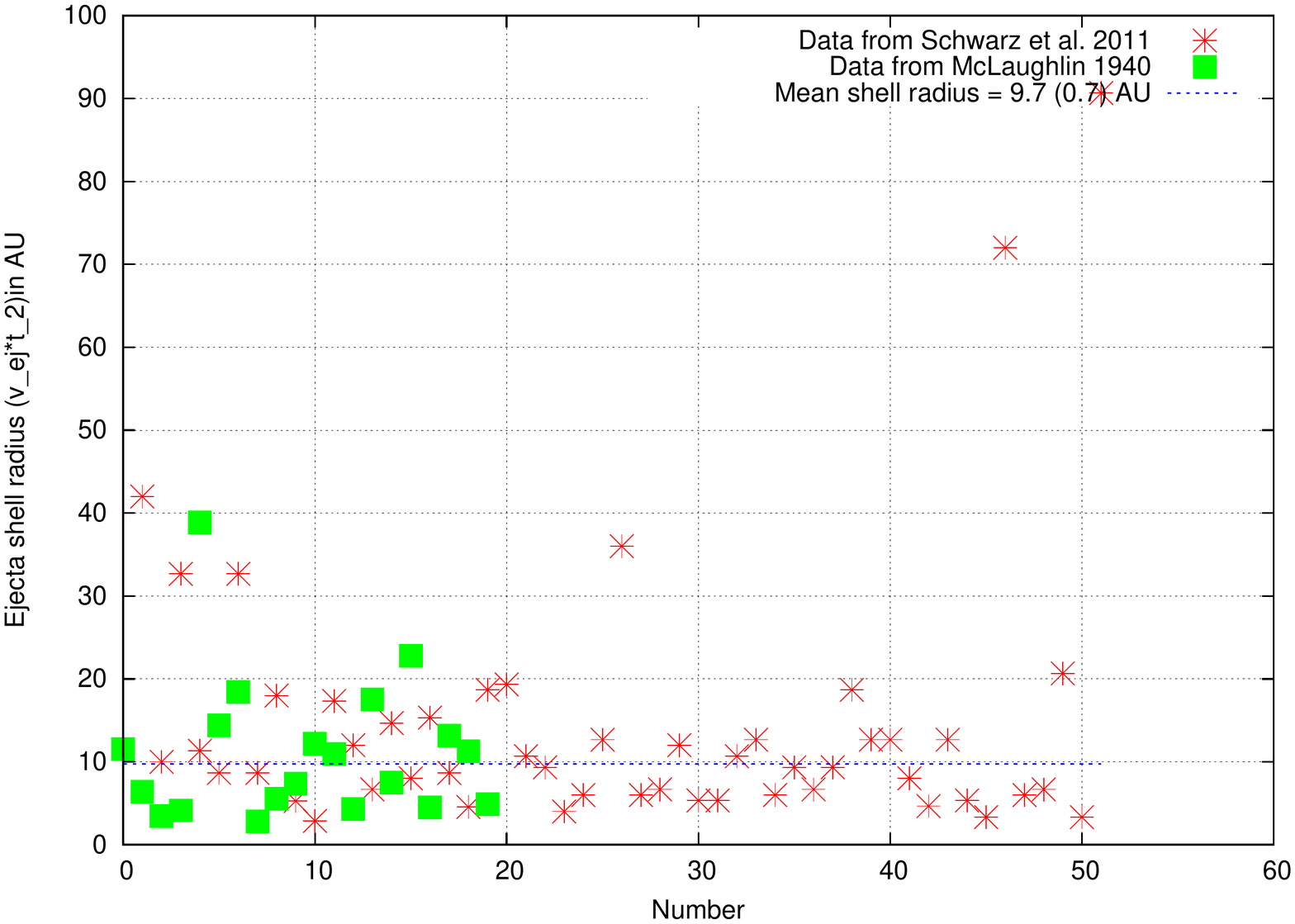}(a)
\includegraphics[width=7cm]{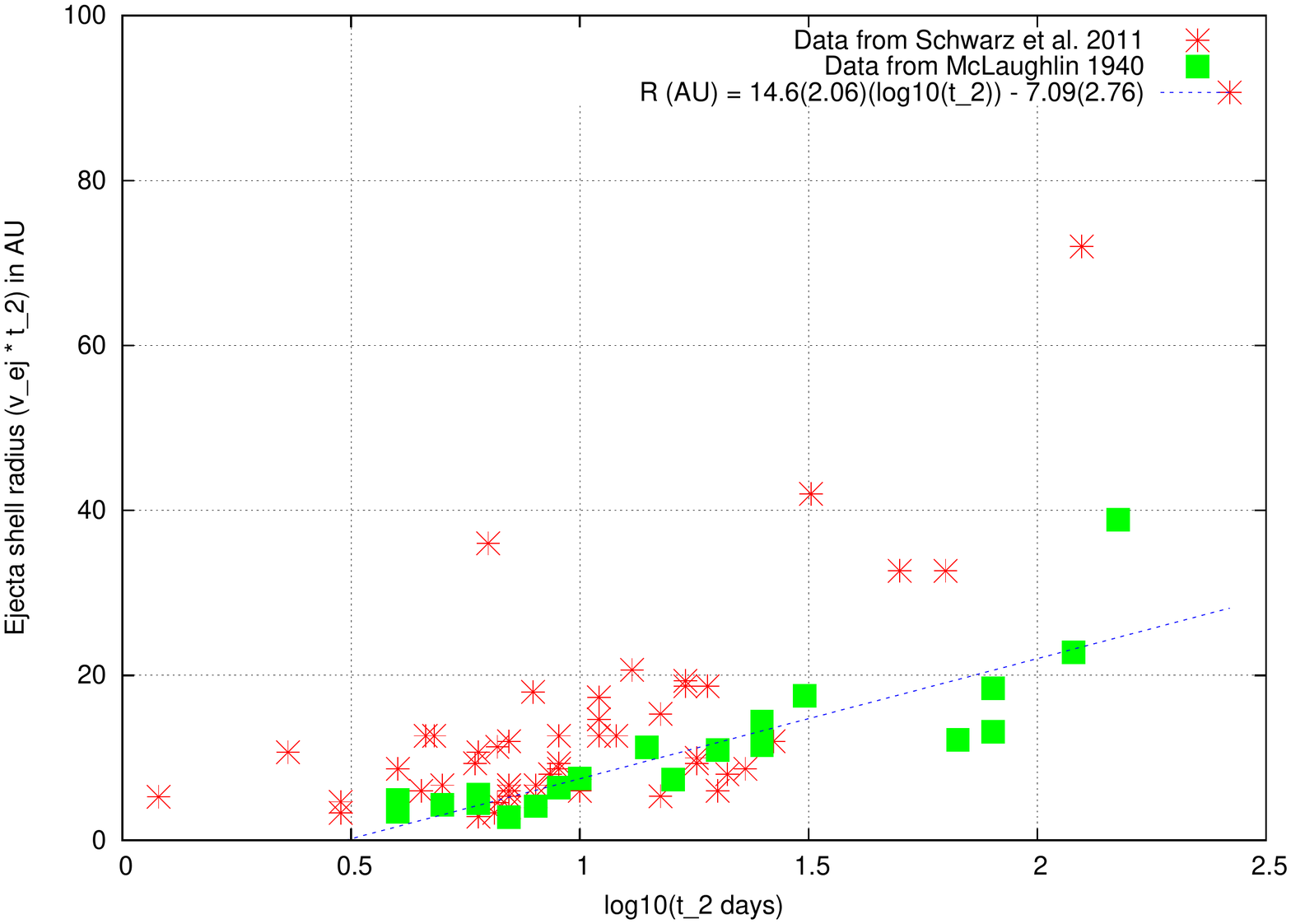}(b)
\caption{Radial separation $r_2$ of the shell from the white dwarf at time $t_2$ estimated
for the sample of novae in \citet{2011ApJS..197...31S,1940ApJ....91..369M}.   
In (a) $r_2$ appears to show a well-defined mean value.  (b) $r_2$ is plotted against
$t_2$ and in which a correlation is detectable which supports a higher mass ejection in slow novae.  }
\label{shellradius}
\end{figure}

\paragraph{Radius of ejecta shell at $t_2$:} We estimated the radius $r_2$ of the ejected shell when the peak
luminosity had declined by two magnitudes using $\rm v_{ej}$ and $\rm t_2$.  These ranged from
3 kpc to 90 kpc (see Figure \ref{shellradius}) for the sample of novae listed in 
\citet{2011ApJS..197...31S}.  Excluding 6 novae with $r_2 > 30$ kpc,
a mean value of $9.7\pm0.7$ AU (see Figure \ref{shellradius}a) is determined. 
The sample of novae listed in \citet{1940ApJ....91..369M} is also overplotted and show the 
same distribution of shell radii. 
$r_2$ shows no correlation with $\rm v_{ej}$ except that the slow novae show a larger 
spread in $r_2$.  In Figure \ref{shellradius}(b), $r_2$ is plotted against $t_2$ for a nova and a
correlation such that slow novae show a larger radial extent of the
shell at $\rm t_2$ seems to be present.  The correlation seems tighter for the sample of
novae taken from \citet{1940ApJ....91..369M} and a fit to this sample 
gives $r_2=\rm 14.6(\pm2.06)~log_{10}~t_2 - 7.09(\pm2.76)$ with $t_2$ in days and $v_{ej}$ in kms$^{-1}$
which is shown by the solid line in Figure \ref{shellradius}b.
The correlation appears to be present in the sample from \citet{2011ApJS..197...31S} but the scatter
appears to be larger and the slope slightly different. 
The radial extent of the shell at $\rm t_2$ quantifies
the required change in the emission measure of the shell to reduce the luminosity
by two magnitudes.
The correlation of the shell extent with $\rm t_2$ such that slow novae need a larger shell extent
to reduce the emission measure sufficiently for the luminosity to change by two magnitudes 
could be suggestive of larger initial densities in the ejecta of slow novae compared to
fast novae.  This could indicate a larger mass ejection in slower novae compared to
faster novae and hence explain their distinct speed classes if the outburst energies are comparable.  
However since the outburst energy from the CNO thermonuclear reaction 
is a sensitive function of temperature beyond $10^8$ K so that it is proportional to $T^{18}$, 
a spread in outburst energies should also be expected.  This makes us cautious and we end
with the comment that while it does seem possible that slow novae eject a larger mass, 
we need to investigate this point further before drawing any conclusions. 

\paragraph{Narrow lines post-outburst:}
Some novae have shown the presence of narrow emisssion or absorption lines of helium (He I or He II) 
and hydrogen located at their rest frequencies within a broader feature in a spectrum taken soon after 
maximum (e.g. Nova Geminorum 1912, Persei 1901, Aquilae 1918, T Coronae Borealis 1946) and 
the origin has been suggested to be near the central star 
\citep[e.g.][]{1947PASP...59...81M, 1949POMic...9...13M}.  Note that these lines were first
detected before it was known that novae were binaries with the explosion occuring on a white dwarf. 
More recently, \citet{2014A&A...564A..76M} detect a narrow He II in KT Eridani.  Other novae which
have shown such narrow components are U Scorpii, DE Circinus and V2672 Ophiuchi. 
Their origin on the white dwarf is supported by our model. 
The outburst leaves behind a hotter white dwarf which will be eventually detected in soft X-rays
when the  ejecta becomes transparent to it.  In the novae where the ejecta becomes optically
thin to the white dwarf radiation in the visible bands
around maximum i.e. the novae wherein the white dwarf contributes $\le 2$ magnitudes after
the pre-maximum halt, high excitation lines forming on the white dwarf, if sufficiently
intense should be detectable soon after the maximum. 
In the novae where the ejecta remains opaque to the optical radiation till much later (for example
some of the novae which show a plateau in the light curve near the transition phase),
these narrow lines should not be detectable close to the maximum.  However if the plateau arises
due to an additional non-white dwarf component to the emission then the correlation will not exist. 
One can speculate that while an absorption line could arise in the photosphere around the white dwarf, 
an emission line could indicate formation in a wind-like component and which could also signify
the restarting of accretion in the system. 
Novae at minimum show a blue continuum which is sometimes superposed by emission lines of
helium and hydrogen \citep[e.g.][]{1938ApJ....88..228H}.
At the end of the outburst, a featureless continuum which is strong in the violet or with
emission lines of hydrogen or helium lines at rest frequencies is often observed. 
The similarity of the narrow lines detected near maximum with the features detected
at minimum support the origin of the narrow lines on the white dwarf and can help
understand the physical properties of the envelope around the white dwarf.

\paragraph{Temperatures of novae in outburst:}
We can estimate various temperatures for a nova in outburst: (1) colour temperature derived from the
continuous spectrum, (2) photoelectric or Zanstra temperature derived from intensities of emission
lines relative to the continuous spectrum,  (3) excitation temperatures derived from the relative
strengths of emission lines of differing excitation potential and (4) electron temperature of
the ejecta.  \citet{1943POMic...8..149M} has discussed these temperatures and states 
that `..the different methods
yield very different temperatures from observations made on the same date,....'.  He also adds
that `...that different investigators derive different temperatures from the same method'.  
While this could be indicative of the wide range of excitation conditions and 
varying abundances in the ejecta,  
it would be advisable to not consider the determined temperatures as exact values but use
the estimated range of values to study trends.  It was believed that the colour,
photoelectric and excitation temperatures determined the temperature of the central star
since the star was believed to be the main source of the continuous emission and 
the main exciting source for the spectral lines
while the electron temperature characterised the ejecta gas.  Thus, astronomers expected
the colour, photoelectric and excitation temperatures to match and it was surprising
that this was not found to be supported by the empirical data. 
The photoelectric temperatures estimated using spectral lines of He II 4686A, H, N III, N IV, 
[O III] differ with the lowest temperatures estimated from the hydrogen
and oxygen lines and systematically higher temperatures from rest of the lines. 
This was believed to reflect the range of excitation conditions which prevailed inside a nova ejecta.
The typical photoelectric temperatures were estimated to range from 25000 K to 75000 K
\citep{1943POMic...8..149M}.  The excitation temperatures estimated from 
the ratio $\rm He II~4686 / H\beta$ was around 70000 K \citep{1943POMic...8..149M}.
Colour temperatures for novae were found to vary from few thousand K measured near optical maximum and
increasing to 20000 K in the decline phase of the light curve \citep[e.g.][]{1943POMic...8..149M}.
Thus the estimated excitation and photoelectric temperatures were systematically higher 
than the colour temperatures.  Electron temperatures using the intensity ratio of the [O III] lines
(i.e. I(5007+4959) / I(4363)) were estimated to range from 6000 to 10000 K showing a tendency
towards a decline as the nova faded \citep{1960stat.conf..585M} and are 
closer in magnitude to the colour temperatures.
This discrepancy, especially the very different values of colour temperatures, were hard to reconcile
with the prevalent model.  The study of temperatures of novae seems not to have been 
pursued to any great extent after this.  

In the updated model, we can explain the aforementioned discrepancy.  
We recall that in the updated model, all the energy upto the pre-maximum halt is believed to 
be due to the explosion owing to the opacity of the ejecta and the earliest contribution of the 
white dwarf to the energy budget can only be near the maximum.  Thus, when measured 
near the maximum, the photoelectric and excitation temperatures will be proxy to the explosion energy.
As the light curve evolves, they will indicate a combination of the explosion energy and the
radiation field of the white dwarf and as the fractional contribution of the white dwarf increases,
they will be indicative of the temperature of the white dwarf as the explosion energy is
not replenished.  Thus the photoelectric and excitation temperatures are always high since they
denote the temperature of the hot white dwarf or equivalent temperature of the explosion 
energy.  Since in the updated model, the optical continuum at maximum is dominated by the 
ejecta with some contribution from the white dwarf, the colour temperature is dominated by
the low temperature ejecta at the maximum and is comparable to its electron temperature.
As the light curve declines, the fractional
contribution of the hot white dwarf to the light curve increases as the ejecta fades and hence the colour
will get bluer i.e. the colour temperature will rise as the light curve declines as has been estimated.  
The electron temperature is expected to decrease as the light curve declines since the ejecta
is cooling and fading.  Thus the
excitation and photoelectric temperatures are indicators of the temperature of the same phenomenon
- explosion energy, white dwarf and a combination of both and hence are expected to be similar.  
On the other hand, the colour temperature determined from the optical continuum will be a combination
of emission from the ejecta and the white dwarf. It will be closer to the electron temperature
of the ejecta near the optical peak and then increase as the light curve declines 
to finally settle down to the quiescent colour temperature of the system. 
Thus the measured temperatures can be explained in the updated model. 
It would be useful to determine these temperatures for newer novae and further verify this.  
We recall a comment by \citet{1943POMic...8..149M} that `...the trends of
temperature are almost exactly the opposite of what would be expected if the nova phenomenon
were due to a simple heating and subsequent cooling of the surface of a star.'

\paragraph{Outburst winds:} While the occurence of a single explosive ejection of matter
is well established from observations, the existing model also includes a subsequent long-lived 
wind phase with dramatically lower mass loss rates.  In the old model, these winds are 
believed to be the site of formation of the diffuse enhanced and Orion line systems 
\citep[e.g.][]{1943POMic...8..149M}.  
In the updated model, we are able to explain most of the observational results with 
phenomena associated with the main ejecta and have not needed to invoke continuing mass loss 
from the white dwarf.  This prompts us to suggest that the energetic explosion is capable
of adiabatically energising and ejecting all the matter above the layer in which the nuclear
explosion occurs and hence there remains no matter to eject as winds. 
It appears certain that the role of white dwarf winds in the post-maximum phase
is non-existent or more restricted in the updated model than it was in the old model. 

\paragraph{Synchrotron radio and hard X-ray emission from novae:} 
While radio thermal emission has been detected in several classical novae, 
synchrotron radio emission has been detected in only a few novae. 
The thermal emission is due to the physical process
of thermal brehmstrahlung whereas the synchrotron radio emission arises from
relativistic electrons accelerated in a magnetic field.  The emission sites are hence
different with the thermal radio emission arising in the main ejecta and the synchrotron
radio appearing to arise in a region which is ahead of the ejecta and might sometimes be
coincident with the ejecta.  An important argument against the bulk of synchrotron radio
emission arising in the ejecta is the differing free-free absorption felt by the emission
from the two processes.  This would indicate different sources of the free-free absorption
and hence support different emitting locations.  The detection times are also different with the 
detection of radio synchrotron preceding detection of the thermal radio emission. 
There does appear to be a higher rate of detection 
of radio synchrotron emission from recurrent novae as compared to classical novae. 
Out of the 11 identified Galactic recurrent novae,  radio synchrotron emission has 
definitely been detected from two - RS Ophiuchi and V745 Scorpii.  The companion
star in both the novae is a red giant.  The classical novae from which synchrotron
radio emission has been reported are V1370 Aquilae 1982 \citep{1987MNRAS.228..329S},
QU Vulpeculae 1984 \citep{1987A&A...183...38T}, GK Persei 1901 
\citep{1984ApJ...281L..33R,1989ApJ...344..805S}, 
V445 Puppis 2000 \citep{2001IAUC.7717....1R,2001IAUC.7728....3R},
V1723 Aquilae 2010 \citep{2011ApJ...739L...6K, 2016MNRAS.457..887W}.
There could be more detections which are hidden in literature. 
%Out of about 400+ known classical novae, $<15$ have been reported to have detected radio synchrotron emission.  
Some of the reasons for the infrequent detection of synchrotron radio in classical novae could be 
a combination of lack of
observations at sufficiently low radio frequencies where its easier to distinguish between
thermal and synchrotron emissions and at appropriate epochs since the
radio synchrotron emission from these systems is not as long lived as thermal radio or optical emissions
probably indicative of the relatively low energy of the explosion. 
In both the synchrotron-emitting recurrent novae, the initial ejecta velocities were
recorded in excess of 10000 kms$^{-1}$ which would have generated a sufficiently large pool of
relativistic electron population with $\gamma \ge 2$.  The relativistic electrons
were likely accelerated in the ambient magnetic field.  The frequency-dependent onset
of the different radio frequencies in both the novae, indicated that the process of 
free-free absorption was delaying the detection of the synchrotron radiation.  
The electrons would start emitting synchrotron radiation soon after the outburst since
in the updated model, electrons should be instantly accelerated to relativistic
velocities alongwith the ejecta especially if expanding with velocities $> 6000$ kms$^{-1}$.
Synchrotron radio emission has been detected in two outbursts in V745 Sco - in 1989 
\citep{1989IAUC.4853....2H} and 2014 \citep{2016MNRAS.456L..49K}
and two outbursts in RS Ophiuchi - in 1985 \citep{1986ApJ...305L..71H} and 2006 
\citep[e.g.][]{2007ApJ...667L.171K} which allowed a glimpse into their evolution.  
In case of both RS Ophiuchi and V745 Scorpii, the onset of radio synchrotron emission at the 
same frequency was found to be earlier in the later outburst.  It was suggested that the free-free 
opacity was caused by the
white dwarf winds which were blowing when the accretion rate exceeded the critical rate in
the recurrent novae \citep{2016MNRAS.456L..49K}.  Since the accretion rates vary, the 
wind rates vary and the free-free absorption
also varies with epoch.  This was shown to well-explain the behaviour of the radio light curves in 
successive outbursts demonstrating how synchrotron emission can be used to understand the
winds and accretion rates \citep{2016MNRAS.456L..49K}.
Radio synchrotron emission from nearby fast classical novae ($v_{ej} \ge 10000$ kms$^{-1}$)
should be detectable immediately after the outburst since foreground obscuration is likely to be low 
owing to the absence of winds from the white dwarf in classical novae wherein the accretion rates
are generally much lower than the critical rates. 

We note that hard X-ray emission is generally detected soon after the outburst and precedes
the detection of soft X-ray emission \citep{2011ApJS..197...31S}.
Moreover hard X-rays are more frequently detected from fast novae as compared to slow novae
\citep{2011ApJS..197...31S}.  Since the presence of relativistic electrons is more likely
in a fast nova wherein the energy imparted to each particle is larger, the above 
suggests that generation of hard X-rays is connected to the
relativistic electron population and the process is likely to be synchrotron or inverse Compton effect.
Hard X-rays are also occasionally detected from novae which have lower
recorded early ejecta velocities.   This could indicate rapid deceleration of the ejecta
or another physical process for their generation.  Clearly the detection of hard X-rays from
a nova is more frequent than of radio synchrotron emission.  If emission in both bands is owing to
the relativistic electrons, then this difference could either be due to different observational
sensitivities at the two bands or the absence of a magnetic field            
and a non-synchrotron origin for the hard X-rays in several cases.   
This remains to be investigated further by experts. 

\paragraph{Recurrent novae:}
The entire evolution is compressed into a shorter time in fast recurrent novae 
whereas the evolution of slow recurrent novae resemble 
classical novae.  The evolution of the last recorded outburst in the fast recurrent
nova V745 Scorpii on 6.7 February 2014 probably just after its optical peak demonstrates the
fast evolution.  There 
is a low significance detection of $\gamma-$rays $> 100$ MeV from
V745 Sco on 6, 7 February 2014 \citep{2014ATel.5879....1C} i.e. near the optical peak in addition
to hard X-rays \citep{2014ATel.5862....1M}.  Radio synchrotron emission was detectable at
610 MHz from day 12 to day 217 and indicated that it turned on between days 3 and 12 
\citep{2016MNRAS.456L..49K}.   Soft X-rays were detected from V745 Sco on
10 February 2014 (day $\sim 3$ after optical peak), peaked on day 5.5 after 
the optical peak and faded by February 22 \citep{2015MNRAS.454.3108P}.  The soft X-rays
were detectable for a total of about 12 days.  We note that the existing explanation for
the fast ejecta observed in some recurrent novae is attributed to the small masses that are ejected
compared to a classical nova.  The energy release will continue to be due to the CNO explosion
and hence comparable to classical novae. 
The smaller mass in the recurrent nova will be ejected more violently
and the larger energy allocation per particle will lead to the generation of a highly 
relativistic population of electrons.   The rest of the observations can be explained
as follows.  The lower energy $\gamma-$ray photons generated in the 
thermonuclear reaction would have been the seed photons which were boosted to $>100$ MeV energies by 
inverse Compton scattering by the relativistic electrons.  The nuclear reactions were quenched 
soon after and hence no $\gamma-$rays were detected beyond day 2. 
The relativistic electrons gyrating in the magnetic field either frozen in the ejecta
or present in the ambient medium start radiating synchrotron emission.  The hard X-rays could 
indicate inverse Compton scattering or synchrotron radiation from high energy electrons. 
The detection of soft X-rays around day 3 after optical peak indicates that the ejecta was
fully ionized by then and the quenching in 12 days indicates that the accretion had resumed
and started forming a low temperature envelope around the white dwarf.
If radio observations sensitive to thermal emission existed early on, it would have been possible
to detect radio thermal before day 3 although the smaller ejecta mass might have led to lower
intensity of radio thermal emission.
That the synchrotron radio was detectable for a long time indicates that a large pool of relativistic
electrons spanning a large energy range were energised in the explosion. 

Fast recurrent novae evolve very quickly and the ejecta velocity also declines rapidly which
is unlike classical novae which evolve slowly and the ejecta velocity remains constant for
several years.  The evolution of fast recurrent novae does not allow
the detection of the detailed evolution of the light curve or spectral stages unlike in
the slower classical novae.   However slower recurrent novae like T Pyx
are seen to show an evolution similar to classical novae supporting similar evolution in classical
and recurrent novae. 
Fast recurrent novae evolve to quiescence within a year of the outburst
while slower recurrent novae like T Pyx take longer.

\paragraph{Dwarf novae:} 
We have suggested that dwarf novae are due to an episode of smaller energy injection into the 
accreted envelope on the white dwarf which then isothermally expands causing it to brighten
($L \propto R^2$).  There have been suggestions of a thermonuclear origin to this energy 
output in literature \citep[e.g.][]{1975MSRSL...8..407S,1978ApJ...222..604P} which we think
observations support, but the
release of gravitational energy has found favour in literature.
The continuous emission increases as the photosphere expands reaching 
a maximum when it is largest and when the energy source is removed, the photosphere contracts
and the dwarf nova goes into decline with the entire outburst spread over a short duration of a
month or less. 
A brightening by $\sim 5$ magnitudes would require the radius of the photosphere to increase by a factor
of $\sim 10$.  We find that observations support this scenario.
Dwarf novae show the presence of emission lines during quiescence and absorption
lines are detected at the outburst maximum \citep[e.g.][]{1995warner.book.....W}.  
If the dwarf nova maximum, as mentioned above, is a result of expansion of the envelope
to $\le 10$ times its quiescent radius then 
this expanded envelope is the photosphere and the absorption lines at maximum arise in this photosphere.
On the other hand
the narrow emission lines detected in quiescence could be signatures of accretion winds around
the white dwarf in the nova.

\paragraph{Symbiotic stars:} Symbiotic stars are binary systems consisting of a white 
dwarf primary and a red giant 
secondary star and the detected light is a combination of the two components.  
Symbiotic stars show non-periodic short duration sequence of
brightenings and dimmings \citep[e.g.][]{1969CoKon..65..395B} but no mass ejection happens.  The 
brightness variations in symbiotic stars are also accompanied by colour variations such that 
it gets redder with decreasing brightness \citep[e.g.][]{1969CoKon..65..395B}.  This supports the
occurrence of the brightness variations on the white dwarf so that when the system brightens,
the contribution of the hot white dwarf increases and the colour is bluer.  When the brightness
declines the white dwarf contribution is reducing and the colour gets redder as the red giant dominates.  
A thermonuclear pulse due to proton-proton reaction on the white dwarf has been found to explain the
frequent brightenings in these systems \citep{1978ApJ...222..604P}. 
The pulse can then lead to an increase in the radius of the white dwarf photosphere like suggested for
dwarf novae or it could just trigger an increase in temperature of the envelope around the white dwarf.  
The spectra of symbiotic stars also follow the light variations such that when the
brightness of the symbiotic system decreases, the late type spectrum strengthens and excitation
degree of the emission spectrum increases \citep{1969CoKon..65..395B}. 
This behaviour is reminiscent of the light oscillations and change in excitation that is
noted in classical novae.  The observed spectrum consists of lines arising in the white dwarf and 
the red giant.  When the brightness of the white dwarf decreases, the late type spectrum due to the
red giant will appear relatively stronger.  If the excitation level of the spectrum increases
then it would mean that the white dwarf radiation is being used in increasing the excitation. 
The different radial velocities of lines from symbiotic stars AG Peg, BF Cyg, RW Hya, R Aqr
showed that these are binary stars.   The radial velocity variation of 
the Fe II line is significantly lower than that traced by lines of He~II, [O~III], N~III and [Ne~III] 
(Figure 12 in \citet{1969CoKon..65..395B}).  The lines of Fe~II arise in 
the red giant whereas the rest of the lines can form on the white dwarf.
Thus we can use the knowledge gained from symbiotic stars in the study of novae. 

\paragraph{[O~III] excitation:} In few novae, the [O~III] morphology is found to be distinct from 
that traced by other lines (e.g. DQ Herculis) indicating either a different elemental distribution 
or different excitation mechanism.  \citet{1988SvA....32..244P} discuss the importance of charge transfer 
such that $O^{+++} + H^0 \rightarrow O^{++} + H^+$ in context of the active nucleus in 
Seyfert galaxies and suggest that this makes an important
contribution to the strength of [OIII] lines and also ionizes hydrogen.  
and in neutral hydrogen regions the charge transfer reaction can enhance the [O III] emission. 
This requires that H, He are not ionized so that $\rm n_e << 10^6~ cm^{-3}$.  
If $\rm n_e > 10^7~ cm^{-3}$,
then collisional deexcitation of the O$^{++}$ ion will happen and no forbidden lines will result.  
They also explain variability in [O III] emission due to the above mechanism.  The upper level
of [O III] transition is populated by electron impact.  They also suggest
that O$^{++}$ ions can be formed by removal of electrons from O I by X-rays 
which is known as the Auger effect.
Since we do not the know the reason for the distinct distribution of [O III] sometimes
noticeable in the nova images, such processes could also be active in
the regions around a nova outburst and need to be further investigated. 

\paragraph{Explaining different light curve shapes:}
Light curves of novae display a large variety and hence have been prone to classification into different
types.  We refer to the classification in \citet{2010AJ....140...34S} where the authors have classified
the light curves into seven types: (1) smooth (S)  (2) plateau (P) (3) dust dip (D) (4) oscillations (O)
(5) cusp (C) (6) flat top (F) and (7) jitters (J).  
We attempt to understand these within the framework of the model presented here.  
We suggest simple explanations and predictions
for the types of light curves based on the discussion so far.  The $S$ type of light curve   
will arise for a nova in which the white dwarf contributes to the light near the optical maximum
and hence should show a pre-maximum halt.  No further significant contribution to the
optical light is expected and hence no plateau accompanies the soft X-ray detection.
The $P$ type should arise for a nova in which their is a delayed contribution to the total
light of the system from the white dwarf and/or irradiated companion star.  
The white dwarf in this case does not contribute near the
optical maximum due to opacity of the ejecta and hence such novae should seldom show a pre-maximum
halt.  The white dwarf contribution is delayed to a later date
giving rise to the plateau in the optical light curve and in most cases is found to coincide
with the soft X-ray emission which heralds the full ionization of the foreground ejecta. 
The $P$ types can be of any speed class although its easier to imagine these to be slower novae.
The $D$ types indicate dust formation in the Orion clumps and the advancing of the clumps to the
front of the ejecta so that they can obscure the continuous emission from rest of the ejecta
and the white dwarf.  When the dust disperses, the light curve revives.  These are slow novae
since only these have sufficient time for the dust to form inside the Orion clumps and 
move forward and influence the light curve.  The $O$ types are the novae wherein the white dwarf
contribution to the visible band is significant and the radiation field of the white dwarf has a
significant impact on the Orion clumps.  These novae should show the pre-maximum halt. 
The $C$ types can be an extension of $P$ types wherein
the eventual contribution to the light from the white dwarf and possibly companion star are
significantly higher than in a $P$ type so that the light curve shows a cusp.  There should
not be a pre-maximum halt for the $C$ type. 
The $F$ type is a slow nova since $t_2$ and $t_3$ are very long.  This could be indicative
of a dense slowly expanding ejecta which remains optically thick for a long time 
and hence the emission remains near maximum for a significant time.  The $J$ types
show jitters of amplitude generally $\le 2$ magnitudes detected soon after the maximum
and could arise due to varying extra contribution by either the white dwarf or hot spot on the companion
heated by the X-ray emission from the white dwarf.  Sometimes these jitters
lead to maxima perched on a flat light curve.  
%The variation could be internal to the white dwarf or could be due to the density inhomogeneities in the ejecta.   

From the tabulated values in \citet{2010AJ....140...34S}, 
we note that while most novae show similar trends in $t_2$ and $t_3$ so that a fast novae has
short $t_2$ and $t_3$, a large variation is seen in $t_6$ such that
some fast novae show a short $t_6$ and some show a long $t_6$.  This kind of behaviour indicates
a change in the rate of decline of the light curve beyond $t_3$ which should help us further understand
the nova.  For starters, it appears that the fast novae with a long $t_6$ 
will be prone to showing a plateau in the light curve which as stated earlier would indicate 
an additional light contribution to the light curve.  A detailed study would be useful before
arriving at any firm conclusions on this. 

\begin{figure*}[t]
\centering
\includegraphics[width=14cm]{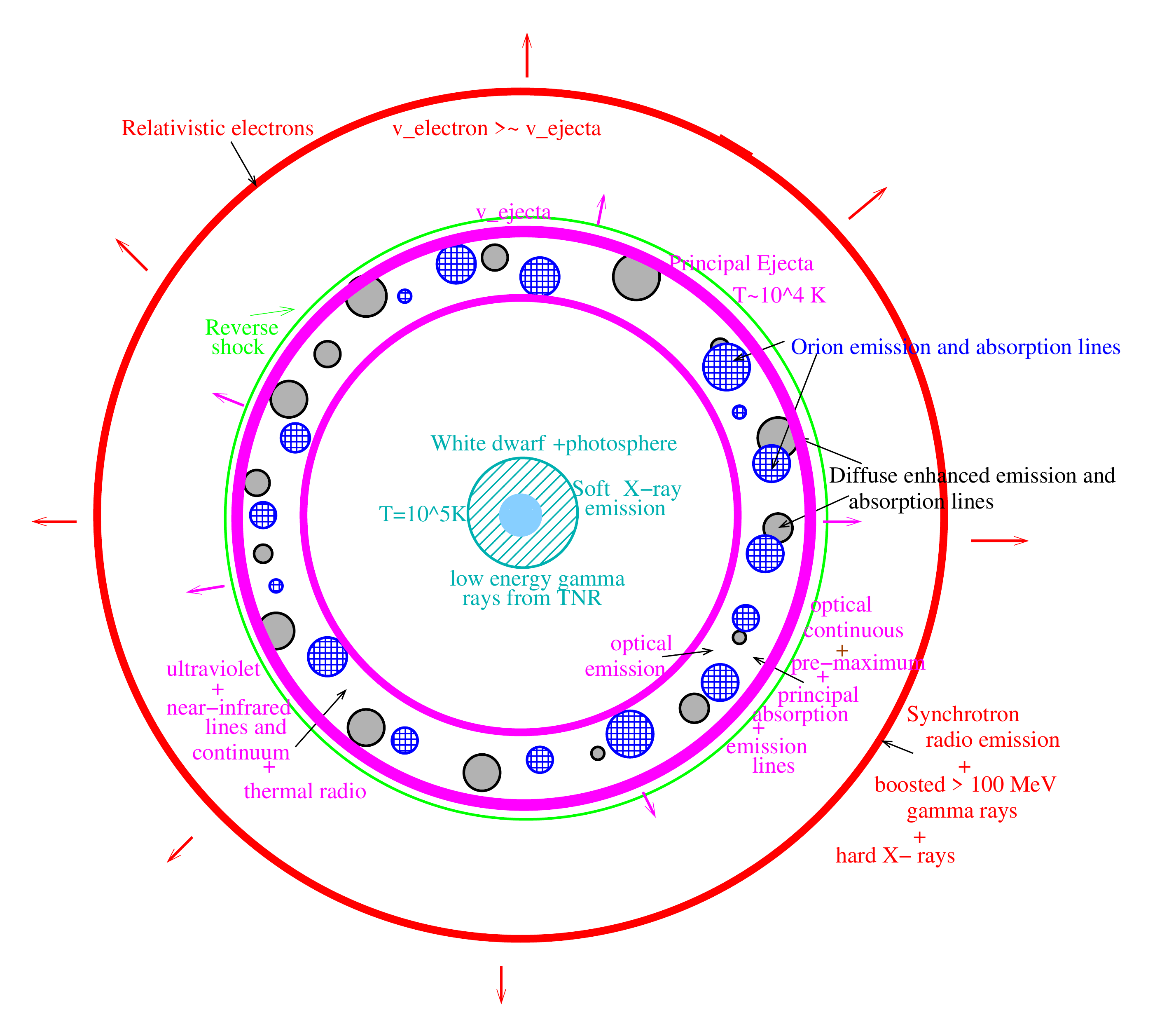}
\caption{The schematic of a nova outburst labelled with the dominant site of origin of 
the various multi-wavelength emissions.  This schematic will apply to 
non-rotating (or weakly rotating) white dwarfs which accrete a spherical envelope.
A rotating white dwarf will acquire a prolate envelope and the eruption will eject
an ellipsoidal envelope which is observed around several novae.  The figure is not to scale.  }
\label{multiband}
\end{figure*}

\subsection{Origin of multifrequency emissions}
We can now schematically summarise the model of a nova outburst which emerges from multifrequency
observations (see Figure \ref{multiband}).  
The deduced sites of origin of the multi-band emission are labelled in the figure.
The schematic is relevant for a spherical ejecta and depicts the evolutionary
states expected for times ranging from just after outburst to $\sim t_6$ i.e. upto about
six magnitudes below optical maximum.  During this period, some of the emissions labelled in the 
schematic will disappear, new components will appear and some will survive.  
About six magnitudes below optical maximum, the principal shell of ejecta will be devoid of 
absorbing clumps and will be emitting nebular lines as it expands around the binary. 
The white dwarf will begin its journey back to its pre-nova state.  

The thermonuclear explosion at the base of the inflated accreted envelope 
leads to ejection of the overlying layers of matter.  The cooler outer parts are 
ejected leaving behind a thin photosphere at high temperature $\ge 10^5$ K which
emits soft X-rays.  The soft X-rays will be visible when the foreground
becomes transparent to it which is generally observed to be around the transition or nebular phase of
the light curve and are extinguished once accretion restarts and a low temperature envelope 
forms around the white dwarf.  The thermonuclear reactions will release
$\gamma-$rays of energies $<20$ MeV.  These $\gamma-$rays will be quenched when the thermonuclear 
reactions stop.  These emissions necessarily arise close to the white dwarf
as labelled in Figure \ref{multiband} and propagate outwards.  Neither of these emissions 
are expected from novae in quiescence nor are they detected. 
Outside the binary lies the expanding ejecta.  The radial separation of the ejecta from
the white dwarf is continuously increasing and the physical parametres of the ejecta like its
density is continuously decreasing as it expands.  The ejecta which is shown enclosed
by the two magenta circles in Figure \ref{multiband} contains most of the ejected mass which is typically
estimated to be $\sim 10^{-5} M_\odot$ in classical novae and $\sim 10^{-7} M_\odot$ in
recurrent novae.   Emission at several wavebands arise in the ejecta -  optical, ultraviolet,
infrared and radio.  These can be described further to be optical continuous emission
and line spectra,  near-infrared continuum and line spectra (including dust emission), 
thermal radio emission and
ultraviolet continuous and line emission.   The various optical line systems - pre-maximum,
principal, diffuse enhanced and Orion arise in this ejecta.  These are labelled in
the figure.  The clumps which form due to mass-based segregation of elements are also shown.
The blue crossed circles indicate the Orion clumps and the grey black circles indicate the
diffuse enhanced clumps.  The clump formation begins after the maximum and they have
formed about a magnitude below the optical maximum especially in slow novae as the detection of diffuse
enhanced lines indicate.  Spectral lines from these clumps are detectable to about 4 magnitudes below 
maximum i.e. till the transition phase of the light curve.  Dust also forms in the clumps
in this period.  The clumps
move forward within the ejecta under the influence of the radiation pressure exerted
by the white dwarf.  In some novae, the presence of dust in the leading parts of the ejecta
obstructs the optical continuous emission leading to a steep drop in the light curve which
eventually recovers in the nebular phase.  The clumps 
disperse and become optically thin as the nova moves to the nebular phase and the 
diffuse enhanced and Orion lines at higher velocities displacements are not detected again.  
The electron temperature of the ejecta is around $10^4$ K.  The ejecta continues to expand and 
the emission lines of the principal
system continue to be detectable.  In many cases, the ejecta shell is detected several years
after the explosion still expanding with similar velocities as inferred from the principal lines.
Synchrotron radio emission is detected from some novae, soon after the outburst and
is inferred to be arising from a region ahead of the ejecta which in literature is referred
to as the forward shock.  It is believed that the electrons are accelerated by the forward
shock and hence are coexistent with it.  However we have shown that electrons in the envelope
will be accelerated to relativistic velocities if equal energy is distributed to all particles
in the ejecta which means that they are accelerated alongwith rest of the envelope and
no separate shock acceleration needs to be invoked.   Observations show that the
epochs of detection of radio synchrotron and radio thermal are distinct and so is the
cause of free-free absorption which indicates that they arise in different locations.
Moreover imaging of radio emission following a nova outburst (e.g. RS Ophiuchi) shows
that the synchrotron emission arises in a region ahead of the main ejecta from which the thermal
radio emission is detected.  Hence it appears that relativistic electrons occupy a region
ahead of the main ejecta as shown by the red circle in Figure \ref{multiband}. 
Relativistic electrons should also be present in the main
ejecta but their number seems to be insufficient to result in detectable synchrotron radio 
emission.  The population of relativistic
electrons in the red circle gives rise to detectable radio synchrotron emission and also hard X-rays. 
The radio synchrotron requires a magnetic field which would be the field frozen in the
ambient medium.  The electrons 
can also inverse Compton boost the low energy $\gamma-$ray photons produced in the
nuclear reactions to $\ge 100$ MeV energies thus explaining the detection of such $\gamma-$rays
from novae as labelled in the figure.  If we refer to the red circle as the 
`relativistic electron ejecta', then it will survive till the electrons lose their energy. 

In case of a prolate-shaped envelope being ejected from a rotating white dwarf in the nova
explosion, the ejecta in the polar regions will appear to expand further than
the equatorial regions.  Many times the polar ejecta is observed to
expand faster than the equatorial part indicating a possibly more energetic explosion at the
polar regions or if the explosion is similar everywhere the non-polar expansion might
have been stunted by the presence of an accretion disk. 
It is difficult to break this degeneracy but we think it should be possible with
a further analysis of observational data. 

It is interesting to note that FH Serpentis 1970 was 
a first timer in several wavebands - thermal radio emission was first detected in this nova 
\citep{1970ApJ...162L...1H},
dust emission was first detected in this nova when it was found that a decline in the optical light curve
was coincident with rise in flux at wavelengths longer than 2$\mu$m \citep{1970ApJ...161L.101G}
and an ultraviolet peak was detected after the optical peak leading to models which suggested a
constant luminosity phase in novae \citep{1974ApJ...189..303G}. 

\section{Case studies} 
We summarise and examine existing multi-frequency observational results on a couple classical and
a couple recurrent novae in context of the model put forward here.   
The highlight of the study is that the observed characteristics are well-explained in
this model within the purview of known physics.  There do remain several details that need to
be worked out but the overall framework consistently explains the observables. 
We also extrapolate some of the model features especially the synchrotron radio emission to 
supernovae (supernova remnants) and active nuclei but this discussion is left to the
next part.  

\subsection{Classical novae}
While the shells of several old classical novae detected in optical 
are spherical in shape e.g. CP Puppis 1942, \citep{1979Msngr..17....1D};
V533 Herculis 1963, V476 Cygni 1920, DK Lacertae 1950 \citep{1995MNRAS.276..353S}
there are also novae which show ellipsoidal or bipolar shells 
e.g. RR Pictoris 1925 \citep{1979Msngr..17....1D}; 
T Aurigae 1891 \citep{1980ApJ...237...55G},
DQ Herculis \citep{1995MNRAS.276..353S}, V959 Mon \citep{2015ApJ...805..136L}.  
The enhanced bipolar or prolate morphology of the ejecta is often deduced 
from the spectral line profiles recorded soon after outburst. Profiles which 
consist of a central component flanked by high 
velocity shoulders or a double profile indicate existence of a faster bipolar outflow 
e.g. RS Oph; \citep{2008ASPC..401..227S} and V1535 Sco \citep{2017arXiv170303333L},
although sometimes it could also result from the equatorial shell if viewed edge-on. 
A dependence of the axial ratio of a nova shell on
the speed class such that slow novae frequently left behind ellipsoidal shells whereas faster novae left
behind spherical shells was noted by \citet{1995MNRAS.276..353S}. 
The occurrence of such aspherical remnants has been explored and the possible reasons 
include a common envelope phase, localised thermonuclear runaways, binary nature of the system
and rotation of the white dwarf.  As discussed in the paper, an ellipsoidal ejecta gives direct
evidence to rotation of the accreting white dwarf which will accumulate the accreted matter
in a prolate envelope around it due to the dependence of the accretion rate on latitude.  
Spherically symmetric matter infall is assumed.  In a spherically symmetric or bipolar 
explosion, the prolate-shaped envelope will form an ellipsoidal
or bipolar ejecta.  While other physical effects might also contribute to the ellipsoidal envelope,
we believe that the effect due to rotation is the primary reason. 

In a few classical novae such as Nova QU Vulpeculae 1984 \citep{1987A&A...183...38T},  V1723 Aquilae
\citep{2011ApJ...739L...6K, 2016MNRAS.457..887W}, a
double peaked radio light curve is observed with the first peak indicating a synchrotron origin
and the second peak being attributed to thermal free-free emission.  
However such cases are rare and most radio-detected novae have shown the presence of thermal
free-free emission, sometimes detectable for several years after the explosion e.g. HR Delphini 1967. 
A frequency-dependent behaviour wherein the higher frequencies
are detected before the lower radio frequency emission is observed in most novae
e.g. T Pyxidis \citep{2014ApJ...785...78N}, V1974 Cygni \citep{1996ASSL..208..317H},
HR Delphini \citep{1979AJ.....84.1619H}. 
Such a behaviour indicates that the radio detection epoch is determined by the optical
depth ($\propto 1/\nu^2$) of the foreground/mixed absorbing material. 
In a few novae, the radio peaks occur simultaneously \citep[e.g. V1723 Aql][]{2016MNRAS.457..887W}
indicating that the delay in the onset of radio thermal emission is not due to free-free absorption.
Radio emission from classical novae was first detected in the novae HR Delphini and FH Serpentis
\citep{1970ApJ...162L...1H}.  Radio imaging observations soon after the outburst
seem to indicate that synchrotron radio emission arises outside the main
ejecta indicating that the relativistic electron population has escaped with larger forward velocities.
The synchrotron radio emission which also shows the frequency-dependent onset 
suffer free-free absorption due to the circumbinary material, mainly the remnant of the
white dwarf winds.  The thermal radio emission which arises in the ejecta 
and is generally detected at a later epoch suffers 
free-free absorption due to the cooler foreground material in the ejecta.  
The occasional simultaneous detection of thermal radio peaks at later times could be interpreted 
to be due to partial ionization of the ejecta and hence
lower emission measures in the initial stages and the detection being delayed
till the entire ejecta was ionized and had sufficient emissivity.   

For dust formation to happen in a nova ejecta, high particle density rich in metals and
not subjected to intense radiation field of the white dwarf is required \citep{1988ARA&A..26..377G}. 
Our model pinpoints the sites of dust formation to the metal-enriched
clumps which are formed due to mass-segregation of elements in the ejecta.  The
diffuse enhanced and Orion lines are also formed in these clumps.  Thus, dust formation in 
predominantly slow novae which also show the presence of diffuse enhanced and Orion lines 
should be fairly common as supported by observations.
Deep minima in the transition phase have been detected in the slow novae such as DQ Herculis 1934 which 
showed a deep minimum which started in the transition phase and ended when the nova entered
the nebular phase.

After this short summary of a few observations which our model consistently explains, 
we detail some of the observational results on two well-studied novae DQ Herculis 1934 and 
V339 Delphini 2013.  Although no efforts have been spared to make the summary extensive, 
it is humanly impossible to do justice to the vast literature that exists so if any significant
observational results have been overlooked, we request the readers to augment this 
and verify if the model presented here can consistently account for them. 

\subsubsection{DQ Herculis 1934}
\begin{figure}[t]
\centering
\includegraphics[width=7cm]{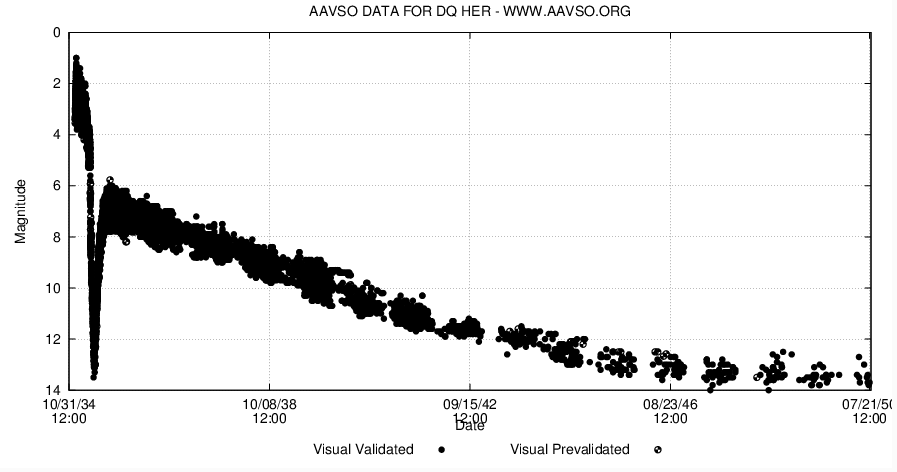}(a)
\includegraphics[width=7cm]{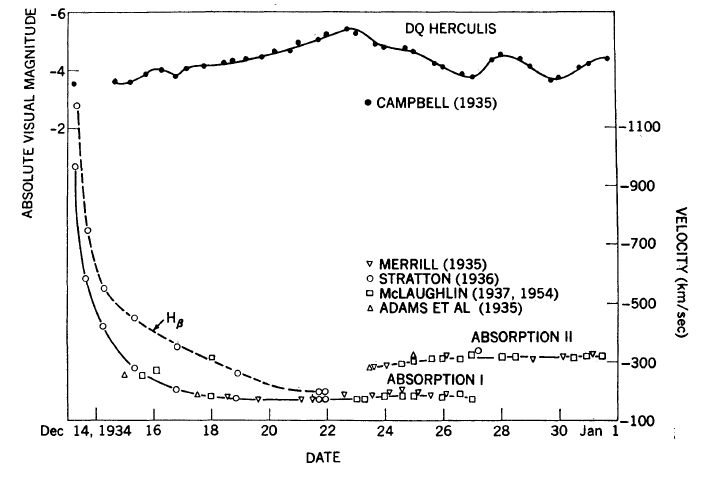}(b)
\includegraphics[width=7cm]{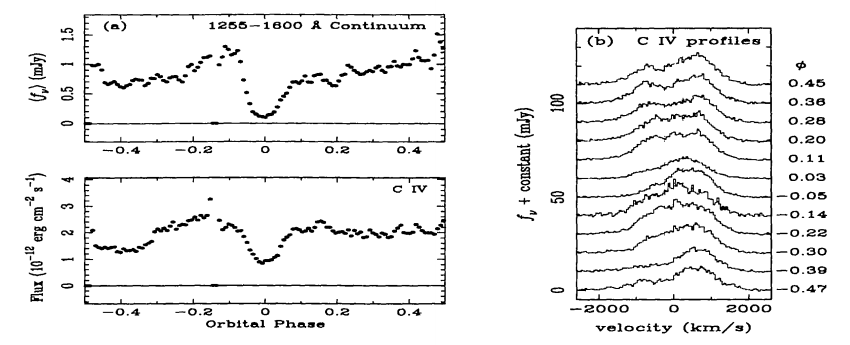}(c)
\caption{(a) Light curve of DQ Herculis 1934 has been copied from the AAVSO website
(https://www.aavso.org/lcg).  The nova
took $\sim 15$ years to decline to minimum.  T Aurigae which showed similar evolution took
30+ years to decline to minimum.  (b) This figure reproduced from \citet{1969ApJ...156..569S}
shows the early light and velocity evolution of the pre-maximum (Absorption I) and principal (Absorption II)
spectral line systems.  (c) This figure reproduced from
\citet{1998ASPC..137..438E} shows the eclipsing ultraviolet continuum and C IV lines as a function
of orbital phase.  }
\label{dqher}
\end{figure}

DQ Herculis was discovered on the morning of 13 December 1934 in the pre-maximum phase and 
it peaked in the optical bands on 23 December 1934.  This remains one of the best studied novae.
In the following, we summarise some of the observational results from literature and our
inferences/comments are included in italics: \\
(1) The visual light curve took almost 15 years to decline to minimum (see Figure \ref{dqher}a).
{\it This would indicate the time required for a quiescent-like envelope to be accreted around the
white dwarf.  Subsequent accretion would increase its density. }  \\
(2) DQ Her is an intermediate polar ie has a magnetic field and rotates. {\it The white dwarf is
rotating and hence the accreted matter was accumulated in a prolate-shaped envelope which was 
ejected in the outburst which continues to be detectable as an elliptical shell. }\\
(3) In July 1935, \citet{1941ApJ....93..133K} found DQ Her was a double star with a separation of 0.2'' 
while the eclipsing binary nature of DQ Herculis with a period of $\rm 4^h59^m$ was
established by \citet{1954PASP...66..230W}.  It is a single line binary and hence the companion spectrum is
not detected.  From the yellow light, it was estimated that the inclination of the system was $77.3^\circ$ and the 
eclipse was 1.4 mag deep in the yellow, 1 mag in blue and 0.8 mag in ultraviolet \citep{1956ApJ...123...68W}.
The radius of the nova was estimated to be $\sim 0.1 R_\odot$, of the companion star to be $\sim 0.11 R_\odot$
and the nova was found to become bluer during the primary eclipse while another binary
UX UMa would found to become redder in the primary eclipse \citep{1956ApJ...123...68W}. 
Maximum blueness is found in the shoulder that precedes the primary eclipse and the colours in
the non-eclipsing part indicated that the eclipsed star is bright in the ultraviolet and
could be a white dwarf \citep{1956ApJ...123...68W}.
\citet{1959ApJ...130..110K} suggested that DQ Herculis hosts a white dwarf as the nova star and 
a companion star which overflows its Roche lobe.  
%Irregular light variations are seen in the white dwarf and
A light oscillation of 71 second was also detected \citep{1956ApJ...123...68W}.  \\
(4) The nova showed several spectral
changes within the first ten days after detection starting with a steep fall in the ejecta velocity
as shown in Figure \ref{dqher}b reproduced from Sparks (1969).  Table 10 in \citet{1960stat.conf..585M}
lists the evolution of the ejecta velocity in DQ Herculis in the pre-maximum phase after detection
with H$\alpha$ showing the highest velocity of $-1984$ kms$^{-1}$ on 13.3 December 1934 which
reduced to $-320$ kms$^{-1}$ on 21.8 December 1934.
{\it This would indicate that the ejecta rammed into dense circumstellar material and quickly 
slowed down as it swept up matter.  This behaviour is more typical of fast recurrent novae
like RS Ophiuchi and V745 Scorpii.  Considering that the shell of DQ Herculis continues to expand
with a velocity $\sim 300$ kms$^{-1}$ suggests the existence of relatively tenuous material it has
since been expanding into. } \\
(5) \citet{1937POMic...6..107M,1954ApJ...119..124M} has extensively discussed the observations
of DQ Herculis and the following is based on his research papers.  
The B-type spectrum (starting with $\rm T_{eff} \le 28000$ K) of 13 December 
which showed emission bands evolved to a
A-type spectrum ($\rm T_{eff} \le 10000$ K) with P Cygni profiles of hydrogen and other 
metals including Fe II on 14 December 1934.
The velocity displacement of the blue-shifted absorption lines was 
$>500$ kms$^{-1}$ on the night of 13 December which declined to $\sim 180$ kms$^{-1}$ on December 18
and which was considered the pre-maximum spectrum velocity.  {\it However it should be noted here that the
explosion energy was much higher than would be indicated by the pre-maximum spectrum velocity.}
Around 21 December before the maximum, the spectrum changed to a F5 type
($\rm T_{eff} \sim 6500$ K).  {\it  The evolution of the
spectrum to late type stellar types would indicate the expansion of the nova ejecta with no 
further significant energy input.  }
During decline from maximum after December 23, a new spectrum 
of class cF5 containing absorption lines of hydrogen, Fe II, O I and other 
metals with a displacement of $-300$ kms$^{-1}$ appeared - this was the principal spectrum and the
ejecta has since continued to expand with this velocity. {\it As suggested in the model, the
optically thick ejecta before maximum will be subjected to an extra push by the
radiation pressure exerted by the hot white dwarf 
left behind by the ejection of the cooler envelope in the eruption and which explains the
slightly higher velocities of the principal spectrum.  If this process makes the ejecta optically thin,
then the final $\le 2$ magnitudes rise in the light curve could indicate the contribution of
the white dwarf.  It could also include emission from 
isothermal expansion of the ejecta thus explaining the change from the spectral type F5 before maximum to
the supergiant cF5 post-maximum. }
The principal spectral lines evolved to double-peaked profiles with peaks
separated by about 600 kms$^{-1}$ with some central emission. {\it These would indicate enhanced
emission from the bipolar ejecta.}
Diffuse enhanced spectrum appeared in January 1935 with lines of hydrogen, Fe II, Ti II,
Cr II, Ca II, Na I at a displacement of $-550$ kms$^{-1}$ with more components appearing at
$-700$ and $-800$ kms$^{-1}$. {\it The diffuse enhanced clumps formed in the ejecta
as heavier elements segregated into clumps.  These optically thick clumps acquired an extra acceleration
as the radiation pressure due to the hot white dwarf acted on them.  Different clumps acquired
different velocities which support the radiation pressure scenario.  The Fe II, Ti II etc lines
could be a combination of the ejecta material and swept-up material which form the clumps
in the ejecta as shown in Figure \ref{multiband}.}  The diffuse enhanced spectrum
faded in mid-February.  The Orion spectral lines appeared in March
with velocity displacements between $-300$ and $-500$ kms$^{-1}$ with more components appearing
at $-1000$ kms$^{-1}$.  All these absorption systems emerged at a secondary minimum in the light
curve.  {\it The Orion lines arose in another set of clumps which formed at the rear part of the ejecta
(see Figures \ref{novanew} and \ref{multiband}) and were also accelerated due to the radiation pressure. 
The excitation and velocity of these clumps are sensitive to the light curve variations so that
light recorded by us decreases when the Orion line-forming clumps absorb it and either increase
its velocity displacement or excitation.  This could be useful in understanding the physical differences
between the diffuse enhanced and Orion clump properties.}
In April 1935, when the light curve was about 3 magnitudes below
maximum, the nova started to fade (see Figure \ref{dqher}a).
During this time all the absorption bands and most of the emission features disappeared.
The emission bands of hydrogen, Fe II and Ca II were detectable and widened
to include the velocities previously occupied by absorption even as the red parts of the emission bands faded.
These lines were the last to disappear \citep[also][]{1945MNRAS.105..275S}. 
{\it Since obscuration is due to dust formation in the Orion clumps which are accelerated
to the outer parts of the ejecta, it will obscure all the optical continuum 
and line emission arising in most of the ejecta and white dwarf.  The longer time taken by the
emission bands of Fe II, Ca II and hydrogen to fade indicate they arise in the outermost part 
of the ejecta which suggest that the Fe II
and Ca II could be from the swept-up material of the companion. The red sides of emission arise on
the far side of the ejecta and hence faded before the blue side. }
The light curve faded to a minimum of 13 magnitudes on 1 May 1935 after which the nova started to
brighten and the nebular spectrum consisting of [O III], hydrogen and '4640' band emission started to appear.  
Absorption lines did not reappear in the spectrum.  The red sides of the emission bands gradually strengthened 
and when the light curve recovered to about 7 magnitudes below maximum the red part was almost as
strong as the violet part of the line.   \citet{1935PA.....43..323M,1937POMic...6..107M}
suggested that dust formation in the principal shell in DQ Herculis was responsible
for the abrupt drop in the light curve which could explain the spectral changes also.  When the dust
cleared, the light curve revived and continued its decline which ended around 1950 when the nova
was at its pre-outburst brightness (see Figure \ref{dqher}a).  
{\it The long time taken to revert to pre-nova condition might indicate low accretion rates. }   \\
(6) In 1940, the central star in DQ Herculis was of photo-visual magnitude 13.4 magnitudes 
whereas its pre-outburst photographic magnitude was fainter at 14.6 magnitudes \citep{1940PASP...52..386B}.
{\it The brighter white dwarf noted in the post-outburst phase gives support to our model
wherein the white dwarf is believed to contribute $\le 2$ magnitudes increase in the light curve and
the hypothesis that the long phase of final decline is due to the time taken by the cooler
envelope to be accreted around the white dwarf with similar properties as the pre-nova phase. }
The light curve was around 11 magnitudes and the brightness of the nebula was 13.2 magnitudes/arcsec$^2$ 
on 1 October 1940 when the photovisual brightness of the star was 13.4 magnitudes  
\citep{1940PASP...52..386B}.  
{\it This observation lends strong support to the dominant contribution to the optical light
curve by the ejecta as expected by the MMRD and not the expanding photosphere of the white dwarf. 
The observation demonstrates that the brightness of the white
dwarf and nebula were slowly declining to quiescent strengths. If more such data wherein
the emission from the white dwarf and ejecta can be separated soon after the outburst are collected then
it would be possible to separate the contributions of the white dwarf and the ejected shell 
to the light curve.  This would help study the decline of light from the nebula
and white dwarf. } \\ 
(7) The expanding shell left behind by the nova outburst which is ellipsoidal is still detectable.
In 1940s, the [O III] shell showed a distinct morphology from the N~II+H$\alpha$ shell. 
While the former was detected along the major axis of 
the shell which was polar, N II+H$\alpha$ were predominantly detected in the equatorial plane 
\citep{1940PASP...52..386B,1942PASP...54..244B,1970Ap&SS...6..183M}.  Even now the [O III] emission
appears to be confined to the end parts of the major axis of the shell while 
$\rm H\alpha$ and [N II] are detected
from the entire shell \citep{2003MNRAS.344.1219H,1995MNRAS.276..353S}.  
It appears as if the polar extent of $\rm H\alpha$
and [N II] has increased more than that of [O III].  However this needs to be verified.  
{\it A differing morphology in [O III] has been detected in several nova ejecta.  We speculate
that this could indicate a pre-outburst mass loss which is excited when the electron ejecta traverses 
through it or could be similar to Auger effect.  Since this has been observed in multiple
novae, it needs to examined in detail. }\\ 
(8) \citet{1960stat.conf..585M} discusses the colour, excitation, photoelectric and 
electron temperatures of DQ Herculis.
The colour temperature decreases from 11000 K to 9000 K as the nova rises to maximum brightness and 
then increases to 18000 K about 4 magnitudes below maximum when the Orion spectrum is strong. 
The photoelectric and excitation temperatures are found to be high at $>30000$ K
whereas electron temperature of the ejecta is estimated to be between 6000 and 10000 K. 
{\it This behaviour is as expected in our model.  We recall that the continuum emission
near the maximum predominantly arises in the hot ejecta and hence the colour temperature 
will be similar to the electron temperature.  As the light curve declines, the fractional
contribution of the hot white dwarf to the optical continuum increases and the colour temperature increases.
The photoelectric and excitation temperatures are a combination of the explosion energy
and the hard radiation of the hot white dwarf and which will remain high till the white dwarf
forms the cool envelope.  At early times these will indicate equivalent
temperature signifying the explosion energies and at later times the white dwarf. }\\
(9) \citet{1936CMWCI.545....1A} describe the spectral state prior to (25 March 1935), 
during (4,15 April 1935) 
the fading and after the light curve recovered (20 May 1935).  We reproduce their summary: \\  
``March 25: Continuous spectrum strong.  Prominent absorption lines of hydrogen and calcium 
showing multiple components and numerours fainter absorption lines of ionized elements.
Emission bands of hydrogen broad and diffuse.  Faint emission lines
of Fe~II and [Fe~II]. \\
April 4: Continous spectrum weak.  No certain absorption lines.  Narrow emission components of hydrogen. 
Very strong emission lines of Fe~II and [Fe~II]. \\
April 15: Spectrum similar to that of April 4, but Fe~II lines fainter and [Fe~II] lines stronger. \\
May 20: Continuous spectrum very weak or absent.  No Fe~II or [Fe~II] lines.  
Nebular spectrum fully developed. '' \\
{\it This summary supports the suggestion that the Fe~II lines arise in the swept up circumstellar
material.  } \\
(10) While the nebular spectrum of DQ Herculis observed around 1940 \citep{1940ApJ....92..295S} was similar to 
that in 1935-1936 \citep{1936CMWCI.545....1A}, significant changes in the dominant spectral lines were 
noted in 1947 and 1949 \citep{1949ApJ...110..475S,1952ApJ...116..229S}.  However there was no
change in the expansion velocities which ranged from 222 to 396 kms$^{-1}$ and several 
lines continued to show double components separated by $\sim 300$ kms$^{-1}$ from the 
centre frequency \citep{1949ApJ...110..475S,1952ApJ...116..229S}
as was observed in the principal spectrum after outburst in 1934.  Table 1 in \citet{1949ApJ...110..475S}
lists the changes between spectra of 1940/1942 and 1947/1949 and the main changes can be summarised to
be that [O III] 4959, 5007 weakened, [Ne III], [Fe VII] disappeared and [O II], He II 4686 strengthened.  
{\it It is not immediately evident what could have triggered these changes and what they indicated but
it is likely that the changes indicate some change in the physical and excitation conditions of
the nebula. } \\
(11) DQ Her is an eclipsing binary and the ultraviolet continuum emission disappears in
the eclipse while emission lines (C IV, N V, Si IV, Ne II, He II) 
show partial eclipse  with C IV being the least obscured (see Figure \ref{dqher}c; 
\citet{1998ASPC..137..438E}).   Origin of these lines have been suggested to be either 
winds or the accretion disk.  The C IV line (see Figure \ref{dqher}c) shows two components about the 
rest frequency with the red-shifted component always being detected 
whereas the blue-shifted component fading twice in an orbital period. 
In a spectrum taken in 1995, it was found that four spectral lines 
namely  C IV, N V, Si IV, He II  have strengthened by $50\%$ compared to 1993
\citep{1996AJ....112.1174S}. 
The velocity of the He~II 4686 line changes with the orbital phase \citep{1958PASP...70..598K}
and a similar behaviour is also exhibited by the higher lines of the Balmer series while it is 
not seen in $\rm H{\beta}$  which prompted the suggestion that the forbidden lines
and lower Balmer lines like $H\beta$ are formed in the nebula around the nova whereas He II and
higher Balmer lines arise in a disk/ring
around the white dwarf which would explain their partial eclipse and changing radial velocities
with orbital phase \citep{1959ApJ...130...99G,1959ApJ...130..110K}.  {\it 
The high ionization lines form close to the white
dwarf since they do show a dependence on the orbital phase and possibly 
form in the inner parts of the accretion disk.  Since the white dwarf
in DQ Herculis is known to be rotating, an accretion disk should exist. 
The width of these lines would indicate the  
rotation velocity of the accretion disk which should be useful in estimating the rotation velocity
of the white dwarf and the partial eclipse of the lines might be useful in estimating
the radial extent of the disk.  Since a cooler envelope should have accumulated on the white dwarf,
the radiation escaping from the white dwarf should be of lower temperature making the
presence of such highly ionized lines puzzling.  }  \\
(12) We used the eclipse parameters in the V band to estimate the luminosities contributed
by the binary components.  For $\rm t_2=67$ days, $\rm m_{V,0}=1.4$ magnitudes, $\rm m_{V,q}=15$ magnitudes 
\citep{1960stat.conf..585M},
we estimate MMRD $\rm M_{V,0}=-6.9$ mag \citep{2017arXiv170304087K} and hence $\rm M_{V,q}=6.7$ magnitudes. 
Using an eclipse depth of 1.3 magnitudes, \citep{1956ApJ...123...68W} 
and $\rm M_{V,\odot}=4.83$
magnitudes we estimate $\rm M_{V,secondary}=8.0$ magnitudes and $\rm M_{V,WD}=7.1$ magnitudes.  Both
objects emit sub-solar V band luminosity.  The companion's luminosity suggests that it is
a main sequence star of spectral type between K5 and M0 based on the table in 
\citet{1973asqu.book.....A}.  \\
(13) The shell of DQ Her is estimated to consist of two components - one with temperature $\sim 500$ K 
as inferred from the Balmer continuum radiation which is confined to a small wavelength range and second 
at $10^4$ K \citep{1978ApJ...224..171W}.  This conclusion is supported by ultraviolet 
observations \citep{1984ApJ...281..194F}.  Several lines detected in the nova shell are similar to
planetary nebulae except for the excess of permitted recombination lines of C,N,O lines.  
The existence of high excitation
lines at such low electron temperatures has been a perplexing problem.  {\it The excess CNO
observed in nova shells as compared to planetary nebulae gives evidence to the underlying
energy source being the thermonuclear runaway CNO reaction in which C,N,O are catalysts and
are retained whereas the hydrogen and in some cases helium accreted from the companion
will be fused to the higher elements and hence be depleted. }\\
(14) Tails extending outwards from the clumps in the polar regions of the shell of 
DQ Herculis are detected in $\rm H\alpha$ and show an increasing radial velocity upto 
$\rm 800-900~ kms^{-1}$ which are interpreted as being caused by the stellar wind 
\citep{2007MNRAS.380..175V}.  {\it There seem to exist at least two observations (this one
and the presence of wide emission lines detected in ultraviolet which show partial eclipse) which suggest
winds blowing from the white dwarf in DQ Herculis.  Since this is a classical nova the accretion
rates are likely to be much lower than critical rates and hence the origin of the winds is not
clear unless excess surface temperature leads to excess exertion of radiation pressure which blows
off the accreted matter.   This needs to be examined further }\\

Thus, although several observational results on DQ Herculis find an explanation in the updated model, 
there still remain several perplexing results which need to be understood. 

\subsubsection{V339 Delphini 2013}  
\begin{figure}[h]
\includegraphics[width=8cm]{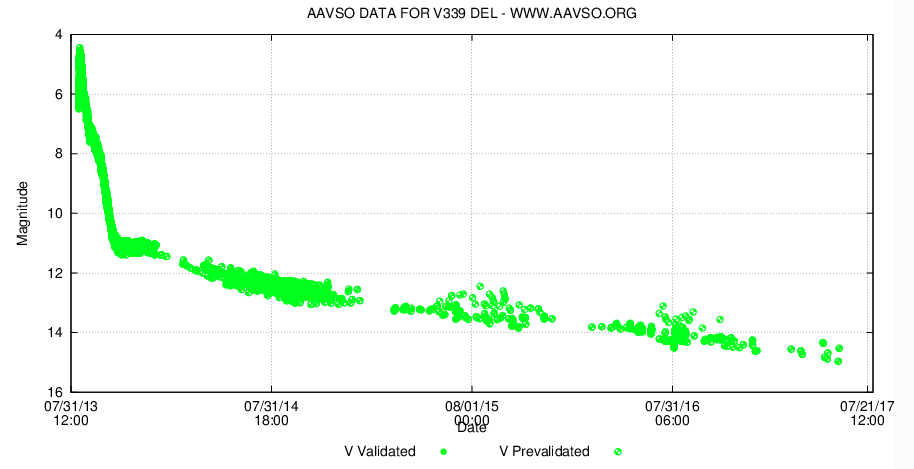}(a)
\includegraphics[width=8cm]{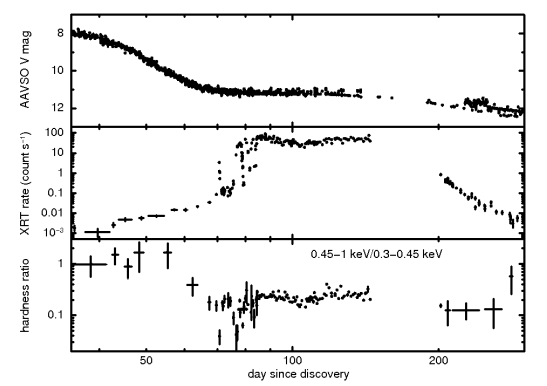}(b)
\caption{(a) Figure showing V band light curve of V339 Delphini is copied from the AAVSO website.
The light curve continues to decline. 
(b) Figure showing the optical and X-ray light curves of V339 Delphini
reproduced from \citet{2016A&A...590A.123S}.  
For comparison, note that radio thermal 
was detected around day 30 and peaked between days 150-200 after optical detection 
and was detectable till day 700 or so (Information from an online talk slide by Linford 2015).}
\label{v339del}
\end{figure}
V339 Del was a fast nova detected on August 14.6, 2013 \citep{2013CBET.3628....1N}
which reached maximum V=4.46 magnitudes on August 16.44 UT and had a $\rm t_2=10.5$ days 
\citep{2013IBVS.6080....1M}.  It was extensively observed across the electromagnetic spectrum.
Here we summarise some observational results from literature and inferences/comments are included in
italics:  \\
(1) The nova was in quiescence with $\rm B\sim17.1$ magnitudes about 14 hours before discovery
\citep{2013CBET.3628....1N} indicating a fast rise.  It was 6.31 magnitudes and 6.18 magnitudes on
August 14.94 and August 15.02 UT 2013 \citep{2013ATel.5288....1T}.
{\it Strong support to instantaneous release of enormous energy in the nova in a single episode. } \\
(2) The colour index $\rm B-V$ changed from about 0.1 to 0.55 between August 15 to August 19.2
which indicated a change in the spectral type from A2 to F8 \citep{2013IBVS.6080....1M}.
The colour index $\rm B-V$ settled
down to near 0 around August 25 \citep{2013IBVS.6080....1M}. {\it Thhe
fractional contribution of the hot white dwarf increases as the ejecta fades and hence
the colour gets bluer as the light curve declines.} \\
(3) There was a short plateau in the light curve from August 17.15 to August 19 UT at V=4.85 magnitudes
when the colour $\rm B-I_c$ changed from 0.7 to 1.17 i.e. the nova got redder at the end of the
plateau and the emission lines got fainter \citep{2013ATel.5304....1M}.
There was a longer plateau of brightness in B and V between days 20 (nova had declined
by 2.8 magnitudes below maximum) and 37 after maximum.  
{\it A plateau in the V band light curve indicates addition of excess V band emission which 
arrests the decline in the light curve.  The plateau indicates new components were added. }
The emission component of the H$\alpha$ line profile
changed from trapezoidal on August 14 to flat-topped Gaussian on August 15.83 to
Gaussian profile on August 16 \citep{2013ATel.5297....1M}. 
{\it This could indicate the changing morphology of the line forming region indicative of
increasing ionization in the ejecta.}  It was a Fe II type nova \citep{2013IBVS.6080....1M}
{\it Although V339 Delphini was a fast nova, the ambient medium due to the main sequence
companion appears to have been sufficiently dense to allow the fast ejecta to sweep 
up sufficient matter for the lines to be detectable.} \\
(4) The absorption component of the H$\alpha$ line was detected at $-1600$ kms$^{-1}$ in
the pre-maximum phase and the displacement of the absorption component declined to 
around $-730$ kms$^{-1}$ in the post-maximum phase on August 19 \citep{2014A&A...569A.112S}.   
The absorption features disappeared by around August 20 while the emission lines got
wider and bands with full width at zero intensity (FWZI) of about 5000 kms$^{-1}$ were 
detectable upto day 40 \citep{2014A&A...569A.112S}. 
{\it The wide emission bands seen after maximum till day 40 probably also included
diffuse enhanced and/or Orion lines.} \\ 
(5) It was surmised that atomic hydrogen was present in the ejecta of V339 Del 
from the detection of the broad emission band at 6825 A which is produced due to Raman-scattering of 
O VI 1032A line photons by neutral hydrogen atoms \citep{2014A&A...569A.112S}.  
The line was detected post-maximum around 17 August and had both an absorption and emission component
till 21 August  after which only an emission component was detected till end September 
i.e. about day 40.  This indicated that the ejecta got completely ionized around day 40 
\citep{2014A&A...569A.112S} when the light curve was about 3.5 magnitudes below maximum. 
We note that radio thermal was detected around
day 23 near 96 GHz {\citep{2013ATel.5382....1C} and soft X-rays started rising around day 40  
(see Figure \ref{v339del}). {\it All three observations are consistently explained since the
detection of soft X-rays is delayed till the atomic hydrogen in the ejecta is ionized and
the increasing ionization leading to increasing emissivity of thermal radio.} 
It was surmised that the optically thick phase in the ejected shell ended around day 77 
\citep{2013IBVS.6080....1M}.  As seen in Figure \ref{v339del} soft X-rays reached maximum around
that time and the optical light curve showed a plateau.  
{\it The plateau in the light curve indicates the addition of an extra component of light 
possibly from the white dwarf as it becomes visible.} \\
(6) Dust formed in the shell after 21 September, 2013 \citep{2013ATel.5431....1S}.
{\it Wide Orion lines were detected before this, indicating the presence of clumps
and the formation of dust in the Orion clumps. }  \\
(7) The nebular phase was established in October 2013 when the strengths of
$\rm H\beta$ and [OIII] were comparable and the
light curve had declined by 6 magnitudes from the peak \citep{2013IBVS.6080....1M}. \\
(8) Mass of the ejecta was estimated to have been $2-3\times10^{-5} M_\odot$.  
The ejected shell appeared to be bipolar in nature
\citep{2016A&A...590A.123S}. {\it This would indicate the rotating white dwarf in V339 Del 
which would lead to the formation of a prolate-shaped accreted envelope and hence the
aspherical morphology of the ejected shell.\\}
(9) Detected in $\gamma-$rays between days 2 and day 11 after the optical peak
\citep{2013ATel.5302....1H, 2014Sci...345..554A}. 
V339 Del was not detected in non-thermal synchrotron on Aug 28 2013 (day $\sim 12$)  
\citep{2013ATel.5376....1R}.  
{\it The detection of energetic $\gamma-$rays for a few days near optical maximum supports the 
hypothesis that low energy $\gamma-$rays from the thermonuclear reaction form
the seed photons which are inverse Compton accelerated to $>100$ MeV energies by relativistic
electrons.  This would confirm the presence of relativistic electrons and the non-detection 
of radio synchrotron could be due to lack of magnetic field or it could be a faint signal.} \\
(10) The light curve of V339 Delphini shows a pre-maximum halt 
around 15 August 2013 before the final rise to maximum.  {\it This would indicate that
the ejecta was transparent to optical emission of the white dwarf which pushed the
light curve to the maximum.  However when combined with the formation of atomic hydrogen
in the ejecta after maximum and the appearance of plateaus in the light curve, it indicates
the changing physical conditions in the ejecta so that a fully ionized ejecta recombined
and its opacity changed.}  \\
(11) Modelling of the evolving spectral energy distribution (SED) of V339 Delphini 
that was carried out by \citet{2014A&A...569A.112S}. 
They fitted the SED from $\sim 3500$ A to $\sim 9500$ assuming that in 
in early times, all the emission was from the
expanding pseudophotosphere of the white dwarf resembling a star of type A to F
and that at a later date some contribution by the emission from the nebula was included. 
They present the evolution of the radius of the white dwarf defined by the extent of
the pseudophotosphere, its effective temperature $T_{eff}$ and its luminosity $L_{WD}$.  
Interestingly, there is a steep increase in $T_{eff}$ and $L_{WD}$ just after the pre-maximum
halt when the light curve embarks on its final rise and which in their modelling also
leads to an increase in the radius of the pseudophotosphere  
\citep[Figure 3 in][]{2014A&A...569A.112S} .  
After the optical maximum, the $T_{eff}$ and $L_{WD}$ decline and then seem to be constant
whereas the radius of the pseudophotosphere continues to rise at least till 20 August. 
{\it In the model presented here, all the energy input prior to the pre-maximum
halt is due to the explosion energy which sets the ejecta in motion and the optical emission
is dominated by the ejecta.  The influence of the white dwarf radiation begins at
the pre-maximum halt when it exerts a radiation pressure on the optically thick ejecta
explaining the higher principal velocities compared to pre-maximum spectrum that are
generally obesrved.  If the
ejecta becomes optically thin in the process, then the white dwarf radiation 
can escape and will contribute to the blue/optical SED thus increasing the observed luminosity
and effective temperature of the nebula.  Interestingly,  an abrupt rise in both $T_{eff}$ and
$L_{WD}$  immediately following the pre-maximum halt is deduced from the evolving SED.   
This gives strong support to the contribution of 
the hot white dwarf to the SED which increase the bluewards emission as is observed. 
We note that the SED indicates a rise in temperature and luminosity after  
the pre-maximum halt which is what is expected in our model. }\\ 

\subsection{Recurrent novae}

From the different onset times of radio synchrotron emission at a given frequency
recorded in two successive outbursts in
RS Ophiuchi and V745 Scorpii, both of which happen to be novae hosting a red giant companion, 
\citet{2016MNRAS.456L..49K} suggested a scenario in which the winds, blown by the
massive white dwarf accreting at super-critical rates, would form a halo around
the nova.   The size and densities of the halo will keep varying with the varying accretion rates 
and the winds are responsible for the free-free thermal absorption of the radio synchrotron emission leading
to the frequency-dependent onsets and varying onset times at the same radio frequency for
different outbursts \citep{2016MNRAS.456L..49K}.   Thus, if the white dwarf in a recurrent nova
experiences super-critical accretion rates between two outbursts, then a dense halo of
ionized material will form around the nova and will delay the onset of synchrotron emission.
When the accretion rate falls then the halo will expand and disperse thus leading to a rapid
onset of synchrotron radio emission.  The onset epoch of the radio synchrotron emission in
successive outbursts helps study the varying accretion rates of the system. 
Due to the multiple processes involved, there will be a time lag between the change in accretion
rates and change in the physical state of the wind halo and hence free-free absorption. 
This would be applicable to all novae with super-critical accretion rates which comprise most
of the recurrent novae. 

The interval between the recorded outbursts in RS Ophiuchi have been  
9, 26, 12, 13, 9, 18, 21 years.   The inter-outburst period seems to have increased in the
last two outbursts indicating a decreasing accretion rate and hence reduced winds.  
The synchrotron radio emission in 2006 at a given radio frequency was detected earlier 
as compared to its previous outburst in 1985
which would indicate that the wind halo is getting tenuous as expected from
the inter-outburst period.  The free-free
absorption is hence decreasing accelerating the onset of radio synchrotron emission.  
The long-term pattern in the inter-outburst
periods suggests that the accretion rates should soon increase and lead to the wind halo
becoming larger, the inter-outburst period reducing (i.e. the next outburst in RS Oph should occur
well before 2032) and the onset of radio synchrotron emission in the next outburst being delayed
compared to 2006.  In the recurrent nova which consistently have a super-critical accretion rate,
the optically thick wind halo extent will keep increasing and will lead to a larger delay in the
onset of radio synchrotron emission with each outburst.  From observations, one can
infer that this is possibly the case for U Scorpii and T Pyxidis with the latter actually
showing the existence of a large halo around it which is optically visible. 
In this case, the radio synchrotron will only be detected at later
dates if the electron population has not aged. 
While it is surmised from observations that accretion rates in novae change, it is not known 
what triggers the variations.  While accretion would be disrupted during the outburst, it is
possible that changes in local physical conditions could also trigger
variations in the accretion rate.  However this needs to be investigated further before 
any firm conclusions can be drawn. 

Recurrent novae generally return to optical quiescence within a year of outburst - much faster 
than classical novae which take at least a few years.  U Scorpii is particularly fast and returns
to the pre-nova phase in about two months and hence there is high likelihood of missing
outbursts when it is in the day sky.  The thermal radio light curves also follow the same trend.
Recurrent novae in quiescence show variability on several timescales with
amplitudes ranging from 1 to 2.5 magnitudes \citep{2010ApJS..187..275S}.
The brightness of some recurrent novae dips a year or so before the outburst while in 
others the brightness dips after the outburst \citep{2010ApJS..187..275S}.
Moreover the range of quiescence luminosities of recurrent novae is
larger than of classical novae \citep{2010ApJS..187..275S} and the contributory
factors could be 
the larger range in types of the companion star - main sequence to red giant in addition
to possible differences between the accreting white dwarfs as shown in Table \ref{tab4}.

\subsubsection{RS Ophiuchi}
\begin{figure}
\centering
\includegraphics[width=6cm]{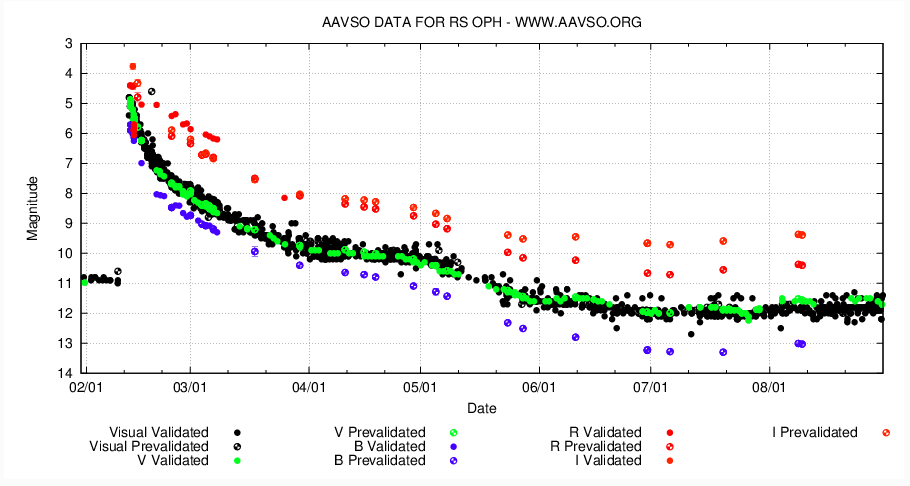}(a)
\includegraphics[width=6cm]{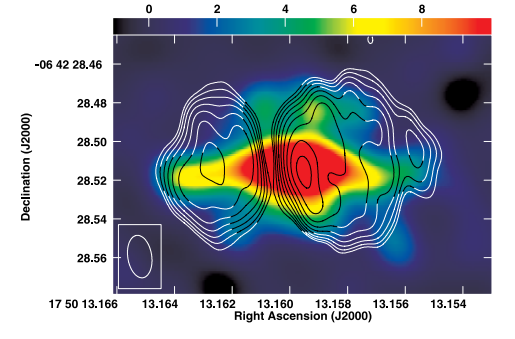}(c)
\includegraphics[width=6cm]{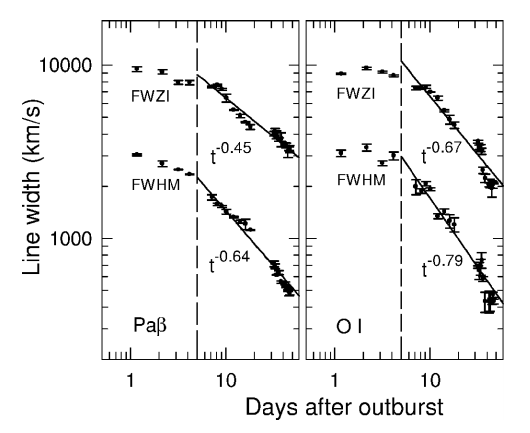}(b)
\includegraphics[width=5cm]{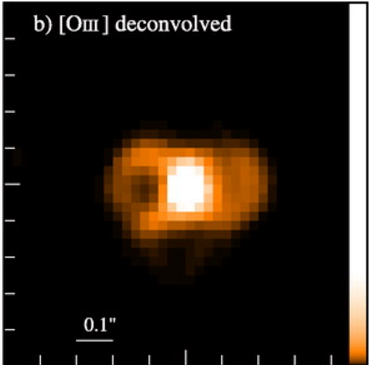}(d)
\caption{(a) Light curve of RS Ophiuchi in its 2006 outburst copied from the AAVSO website.
Note the rapid rise and the rapid initial decline. 
(b) Radio image of RS Ophiuchi at 43 GHz (colour) and at 1.7 GHz (contours) about 7-8
weeks after outburst reproduced from \citet{2008ApJ...685L.137S}. Thermal emission
in central parts is surrounded by non-thermal emission.
(c) Figure showing the rapid evolution in the line widths of $\rm Pa\beta$ and
$\rm OI$ reproduced from \citet{2006ApJ...653L.141D}.  This behaviour is very different from
classical novae which show constant widths of the
principal spectrum detectable from near optical maximum to nebular phase and beyond.
(d) Image of the [OIII] 5007 emission on day 155 in shown in this figure reproduced 
from \citet{2007ApJ...665L..63B}.  Note the double ring structure and the extent is
similar to the synchrotron radio emission.} 
\label{rsoph}
\end{figure}
The last two outbursts in RS Ophiuchi occurred in 1985 and 2006 and have been well studied
at multiple wavelengths.  RS Ophiuchi has also been studied in the inter-outburst
period.  We examine the observational results: \\ 
(1) The optical light curves are similar in all outbursts in terms of maximum luminosity
and decline characteristics (see Figure \ref{rsoph}a).  \\
(2) Radio emission at early times shows a bipolar morphology 
extending along the east-west which indicates an expansion rate of about 1.3 mas/day as noted
in both the 1985 and 2006 outbursts
\citep{1986ApJ...305L..71H, 1989MNRAS.237...81T, 2006Natur.442..279O, 2008ApJ...685L.137S}.
The outer parts of the east-west emission is attributed to the synchrotron process and this
is substantiated from the spectral index of the eastern lobe between 1.7 and 5 GHz which was estimated
to be $\sim 0.5$ ($S \propto \nu^{-\alpha}$) \citep{2008ApJ...685L.137S}.  The
extended east-west emission fades faster at higher radio frequencies while the
central part of the radio emission is thermal and longer lived so that at later dates, only the
central emission is detectable \citep{2008ApJ...685L.137S} (see Figure \ref{rsoph}b) and this
should arise in the main ejecta of the explosion.  
{\it The bipolar-shaped synchrotron emission indicates a rotating 
white dwarf in RS Ophiuchi which would have accreted a. 
prolate-shaped envelope and formed an accretion disk in the equatorial plane.  The origin
of radio synchrotron at a further distance from the binary compared to thermal emission indicates
a faster ejection speed for the relativistic electrons especially from the poles. 
If the accretion disk is destroyed in the outburst then
the radio synchrotron emission even if present in the equatorial plane might remain undetectable
due to enhanced free-free absorption.\\}
(3) There was a rapid decline in the ejecta velocity as deduced from the fast evolving widths of
the Pa$\beta$ and O I lines \citep{2006ApJ...653L.141D}.   The initial FWHM of few thousand
kms$^{-1}$ had dropped to a few hundred kms$^{-1}$ around day 50 (see Figure \ref{rsoph}c).  
A triangular-shaped H$\alpha$ with FWZI of about 7600 kms$^{-1}$ was detected on day 1.38 
which narrowed down to a FWHM of 420 kms$^{-1}$ on day 57 and was $\sim 100$ kms$^{-1}$
on day 209 after outburst \citep{2008ASPC..401..227S}. 
{\it This rapid decline in high initial ejecta velocities is commonly observed in recurrent
novae and rarely in classical novae.  This supports the presence of high densities in the 
circumstellar environment due to the winds blown by the white dwarf which leads to
deceleration of the ejecta.} \\ 
(4) Emission in [O III] 5007A and [Ne V] 3426A was resolved into 
a east-west oriented double ring around the nova following its 2006 outburst (see Figure \ref{rsoph}d) and 
their extent indicated an expansion rate of about
1.2 mas/day  \citep{2007ApJ...665L..63B}.  The extent of the [O III] 
rings appear to be coincident with the radio synchrotron emission (see Figure \ref{rsoph}b,d).  
The double ring structure in [O III] imaged on days 155 and 449 after outburst  
is well fitted by a model containing an outer faster moving bipolar outflow and an 
inner slower moving denser ejecta with more emission arising in the inner 
ejecta \citep{2009ApJ...703.1955R}.
The image of day 449 is consistent with a linear expansion of the bipolar outflow from day 155 and 
no expansion of the inner ejecta \citep{2009ApJ...703.1955R}.  
{\it The central region is the main ejecta from the explosion as traced by the radio thermal
emission also.  The synchrotron emission locates the relativistic electrons which precede
the main ejecta and could have been swept out by the blast wave after the explosion.
The co-existence of [O III] emission with synchrotron emission 
can be due to either pre-existing matter excited by the electrons or some of the material which
was also carried out by the blast wave.} \\
(5) Soft X-rays were strong around day 55 after the
outburst in 1985 and declined soon after that  \citep{1987rorn.conf..167M}.
Soft X-rays were detected on day 26 after the outburst in 2006 when the light curve had
declined by more than 4 magnitudes, strengthened on
day 29 and started to decline around day 60 \citep{2006ApJ...652..629B, 2011ApJ...727..124O}.
{\it The similar behaviour of X-rays in the 1985 and 2006 outbursts is expected from similar
ejecta properties of both outbursts which is expected from the similar behaviour of optical light curves.
One can infer that the ejected mass is comparable in all outbursts and the different accretion
rates only lead to a varying inter-outburst period and wind halo extents.  In both epochs,
accretion restarted before day 60. \\} 
(6) Hard X-rays (14-25 keV) were detected for only six days after discovery in 
2006 \citep{2006ApJ...652..629B}.
Radio synchrotron emission was detected in the first observation at 5 GHz on day 4 after outburst in
2006 when it was dominated by synchrotron emission
and by day 13, the
radio emission was a combination of thermal and non-thermal emissions \citep{2009MNRAS.395.1533E}.
This is also inferable from the result that the east-west size of the radio source
was about 200 mas on day 63 at 1.7 GHz whereas it was 90 mas on day 59 at 22 GHz \citep{2009MNRAS.395.1533E}.
{\it Hard X-rays if mainly due to the relativistic electrons will show some correlation with
radio synchrotron detection.  The radio size differences indicate the presence
of two distinct emitting regions as mentioned above and which also fits the optical observations. \\}
(7) No evidence of dust formation is found in the ejecta.
Features due to silicate dust detected between days 208 to 430 after outburst 
were interpreted as being from the circumnova environment (Evans et al. 2007). 
{\it Since RS Ophiuchi is a fast recurrent nova, its evolution is rapid and 
there might not be sufficient times for clumps and subsequently dust to form. \\}
(8) From the radio (1.5 to 23 GHz) study of RS Ophiuchi during its outburst in 1985
\citep{1986ApJ...305L..71H} and study of classical novae \citep[e.g.][]{1979AJ.....84.1619H}, 
four main differences between the two were pointed out by \citet{1986ApJ...305L..71H}: 
(a) brightness temperatures for RS Ophiuchi $>>10^4$ K
with it being highest at 1.5 GHz,
(b) light curve had timescales $< 1$ yr whereas it took several years or decades for classical
novae to go to quiescence, (3) decay power laws with index in the range $-1$ to $-2$ rather 
than the $-2.3$ to $-2.5$ generally
noted for classical novae and (4) spectral index of radio emission during the decay was different
from $-0.1$ which is generally estimated for classical novae.   \citet{1986ApJ...305L..71H} inferred that
the radio emission from RS Ophiuchi likely consisted of two components - a synchrotron emission
and a gyrosynchrotron emission component. {\it These were interesting first results.
Subsequent studies have brought out the different
locations of the two components and also determined them to be synchrotron and 
thermal emission.} \\

\subsubsection{T Pyxidis}
\begin{figure}[t]
\includegraphics[width=8cm]{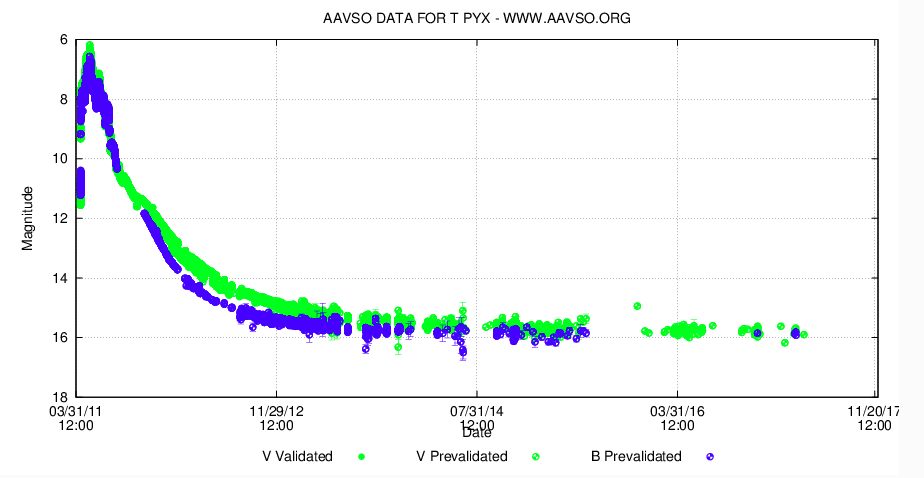}(a)
\includegraphics[width=8cm]{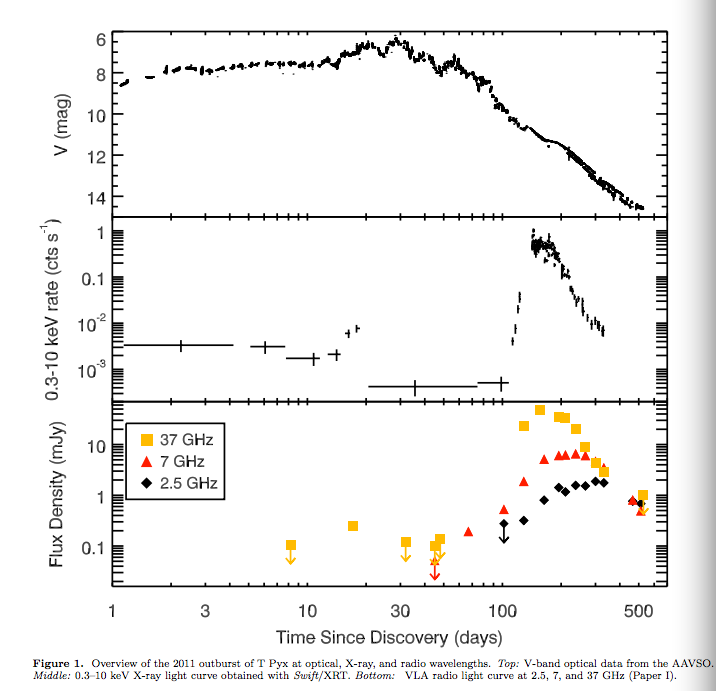}(b)
\caption{(a) Light curve of T Pyx copied from the AAVSO website.
(b) Figure showing the optical, X-ray and radio light curve of T Pyx during its 2011 outburst
reproduced from \citet{2014ApJ...788..130C}.  Note the tight correlation between 
onset of soft X-rays and thermal radio emission.}
\label{tpyx}
\end{figure}
The last outburst in T Pyx which was the first nova in which recurrent outbursts were recorded,
was detected in the pre-maximum phase on 14 April 2011 and reached maximum brightness around 
12 May 2011.  The previous outbursts in T Pyx were recorded in 1890, 1902, 1920, 1944, 1966 
with inter-outburst periods of 12, 18, 24, 22, 45 years.   
Some of the main observational features of T Pyx which have been surmised are: \\
(1) T Pyx is a slow nova and had  a long pre-maximum halt from day 3.3 to day 13 after detection
on 14 April 2011 before it peaked at 6.33 magnitudes on 12.22 May 2011 
(day 27.9 after discovery) \citep{2014AJ....147..107S}.  It 
spent almost three months near maximum emission (see Figure \ref{tpyx}).  The four main
phases of a typical light curve evolution (Figure \ref{lightcurve}) have been identified
in the light curve of T Pyx and the detailed spectral evolution from the initial rise
to transition stage has been presented in \citet{2014AJ....147..107S}.  The nova is 
a Fe II type.  Since T Pyx was detected in its initial rise which is
rare it gave a glimpse into the elemental composition and the spectrum prior to the pre-maximum
halt.  The spectra taken between days 0.8 and 2.7 showed the presence of high excitation
emission lines of C~III, N~III, Ne~II, O~II, N~II, He~I and Ne~I which disappeared between 3 to 14
days (i.e. in the pre-maximum halt phase) and the Fe II lines started getting stronger
\citep{2014AJ....147..107S}.  {\it This interesting
result which is rarely studied due to few novae being detected in the initial rise phase would
indicate the composition of the ejecta before the heavier elements started sinking to the rear
part of the ejecta and the ejecta had swept up sufficient circumstellar material to show
their spectral signatures.  }
The ejecta velocity derived from the  P Cygni profiles of Balmer lines and Fe II lines
does show a tendency for the latter estimates to be lower than the former between days 3.3 to about
day 20.  Near maximum the ejecta velocities appear well-matched 
and post-maximum, the Balmer line velocities increase \citep[Figure 10 in][]{2014AJ....147..107S}.  \\
(2) T Pyx shows a light curve and spectrum
evolution typical of classical novae \citep{1960stat.conf..585M,2014AJ....147..107S}. \\
(3) Following its outburst in 1920, three strongly displaced absorption systems with velocities
between $-1000$ and $-1900$ kms$^{-1}$ were detected from T Pyx (Adams and Joy 1920) and which
corresponded to the principal, diffuse enhanced and Orion systems \citep{1960stat.conf..585M}. \\
(4) In its 2011 outburst, the ejection velocity estimated from the Balmer lines of hydrogen was 
$\sim 4000$ kms$^{-1}$ about 0.8 days after detection, which declined to $\sim 2000$ kms$^{-1}$ 
after 2.7 days and then remained constant $\sim 1500$ kms$^{-1}$  till near maximum
(i.e. 28 days after discovery) after which it increased \citep{2014AJ....147..107S}.
{\it The dramatic reduction in the initial ejecta velocities is commonly observed in recurrent
novae likely indicative of a dense circumstellar medium due to the winds blowing from the
white dwarf.  One would note 1500 kms$^{-1}$ as the
pre-maximum velocity and the increase at maximum as the principal spectrum velocities.  The
original expansion velocities were much higher as expected for a massive white dwarf}   \\
(5) The outburst showed evidence for a bipolar ejecta \citep{2011A&A...534L..11C}. 
{\it The white dwarf in T Pyx is rotating, has an accretion disk and a prolate shaped accreted envelope 
which gives rise to the bipolar ejecta. \\}
(6) Soft X-ray emission was detected around 7.5h after discovery in the initial rise
\citep{2011ATel.3285....1K} which
faded around day 12 (see Figure \ref{tpyx}b).  Interestingly this early detection of
faint soft X-rays coincides with the detection of
high ionization lines described in point 1 above \citep{2014AJ....147..107S}. 
{\it This indicates that the white dwarf had ejected
the accreted envelope soon after the explosion and exposed the hot underlying surface which was emitting
in soft X-rays.  The pre-maximum ejecta was fully ionized and transparent to soft X-rays upto about
day 12 which also increased the excitation of the ejecta.}  
T Pyx started brightening again in
soft X-rays from day 117 \citep{2011ATel.3549....1O}.  This renewed X-ray activity 
consisted of supersoft X-rays of energy 30-50 eV \citep{2014ApJ...788..130C} (see Figure \ref{tpyx}b).  
This also coincided with a period of reduced rate of decline in the light curve at the end of
which the soft X-rays started declining.  Thermal radio emission was detected around day 69 
and was detectable for more than 500 days \citep{2014ApJ...785...78N}.   In fact the steep rise in
soft X-rays coincided with a steep increase in the radio thermal emission \citep{2014ApJ...788..130C}
as shown in Figure \ref{tpyx}b.  {\it The near coincidence of the increase in thermal radio emission 
from the ejecta and X-ray emission from the white dwarf surface is expected in the model since
X-rays ionize the ejecta and propagate outwards while the increasing
ionization leads to increasing free-free emissivity and detectability in thermal radio. \\}
(7) Hard X-rays were detected on days 14-20 after discovery \citep{2014ApJ...788..130C} and there was 
an isolated radio detection at 33 GHz on day 17 \citep{2014ApJ...785...78N}.  
T Pyx was not detected in $\gamma-$rays \citep{2014ApJ...788..130C} or in radio
synchrotron \citep{2012ATel.4452....1R}.
{\it  The non-detection of $\gamma-$rays and radio synchrotron emission could either be due to
lack of relativistic electrons or deep absorption in the large dense nebula 
surrounding the nova.  If hard X-rays is due to a leptonic process such as synchrotron or
inverse Compton also then their detection indicates that relativistic electrons did exist. \\}
(8) The Balmer lines of hydrogen H$\alpha$, H$\beta$, H$\gamma$ displayed P Cygni profiles in
the spectra taken soon after discovery to about 48.6 days \citep{2014AJ....147..107S}.
Around day 42, the Balmer lines started showing a double component profile and the emission lines
of Fe II and Ca II disappeared around day 48 and the Orion line features started appearing
around day 70 \citep{2014AJ....147..107S}}.
The 4640 emission which is generally seen towards the end of the Orion stage was apparent between
70 and 80 days \citep{2014AJ....147..107S}.  
The H$\beta$ lines continued showing P Cygni profiles at least upto day 69
\citep{2014ApJ...785...78N}.  Alongside the existing absorption near $-1900$ kms$^{-1}$ a new 
absorption feature $\sim -3000$ kms$^{-1}$ is seen to be present in a spectrum from 
day 69 \citep{2014ApJ...785...78N} when the optical light curve was about 2 magnitudes 
below maximum and would indicate the appearance of the Orion spectral component.  
The spectrum observed on day 155 was nebular with detection of several coronal and forbidden lines 
and the light curve had faded by five magnitudes  
\citep{2014AJ....147..107S}.  {\it  The higher velocity components would be diffuse enhanced or Orion
features.  Lines with double components is typically observed in diffuse enhanced lines 
and the highly displaced component at $-3000$ kms$^{-1}$
detected on day 63 was probably an Orion line since the 4640 band was detected between
days 70 and 80 when the light curve had faded by about 2 magnitudes from maximum. } \\
(9) An expanding shell of radius $\sim 5''$ surrounded by a faint halo of radius $10''$ has been 
detected around T Pyx \citep{1979Msngr..17....1D,1997AJ....114..258S,2010ApJ...708..381S}.
Its spectrum resembles that of a planetary nebula with solar abundances 
\citep{1982ApJ...261..170W} and does not show CNO enhancement.   We note that CNO enhancement
is commonly observed in nova shells.  The shell is hot and consists of numerous knots with
expansion velocities between 500 and 715 kms$^{-1}$ for a distance of
3.5 kpc to the nova \citep{2010ApJ...708..381S}.  It has been suggested that the shell and halo 
are remnants of previous outbursts.  
{\it As noted earlier, the supercritical accretion rates on the white dwarf in recurrent novae
can set up vigorous winds and the shell+halo might be a result of the matter blowing out as winds.
The matter in the wind would then principally consist of matter accreted from the companion star which
has not undergone any processing on the white dwarf and hence should result in solar-like abundance as 
observed.  The extended emission around T Pyx of radius  10" corresponds to a size 
of 0.17 pc at a distance of 3.5 kpc.  If the winds started blowing in 1890 at a mean speed of
$\sim 1000$ $\rm kms^{-1}$ then they would have travelled
out to about 0.1 pc in the 120 years between 1890 and 2010.  Alternatively the winds could
have been quasi-continuous with varying speeds. 
The radial inhomogeneties observed in the shell would support the
varying wind rates and hence varying accretion rate of T Pyx, mostly hovering near the
supercritical rates. }  \\
(10) In the 1966 outburst, a slowly evolving light curve and line velocities of 900 and 2000 kms$^{-1}$ 
were reported and the spectral development through diffuse enhanced and Orion spectra was also noted
\citep{1969MNRAS.142..119C}.   \\
(11) The quiescent brightness of T Pyx has been decreasing so that it is fainter by 1.9 magnitudes
in B band than it was 120 years ago and this has been attributed to decreasing accretion
rates \citep{2010ApJ...708..381S}.  {\it A possible reason for fading could be the winds blowing
from the white dwarf.  
If there was dust formation in the winds then it could lead to fading of B band emission
but brightening in the infrared.  \\}
(12) T Pyx is not a soft X-ray source in quiescence and its optical emission spectrum
is of low excitation compared to several other novae like V Sge \citep{2008A&A...492..787S}.
{\it Novae in quiescence are not expected to be soft X-ray emitters due to the accreted
envelope which will be cooler ($<10^5$K) and hence not emit X-rays.  }\\
(13) Ejected mass in the 2011 outburst is estimated to be $10^{-4}$ to $10^{-5}$ M$_\odot$
\citep{2014ApJ...788..130C} which is similar to classical novae but about two orders of magnitude 
higher than generally surmised in recurrent novae.  For an inter-outburst period of
45 years, this translates to an accretion rate of $2.2\times10^{-6}$ to
$2.2\times10^{-7}$ M$_\odot~yr^{-1}$.  {\it Such high accretion rates are close to or exceed
critical rates for the massive white dwarf and support blowing of intense winds 
which accumulate around the white dwarf as the halo.} \\

\subsection{Evolution of novae to SN 1a}
This section speculates on the different possibilities regarding evolution of the binary stars in novae.
Let us consider the binary system in a nova which consists of a white dwarf primary and
a Roche lobe-filling main sequence companion.  If the separation between the stars has not
changed significantly since their birth, then it would imply that in its red giant phase,  
the white dwarf engulfed the companion star even as it evolved to a white dwarf and this left the companion
star unaltered.  The alternate scenario is 
that the two stars were at a larger separation when the white dwarf was in its red giant phase
and have subsequently lost orbital angular momentum which has reduced the separation.  While
the mass loss in the red giant phase can carry away some of this angular momentum, the rest
could have been converted into a spin component of the white dwarf increasing its rotation
speed as the binary has become compact. 
The next issue to address would be the evolution of the binary when the secondary star will
enter the red giant phase and engulf the white dwarf assuming that the separation will
remain the same.  This could enhance the
accretion rates on the white dwarf as the giant star loses mass and could either
initiate coalescence of the cores to a neutron star or some of the accreted matter can enter
the hot degenerate core and explosively detonate.  In the latter case, the energy input
to the core would lift the degeneracy 
and disrupt the white dwarf which we would view as a supernova type 1a. 
Alternatvely the white dwarf might accrete the entire outer envelope of the secondary
thus forcing the secondary to enter the white dwarf phase without passing through a giant phase.   
In this case a double degenerate system can result and if these spiral in then again a
supernova type 1a might result. 

Then there are novae which host a red giant companion and the orbital separation is large.   
If there has been no change in the separation between the binary members then the white dwarf in
its red giant phase would have filled up its Roche lobe while the companion star which
would have been a  main sequence star would have had a smaller Roche lobe.  The white dwarf
would have gradually lost its outer envelope even as the secondary evolved to a red giant
phase so that the binary as detected by us show the presence of a white dwarf and red giant
companion.  If the white dwarf is massive then it is possible that some of the accreted
matter will seep into the degenerate core and explode as a SN 1a.   
On the other hand it is possible that the red giant companion will lose its 
outer envelope accelerated due to accretion by the white dwarf.  Eventually the
secondary will settle down as a white
dwarf giving rise to a wide double degenerate system.  It is difficult to comment on further
evolution of such a system. 

Thus while it appears possible that some novae might evolve to supernova explosions, we
need to examine relevant observational data spanning novae to other binaries 
to be able to derive any firm conclusions. 

\section{Conclusions}
This study was aimed at understanding the 
nova outburst that optically brightens the binary system consisting of a white dwarf and a 
gaseous companion star by 8-20 magnitudes.  With the availability of multi-band observing
facilities, the multi-wavelength observational results on novae can be used to build a
consistent model.  Using all these data, the existing model commonly
used to explain nova outbursts has been updated and is presented in the paper. 
This study has also resulted in several significant 
inferences which are applicable to other astrophysical systems. 
One of the significant results is the pointing out that the explosion should adiabatically energise the
overlying layers increasing their internal energy and that equal distribution of this energy to all particles 
which includes electrons will result in the electrons acquiring
relativistic velocities due to their small mass.  No shock acceleration needs to be invoked for this.

The main conclusions are: 

\noindent
1. Several old classical novae ($\rm <M_{V,q}> \sim 2.6$ magnitudes) are brighter than the combination
of an isolated white dwarf 
($\rm M_V \sim 11$ magnitudes) and a late type main sequence star ($\rm M_V > 4.6$ magnitudes).
It is suggested that this is expected since an accreting white dwarf should be surrounded 
by a cooler accreted envelope.  If this envelope is
$\ge 10$ times the radius of the core of a white dwarf and radiating black body emission
then it can enhance the black body emission from the white dwarf by a factor $\ge 100$ and hence explain the
discrepancy.  If this is a realistic scenario, then no soft X-rays should be detectable from
old novae which is indeed the case.  Soft X-rays are detected during the outburst which
is expected since the outer cooler envelope is ejected in the explosion exposing the hot
($\sim 10^5$ K) surface of the white dwarf.
The soft X-ray phase ends in different novae at different times which we suggest can
be consistently explained as the time taken by the white dwarf to accrete sufficient matter to
form a cooler envelope so that it no longer shines in soft X-rays.
Thus, a large accreted cooler envelope around the white dwarf defines its physical state in
quiescence.  Further evidence to the presence of a brighter white dwarf in a nova is presented using data 
on eclipsing novae. 

\noindent
2. The updated model to explain novae in a nutshell is:
The base of the accreted envelope will be compressed and heated.  When it reaches temperatures
$\ge 10^8$ K then an energetic CNO explosion will be detonated, the huge energy release will
be adiabatically transmitted to the overlying layers of normal matter which will acquire
radial velocities in excess of the escape velocity of the white dwarf and be violently ejected. 
The ejected matter will consist of electrons, ions and atoms and assuming
that each particle is equally energised, the light electrons would have acquired relativisitic
random velocities especially if the ejecta expansion velocity $\ge 10000$ kms$^{-1}$.
Detection of radio synchrotron outside the main ejecta indicates that relativistic electrons
could have been ejected with the blast wave or could have acquired higher expansion velocities
thus preceding the main ejecta.  The energy input to the envelope would effectively result
in two velocity components per a particle species namely an outward expansion velocity and a random component.
The explosion should result in a blast wave and a shock wave should set up by the supersonic
ejecta which is often referred to in literature as the reverse shock.  The location of the relativistic
electrons would define the position of what is known in literature as the forward shock.

The rapid rise in the optical emission is owing to the instantaneous energising of the main
hydrogen-rich ejecta which leads to rapidly decreasing optical depths due to expansion, 
heating and ionization thus triggering emission by the free-free and free-bound processes. 
An important change from the previous model is that in this model the
optical continuum emission near the maximum is predominantly from
the ejecta.  This is strongly supported by the empirical result that
most novae follow the maximum magnitude relation with decline time (MMRD) which strongly
advocates the common origin of the optical continuous and spectral line emission.  The spectral lines
are known to arise in the ejecta. 

In the updated model, all the energy input till the pre-maximum halt is from the explosion which is often
characterised by the velocity displacement of the pre-maximum spectrum and the brightness of the nova.  
However it is important to keep in mind that several novae show a decline in the expansion velocity
in the initial rise phase indicating that the explosion energy was higher by a factor of few than 
that estimated from the pre-maximum velocity displacement.
Later additions from the radiation field of the white dwarf in form of radiation
pressure and excitation of lines or continuing winds constitute a trivial fraction of the total energy.   
Observations indicate that the radiation field of the white dwarf contributes the
last $\le 2$ magnitudes rise to the optical maximum before the light curve begins its decline.  
The ejecta fades as it expands, the uniform density component
goes optically thin and its emission measure drops.    
Another important difference in our model is that the diffuse enhanced and Orion systems of
spectral lines are shown to form in clumps which coalesce due to mass-based segregation in the ejecta. 
The optically thick clumps are formed in the inner parts of the ejecta and are subject
to the radiation field of the white dwarf.  
The higher velocity displacement and wider emission lines of these systems are well explained
by the action of radiation pressure on the optically thick clumps accelerating them
to a range of velocities while the higher excitation is
explained as being caused by the radiation field of the white dwarf shining on the clumps. 
The insides of these clumps are shielded from the harsh radiation field of the white 
dwarf and contain metals making them ideal formation sites for dust. 
These clumps are pushed forward in the ejecta with 1.5 to 2 times the principal velocities due
to the radiation pressure exerted by the hot white dwarf. 
In slow novae where there is sufficient time to form dust inside these clumps, the dust can
obscure the optical emission from the entire ejecta and white dwarf and lead to a deep minimum
indicating that the clumps have moved to the front of the ejecta.  The dust disperses as
the ejecta enters the nebular phase and the light curve revives 
continuing its expected slow decline to minimum.  As the light curve declines, the fractional
contribution of the hot white dwarf radiation field goes up whereas that of the cooler ejecta
goes down which explains the increasing colour temperature
during the decline.  The high excitation and photoelectric
temperatures are indicative of the explosion energy in the early times and
of the hot white dwarf at later times and hence are always high.  

Most classical novae take several years to go back to their
pre-nova  brightness which as pointed out in the previous point indicates the time it
requires to accrete an envelope similar to their pre-eruption size. 
It is also suggested that the envelope size has be a combined function of the mass
of the white dwarf and the orbital separation but which in some cases can be modified
by an outburst leaving behind a dwarf nova.  It is suggested that dwarf novae outbursts are powered by 
a low energy thermonuclear explosion on the surface of the accreting white dwarf which
isothermally inflates the accreted envelope to $\le 10$ times the radius of the core of the white dwarf
so that the nova brightens and then declines when the envelope deflates.   The quiescent and
maximum size of this envelope will be a function of the orbital separation in dwarf novae,
which is supported by the observed correlations. 

\noindent
3. An ellipsoidal or bipolar shell left behind by the nova outburst is often observed.  We explain this
as proof of a rotating, accreting white dwarf in the nova.  Rotation of the white dwarf leads to
a latitude-dependent effective potential due to the combined effect of the gravitational and centrifugal 
forces that will be felt by the infalling particles.   This translates to a latitude-dependent
accretion rate such that it is maximum at the poles and minimum at the equator. 
This will lead to the formation of a prolate-shaped accreted envelope around the white
dwarf (see Figure \ref{prolate}) due to higher mass accretion at the poles as compared to
the equator.  This aspherical envelope will be ejected in a nova explosion.  
The lower mass accretion rates in the non-polar regions will lead to the infalling mass accumulating
in an accretion disk around the white dwarf.  This explains the 
formation of the accretion disk around rotating accreting objects and the tight correlation
observed between the presence of an accretion disk and bipolar ejection.  Both the formation of
an accretion disk and bipolar ejection from compact objects result from latitude-dependent accretion rates.    
The existing explanation for the formation of an accretion disk attributed to the angular momentum
of the infalling particles is inadequate. 

In case of a non-rotating accreting spherical white dwarf, the potential felt by the infalling particle
will be the same over the entire surface and hence a spherical envelope will form around the
white dwarf.  The nova outburst will lead to a spherical ejecta. 

This treatement will be applicable to several other astrophysical systems which form accretion disks. 

\noindent
4. Astronomical objects, although they reside at humongous distances,
give us a near-perfect demonstration of the validity of the laws of physics that have been
derived over centuries.  These, alongwith the electromagnetic
signals that these objects emit and their gravitational footprint allow us to
probe the universe in which we reside.

\section*{Acknowledgements}
I gratefully acknowledge using ADS abstracts, arXiv e-prints, AAVSO data, gnuplot, LaTeX, 
xfig, Wikipedia and Google search engines enabled by the internet and the world wide web,  
in this research.  If you happen to use any of the figures copied from literature, please
credit the original reference since I have only used these to better demonstrate a point. 

\bibliography{novae2016}

\end{document}